\pdfoutput=1
\documentclass[11pt,twoside,a4paper,cmspaper,final,collab]{cms-tdr}
\def\svnVersion{3efbb10}\def\svnDate{2025/08/20}\def\cmsCernNoTag{CERN-EP-2024-284}\def\cmsCernDate{\today}\def\cmsMessage{Published in the Journal of High Energy Physics as \href{http://dx.doi.org/10.1007/JHEP08(2025)006}{\doi{10.1007/JHEP08(2025)006}.}}
\begin{document}\cmsNoteHeader{HIN-21-015}

\newlength\cmsTabSkip\setlength{\cmsTabSkip}{1ex}

\providecommand{\cmsTable}[1]{\resizebox{\textwidth}{!}{#1}}

\newcommand{\photos}{\textsc{photos}}
\newcommand{\gammaUPC}{{\texttt{gamma}-\textsc{upc}}}
\newcommand{\starlight}{\textsc{starlight}}
\newcommand{\superchic}{\textsc{superchic}}
\newcommand{\gaga}{\PGg\PGg}
\newcommand{\mgg}{\ensuremath{m^{\gaga}}}
\newcommand{\Aco}{\ensuremath{A_{\phi}}}
\providecommand{\mub}{\unit{\ensuremath{\mu}b}\xspace}

\cmsNoteHeader{HIN-21-015}
\title{Measurement of light-by-light scattering and the Breit--Wheeler process, and search for axion-like particles in ultraperipheral PbPb collisions at \texorpdfstring{$\sqrtsNN = 5.02\TeV$}{sqrt(s[NN]) = 5.02 TeV}}
\author{The CMS Collaboration}

\date{\today}

\abstract{
Measurements of light-by-light scattering (LbL, $\gaga\to\gaga$) and the Breit--Wheeler process (BW, $\gaga\to\EE$) are reported in ultraperipheral PbPb collisions at a centre-of-mass energy per nucleon pair of 5.02\TeV. The data sample, corresponding to an integrated luminosity of 1.7\nbinv, was collected by the CMS experiment at the CERN LHC in 2018. Events with an exclusively produced $\gaga$ or $\EE$ pair with invariant masses $m^{\gaga,\Pe\Pe}> 5\GeV$, along with other fiducial criteria, are selected. The measured BW fiducial production cross section, $\sigma_\text{fid} (\gaga\to\EE)= 263.5 \pm 1.8\stat \pm 17.8 \syst\mub$, as well as the differential distributions for various kinematic observables, are in agreement with leading-order quantum electrodynamics predictions complemented with final-state photon radiation. The measured differential BW cross sections allow discrimination between different theoretical descriptions of the photon flux of the lead ion. In the LbL final state, 26 exclusive diphoton candidate events are observed compared with $12.0 \pm 2.9$ expected for the background. Combined with previous results, the observed significance of the LbL signal with respect to the background-only hypothesis is above five standard deviations. The measured fiducial LbL scattering cross section, $\sigma_\text{fid} (\gaga\to\gaga)= 107 \pm 24 \stat \pm 13 \syst\unit{nb}$, is in agreement with next-to-leading-order predictions. Limits on the production of axion-like particles coupled to photons are set over the mass range 5--100\GeV, including the most stringent limits to date in the range of 5--10\GeV.}

\hypersetup{
pdfauthor={CMS Collaboration},
pdftitle={Measurements of the light-by-light scattering and the Breit--Wheeler processes, and searches for axion-like particles in ultraperipheral PbPb collisions at sqrt(s[NN]) = 5.02 TeV},
pdfsubject={CMS},
pdfkeywords={Light-by-light, ALPs, QED, CMS, UPC, photon-photon collisions, PbPb}}

\maketitle

\section{Introduction}

The electromagnetic field of any charged particle at high energies can be interpreted, in the equivalent photon approximation (EPA)~\cite{vonWeizsacker:1934nji,Williams:1934ad,Fermi:1925fq}, as a flux of quasireal photons~\cite{Brodsky:1971ud,Budnev:1975poe} whose longitudinal energy is proportional to the beam Lorentz factor, $\gamma_\mathrm{L}$, and whose intensity is proportional to the square of the radiating electric charge, $Z^2$. The study of high-energy photon-photon ($\gaga$) processes started in $\EE$ and electron-proton collisions decades ago~\cite{Vermaseren:1982cz,Morgan:1994ip,Whalley:2001mk}, but has received a strong boost in the last decade thanks to the greatly increased centre-of-mass (c.m.) energies and luminosities accessible in collisions with hadron beams at the CERN LHC. Furthermore, the possibility of accelerating not just protons but heavy ions with charges up to $Z=82$ for lead (Pb) ions, has enabled a multitude of novel $\gaga$-collision measurements in proton-proton, proton-nucleus, and nucleus-nucleus ultraperipheral collisions (UPCs), as anticipated in Refs.~\cite{Krauss:1997vr,Baltz:2007kq,dEnterria:2008puz,deFavereaudeJeneret:2009db}. 

Since the photons are coherently emitted by the whole charge distribution of the Pb ion, their EPA fluxes have very low virtuality, $Q^{2} < 1/R^{2} \approx 10^{-3}\GeV^2$ for a Pb radius of $R\approx 7\unit{fm}$, and reach longitudinal photon energies $E^\PGg=\gamma_\mathrm{L}/R\approx 100\GeV$ at the LHC. The photon-photon luminosities associated with PbPb UPCs are enhanced by factors of up to $Z^{4}\approx 5\times 10^{7}$ compared with similar proton-proton or electron-positron interactions. These facts open up new possibilities, beyond the typical heavy ions research topics~\cite{CMS:2024krd}, to study very rare standard model (SM) photon-photon processes, such as light-by-light (LbL) scattering ($\gaga\to\gaga$)~\cite{d'Enterria:2013yra}, and enable searches for new particles beyond the SM (BSM) that couple preferentially to photons~\cite{Bruce:2018yzs,dEnterria:2022sut}.

\begin{figure}[hbtp!]
\centering
\includegraphics[width=0.75\textwidth]{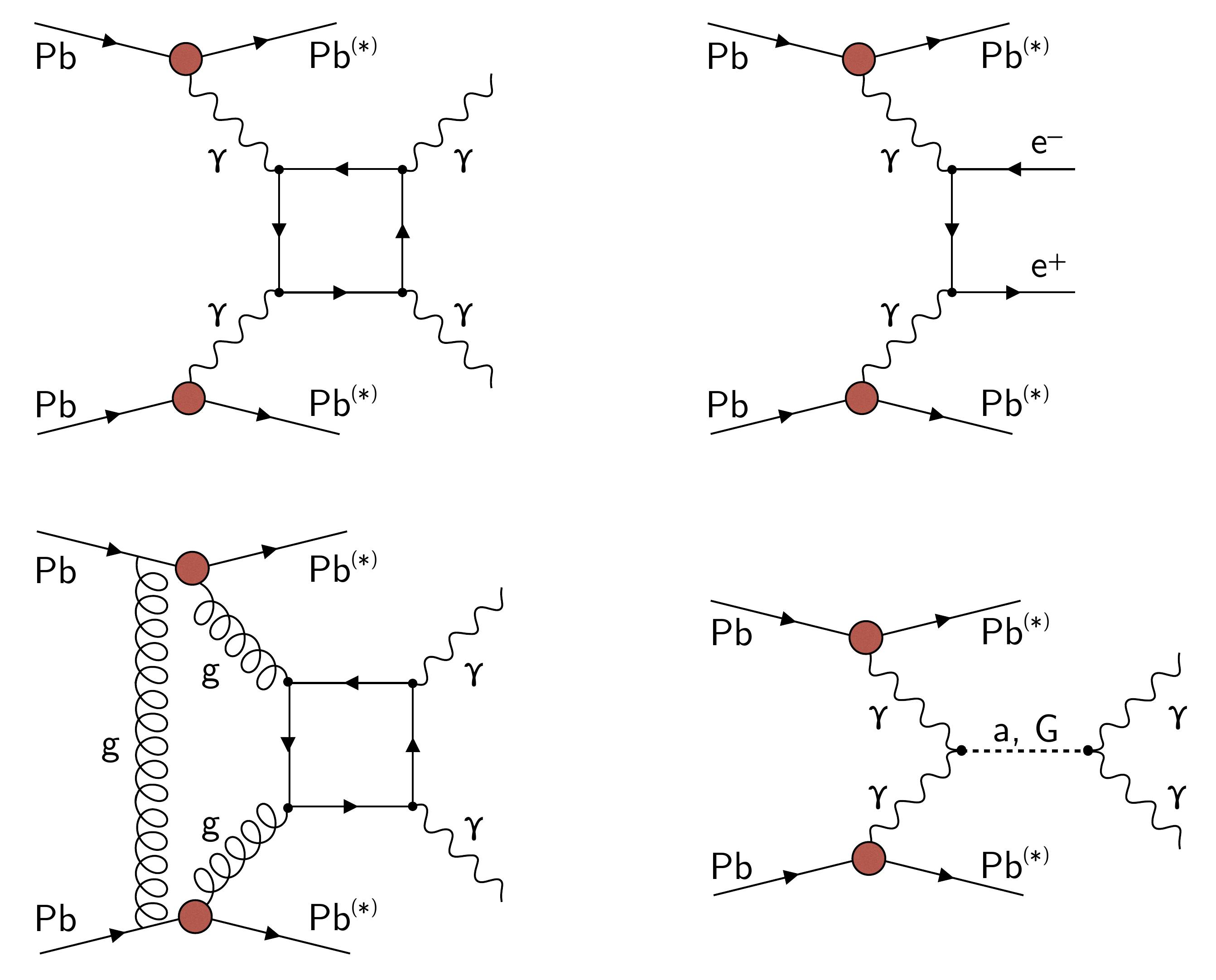}
\caption{Schematic diagrams of light-by-light scattering ($\gaga\to\gaga$, upper left), the Breit--Wheeler process ($\gaga\to\EE$, upper right), central exclusive diphoton production ($\Pg\Pg\to\gaga$, lower left), and axion- or graviton-like particle production ($\gaga\to \Pa,\cPG \to\gaga$, lower right) in ultraperipheral PbPb collisions. The $(*)$ superscript indicates a potential electromagnetic excitation of the outgoing Pb ions.
\label{fig:feynman}}
\end{figure}

The LbL scattering process proceeds at leading order (LO) in the quantum electrodynamics (QED) coupling $\alpha\approx 1/137.04$ via virtual box diagrams containing charged particles (Fig.~\ref{fig:feynman}, upper left). In the SM, the box diagram involves contributions from charged fermions (leptons and quarks) and bosons (\PWpm). Because of its minuscule cross section, $\sigma_{\gaga}\propto\alpha^{4}\approx 3\times 10^{-9}$, the first evidence and observation of LbL scattering were only achieved recently by exploiting the very large fluxes of quasireal photons emitted in PbPb UPCs at the LHC~\cite{ATLAS:2017fur,CMS:2018erd,Aad:2019ock}. The study of the $\gaga\to\gaga$ process, whose cross section is known at next-to-leading (NLO) accuracy in QED and quantum chromodynamics (QCD) with finite fermion masses~\cite{H:2023wfg,H:2023znv}, has been proposed as a particularly clean channel to study BSM physics coupled to photons~\cite{d'Enterria:2013yra}. Modifications of the LbL scattering rates with respect to the SM predictions can occur if, \eg\ new heavy particles, such as  spin-0 axion-like particles (ALPs)~\cite{Knapen:2016moh,dEnterria:2021ljz} or spin-2 massive gravitons~\cite{Sun:2014qba,dEnterria:2023npy} (Fig.~\ref{fig:feynman}, lower right) contribute to the LbL continuum or appear as new diphoton resonances on top of it. In addition, LbL scattering cross sections probe nonlinear Born--Infeld extensions of QED~\cite{Ellis:2017edi} and, generically, anomalous quartic ($4\PGg$) gauge couplings~\cite{Fichet:2014uka}.

Before measuring the very rare LbL process and searching for new BSM phenomena, it is beneficial to study first photon-fusion processes with larger cross sections and well-known properties to use them as references for the more elusive signals. Arguably, the simplest SM photon-photon collision process is the $t$-channel production of an electron-positron pair, $\gaga\to\EE$, by which pure light is transformed into matter (Fig.~\ref{fig:feynman}, upper right). Such a diagram, often called the Breit--Wheeler (BW) process~\cite{Breit:1934zz}, was amongst the first ones studied in QED. The simplicity and large cross section of the BW process have facilitated its measurement in UPCs at fixed-target and collider energies by the WA93~\cite{Vane:1992ms}, CERES/NA45~\cite{CERESNA45:1994cpb}, STAR~\cite{STAR:2004bzo,STAR:2019wlg}, PHENIX~\cite{PHENIX:2009xtn}, CDF~\cite{CDF:2006apx,CDF:2009xwv}, ALICE~\cite{ALICE:2013wjo}, CMS~\cite{CMS:2012cve,CMS:2018erd,CMS:2018uvs}, and ATLAS~\cite{ATLAS:2015wnx,ATLAS:2017fur,ATLAS:2020mve} experiments. These results were found in overall agreement with the theoretical EPA predictions at LO accuracy in QED. Having a good control of the BW process, where the $\Pepm$ radiate a hard bremsstrahlung photon, is also of relevance as a potential background for the LbL scattering measurement.

Whereas the BW process is free from any significant irreducible background, the LbL scattering shares the same final state as the so-called ``central exclusive'' production (CEP) process~\cite{Khoze:2004ak} where a pair of photons are produced in the exchange of gluons in a colour-singlet state (Fig.~\ref{fig:feynman}, lower left). Such a process is short-range, not ultraperipheral, but also leads to a final state with just two photons produced in an otherwise empty detector. 

This paper reports on a measurement of LbL scattering and the BW process using PbPb collision data recorded by the CMS experiment in 2018 at a nucleon-nucleon c.m.\ energy of $\sqrtsNN = 5.02\TeV$, corresponding to an integrated luminosity of $\mathcal{L}_\text{int} = 1.70 \pm 0.03\nbinv$~\cite{CMS:2025rzq}. This analysis is an extension of the previous study of both processes carried out in 2015 in PbPb collisions at the same c.m.\ energy but with about a fourth as much integrated luminosity~\cite{CMS:2018erd}.
The final-state signature of interest is the exclusive production of two photons, $\text{PbPb}\xrightarrow{\gaga}\text{Pb}^{(*)}\gaga \text{Pb}^{(*)}$, or an $\EE$ pair, $\text{PbPb}\xrightarrow{\gaga}\text{Pb}^{(*)}\EE \text{Pb}^{(*)}$, where the diphoton or dielectron final state is measured in the central part of the detector and the outgoing Pb ions survive the interaction and escape undetected  at very low angles with respect to the beam. The lead ions can potentially exchange soft photons and be excited above their ground state, as denoted by the $(*)$ superscript. Such an excitation is often followed by the emission of one or a few neutrons collinear to the beam(s), a process called electromagnetic dissociation (EMD)~\cite{Baur:2003gv}.

A set of criteria are applied to the data sample to select events with just a pair of exclusively produced photons or $\EE$ and to reduce backgrounds. Fully corrected differential cross sections as a function of various kinematic variables are presented for both processes and compared with theoretical predictions. The measured invariant mass spectrum of exclusive diphotons is also exploited to set limits on the yields of ALPs, produced resonantly via the $\gaga\to \Pa \to\gaga$ process on top of the LbL continuum, over the $\mgg = 5$--100\GeV mass range.

The paper is organized as follows. Sections~\ref{sec:CMS}, \ref{sec:MC}, \ref{sec:reco}, and \ref{sec:evt_sel} present the CMS detector, the Monte Carlo (MC) event generators, the reconstruction algorithms, and the efficiencies of the event reconstruction and selection, respectively. The results of the measurement of the BW process and LbL scattering are described in Sections~\ref{sec:QEDee_results} and~\ref{sec:LbL_results}, respectively, and the search for ALPs coupled to photons is discussed in Section~\ref{sec:ALPs}. The paper ends with a summary in Section~\ref{sec:summ}. Tabulated results can be found in HEPData~\cite{hepdata}.

\section{The CMS detector}
\label{sec:CMS}

The central feature of the CMS apparatus is a superconducting solenoid of 6\unit{m} internal diameter, providing a magnetic field of 3.8\unit{T}. Within the solenoid volume are a silicon pixel and strip tracker, a lead tungstate crystal electromagnetic calorimeter (ECAL), and a brass and scintillator hadron calorimeter (HCAL), each composed of a barrel and two endcap sections. Forward calorimeters extend the pseudorapidity coverage provided by the barrel and endcap detectors. Muons are reconstructed in gas-ionization detectors embedded in the steel return yoke outside the solenoid. The silicon tracker, consisting of 1856 pixel and 15\,148 strip detector modules, measures charged particle trajectories within the pseudorapidity range $\abs{\eta}< 2.5$. For charged particles with $1 < \pt < 10\GeV$ and $\abs{\eta} < 1.4$, the track resolutions are typically 1.5\% in \pt~\cite{CMS:2014pgm}.

The ECAL consists of nearly 76\,000 lead tungstate crystals, which provide coverage in pseudorapidity  $\abs{\eta} < 1.479$ in the barrel region (EB) and $1.479 < \abs{\eta} < 3.0$ in two endcap regions (EE). A preshower detector consisting of two planes of silicon sensors interleaved with a total of three radiation lengths of lead is located in front of the EE. The HCAL provides coverage in pseudorapidity $\abs{\eta} < 1.3$ in the barrel region (HB) and $1.3 < \abs{\eta} < 3.0$ in two endcap regions (HE). 

Extensive forward calorimetry, based on Cherenkov-light detectors, complements the coverage provided by the barrel and endcap detectors. The forward hadron (HF) calorimeter uses steel as an absorber and quartz fibres as the sensitive material. The two halves of the HF are located 11.2\unit{m} from the interaction region, one on each end, and together they provide coverage in the range $3.0 < \abs{\eta} < 5.2$. Two zero degree calorimeters (ZDC), made of quartz fibres and plates embedded in tungsten absorbers, located at $\pm140~\unit{m}$ from the collision point, measure neutrons and photons emitted at $\abs{\eta} > 8.3$~\cite{Grachov:2006ke}.

Data are collected with a two-level trigger system. The first level (L1) of the CMS trigger system, composed of custom hardware processors, uses information from the calorimeters and muon detectors to select the most interesting events within a given bunch crossing in a fixed time interval of less than 4\mus~\cite{CMS:2020cmk}. The second level, known as the high-level trigger, consists of a farm of processors running a version of the full event reconstruction software optimized for fast processing, and it is capable of reducing the event rate from around 100\unit{kHz} to less than 1\unit{kHz}, before data storage~\cite{CMS:2016ngn}. A more detailed description of the CMS detector, together with a definition of the coordinate system used and the relevant kinematic variables, can be found in Refs.~\cite{CMS:2008xjf,CMS:2023gfb}.

\section{Monte Carlo event simulation}
\label{sec:MC}

The BW process is generated at LO QED accuracy using different MC event generators:\\
\textsc{starlight}~3.13~\cite{Klein:2016yzr}; \superchic~3.03~\cite{Harland-Lang:2018iur} combined with \photos++~3.61~\cite{Davidson:2010ew} for the final-state radiation (FSR) of photons; and \gammaUPC~1.0~\cite{Shao:2022cly} including the initial photon transverse momentum $\kt$~\cite{Shao:2024dmk} combined with \MGvATNLO~\cite{MadGraph5}, and \PYTHIA8~\cite{Sjostrand:2014zea} (shortened also as \gammaUPC/\textsc{mg5}$\,+\,$FSR(\textsc{py}8) below) to generate photon FSR. The FSR corrections from \photos++ and \PYTHIA8 are very similar. The LbL scattering signal is generated with \superchic\ at LO and with \gammaUPC\ with QED and QCD corrections at NLO accuracy (called \gammaUPC@NLO)~\cite{H:2023wfg,H:2023znv}. The three photon-photon generators employed are based on the EPA, but have different implementations of the Pb photon fluxes and/or treatment of the nuclear survival probability. The \superchic\ and \gammaUPC\ codes share the same photon flux derived from the charged form factor of the Pb ion, whereas \starlight\ uses the electric-dipole flux, which leads to 10--15\% lower $\gaga$ effective luminosities~\cite{Shao:2022cly,Shao:2024dmk}. The \superchic\ and \starlight\ codes use optical Glauber expressions to compute the PbPb overlap probability, whereas \gammaUPC\ uses parameterized overlap functions from a Glauber MC model~\cite{Loizides:2017ack}, for a better description of the very peripheral collisions~\cite{dEnterria:2020dwq}. The central exclusive production process, $\Pg\Pg\to\gaga$, as well as the production of ALPs are simulated with \superchic. Whereas the \gammaUPC-based predictions are mostly used to compare with the final total and differential cross sections, the generated \starlight\ and \superchic\ events are also passed through the \GEANTfour~\cite{Agostinelli:2002hh} detector simulation, and the events are reconstructed with the same software used with the collision data. The simulation describes the tracker material budget with an accuracy better than 10\%, as established by measuring the distribution of reconstructed nuclear interactions and photon conversions in the tracker~\cite{CMS:2014pgm,CMS:2018wqs}.

\section{Event reconstruction}
\label{sec:reco}

Photons and electrons are reconstructed using an algorithm based on the particle-flow (PF) global event description~\cite{Sirunyan:2017ulk}. The PF algorithm uses information from each subdetector system to provide charged particle tracks, calorimeter clusters, and muon tracks. Electromagnetic showers from photons and electrons deposit 97\% of their incident energy into an array of $5{\times}5$ ECAL crystals. The tracker material can induce photon conversion and electron bremsstrahlung and, because of the presence of the strong CMS solenoidal magnetic field, the energy reaching the calorimeter is thereby spread in azimuthal angle. The spread energy is recovered through a collection of adjacent clusters, or ``supercluster''~\cite{CMS:2020uim}. The PF algorithm allows for an almost complete recovery of the energy of the photons and electrons, even if they initiate an electromagnetic shower in the material
in front of the ECAL.

Photons are identified as ECAL energy clusters not linked to the extrapolation of any charged particle trajectory to the ECAL. Electrons are identified as a primary charged particle track and potentially many ECAL energy clusters corresponding to this track extrapolation to the ECAL and to possible bremsstrahlung photons emitted along the way through the tracker material. The default CMS photon reconstruction algorithm is optimized for $\PGg$ and $\Pepm$ with transverse energies $\et = E \sin \theta > 10\GeV$ (where $\theta$ is the polar angle). However, the cross section for photons and electrons from exclusive production peaks at the chosen selection threshold of $\et \approx 2\GeV$. The parameters of the PF algorithm have been retuned in this analysis for $\Pepm$ and $\PGg$ reconstruction in the low-\et range~\cite{CMS-DP-2022-006}. Thanks to the clean UPC environment, the $\et$ threshold for photons, electrons, and superclusters can be lowered to 1\GeV, instead of the 10--15\GeV values used in standard CMS analyses~\cite{CMS:2020uim}, and different calorimeter shower parameters are reoptimized to improve the reconstruction of softer photons and electrons.

For photon candidates, the reconstructed energy of the ECAL supercluster is used to define their energy. A dedicated regression procedure starting from $\et = 2\GeV$ is applied to optimize the $\PGg$ energy scale and resolution, which is validated using MC simulations and control samples in data. The reconstructed supercluster energy and generated $\PGg$ energy agree within a few percent, confirming that the former is well calibrated. The final photon energy resolutions achieved are ${\approx}20\%$~(${\approx}7\%$) for selected photons of $\et = 2~(10)\GeV$, which translate into ${\approx}1.4~(1.9)\GeV$ absolute diphoton mass resolutions at $\mgg\approx 5~(20)\GeV$ masses. Particle identification (ID) criteria are applied to remove converted photons, photons produced in neutral meson decays, and clusters from other neutral hadrons. Those ID requirements are based on the electromagnetic shower properties such as its width along the $\eta$ direction, timing information, ratio of energy deposits in the ECAL and HCAL, and others~\cite{CMS:2020uim}. In order to minimize the contamination from exclusive $\EE$ events in the LbL final state, the photons are required to be unconverted in the tracker.

Additional identification criteria (isolation, number of tracker hits, HCAL/ECAL energy deposit ratio) are applied to the electron candidates, as discussed in Ref.~\cite{CMS:2012cve}. The energy of electrons is determined from a combination of the track momentum, the corresponding ECAL cluster energy, and the energy sum of all bremsstrahlung photons attached to the track. The electron energy scale is verified using a sample of $\gaga\to\EE$ events, comparing the energy of the supercluster $E$ to the momentum of the track $p$. The electron $E/p$ ratio is within 5\% of unity in the barrel and 15\% in the endcaps. The momentum resolution for electrons above a few \GeV is 1.5\%. It is generally better in the barrel region than in the endcaps, and also depends on the bremsstrahlung energy emitted by the electron as it traverses the material in front of the ECAL~\cite{CMS:2020uim,CMS-DP-2020-021}. A good agreement is found between data and simulation, both in the energy scale and resolution.

\section{Event selection and experimental corrections}
\label{sec:evt_sel}

The exclusive diphoton and dielectron candidate events are selected at the trigger level with a dedicated L1 algorithm that requires at least two electromagnetic clusters (L1 EG) with \et above 2\GeV and at least one of the HF detectors with total energy below the noise threshold. The HF veto requirement rejects events with significant particle production, typical of hadronic PbPb interactions.
Data are also recorded with other single-photon triggers with \et thresholds above 3 and 5\GeV, with or without the HF veto. These triggers are used to estimate the efficiency of the main analysis trigger with a ``tag-and-probe'' (TnP) technique~\cite{CMS:2010svw}, as described below.

In the offline analysis, events are selected with exactly two well-reconstructed photons or two electrons, each with  $\et > 2\GeV$ and $\abs{\eta} <2.2$. Events where photons or electrons fall in a few inactive or noisy ECAL areas are removed from the selection. Then, charged and neutral exclusivity selection criteria are applied to reject events having charged or neutral particles produced over the $\abs{\eta} < 2.4$ and 5.2 ranges, respectively. The charged exclusivity condition first removes from further analysis all diphoton or dielectron events that have any additional reconstructed charged particle with $\pt > 0.3\GeV$ and a minimum of four valid associated hits in the tracker. In addition, events are required to have no neutral particles depositing energy in individual EB, EE, HB, HE, and HF calorimeter readout towers, other than those associated with the diphoton or dielectron candidates. This neutral exclusivity condition rejects events with additional neutral particles produced and detected above noise thresholds over $\abs{\eta}<5.2$ and full $\phi$. For events with electron candidates, towers in the region $\abs{\Delta\eta}<0.15~(0.15)$ and $\abs{\Delta\phi}<0.7~(0.4)$ around the electron in the EB~(EE) were not included in this neutral exclusion, while all of the towers in the HB, HE, and HF were included. For events with photon candidates, the criterion is the same as for electrons, but with a narrower $\phi$ window of $\abs{\Delta\phi}<0.15$ around it, for both the barrel and endcap. The chosen noise thresholds are determined from a study of the activity of calorimeters in empty bunch crossing events and are fully efficient to reject events with single neutral particles produced with $\et>1\GeV$ over $\abs{\eta}<5.2$. In addition to requiring no towers above the noise threshold in the central calorimeters, a condition was imposed on the energy deposit in the ZDCs to remove potential events where the diphoton or dielectron system is produced in peripheral nuclear interactions with minimum central activity, instead of through UPC $\gaga$ collisions. 
A loose selection was applied to reject events where there was more than 7\TeV of energy deposits on either ZDC side, 
which is equivalent to the concurrent emission of three or more neutrons in either direction~\cite{CMS:2020skx}.
While neutron emission is not directly simulated in the available MC generators, the impact of this requirement on the signal 
cross sections for the BW and LbL scattering processes was found negligible according to gamma-UPC simulations~\cite{Crepet:2024}.

Nonexclusive backgrounds are typically characterized by a final state with larger pair \pt and larger acoplanarity, $\Aco^{\gaga,\Pe\Pe} = (1-\Delta \phi^{\gaga,\Pe\Pe}/\pi)$, than the back-to-back exclusive $\gaga$ and $\EE$ signal events. To further reduce these backgrounds, the transverse momentum of the reconstructed pairs is required to be $\pt^{\gaga,\Pe\Pe}< 1\GeV$, and the acoplanarity of the pair to be $\Aco^{\gaga,\Pe\Pe} < 0.01$.

Table~\ref{tab:fiducialregion} lists the fiducial phase space where the BW and LbL scattering yields have been extracted. Whereas the exclusive dielectron sample has negligible physical background left after all the criteria above have been applied, the LbL scattering process has at least two background sources remaining in the signal region, both of which are discussed below: (i) $\gaga\to\EE\PGg(\PGg)$ with one or two hard photons radiated by the produced $\EE$ pair, and/or both electrons being misidentified as photons, and (ii) the CEP $\Pg\Pg\to\gaga$ process. 

\begin{table}[htpb!]
  \centering
  \topcaption{Definition of the fiducial phase space for the BW and LbL scattering processes, used in their respective cross section measurements.}
  \begin{tabular}{lc}\hline
  \multirow{ 2}{*}{Diphotons or dielectrons}
                    & $\et>2\GeV$, $\abs{\eta}<2.2$, for each single photon or electron\\
                    & $m^{\gaga,\Pe\Pe} > 5\GeV$, $\pt^{\gaga,\Pe\Pe}<1\GeV$, $\Aco^{\gaga,\Pe\Pe}<0.01$, for the pair\\ [\cmsTabSkip]
  \multirow{3}{*}{Exclusivity}
                    & No additional neutral particles with $\et>1\GeV$ and $\abs{\eta}<5.2$  \\ 
                    & No additional charged particles with $\pt>0.3\GeV$ and $\abs{\eta}<2.4$  \\
                    & Less than 3 neutrons in both ZDCs: $\et<7\TeV$ for $\abs{\eta}>8.3$  \\ \hline
  \end{tabular}
\label{tab:fiducialregion}
\end{table}

The experimental selection of signal events is subject to detector inefficiencies in the trigger, energy reconstruction, photon/electron identification, and exclusivity conditions. In order to transform measured kinematic quantities for exclusive diphoton and dielectron events into physical observables, one needs to account for such inefficiencies. Schematically, a data-driven efficiency factor $C^{\gaga,\Pe\Pe}$ is derived from control samples in data through the factorized expression,
\begin{linenomath*}
\begin{equation}
C^{\gaga,\Pe\Pe} = \varepsilon^{\gaga,\Pe\Pe}~(\text{SF}^{\PGg,\Pe,\text{reco}})^{2}~(\text{SF}^{\PGg,\Pe,\text{ID}})^{2}~\text{SF}^\text{trig}~\text{SF}^\text{ch.excl}~\text{SF}^\text{neut.excl},
\label{eq:corr_fac}
\end{equation}
\end{linenomath*}
where $\varepsilon^{\gaga,\Pe\Pe}$ is the overall exclusive diphoton or dielectron efficiency (accounting for event selection and reconstruction effects) derived from the MC simulations, and $\text{SF}^\text{reco}$, $\mathrm{SF}^\text{ID}$, $\mathrm{SF}^\text{trig}$, $\mathrm{SF}^\text{ch.excl}$, and $\mathrm{SF}^\text{neut.excl}$ are scale factors derived from control regions (CRs) in data, such that $\text{SF} \equiv \varepsilon^\text{data}/\varepsilon^\text{MC}$ accounts for the differences for each individual efficiency between the actual data and MC simulation for photon/electron reconstruction, identification, trigger, and exclusivity criteria, respectively. The efficiencies and SFs are obtained from detector simulations based on events generated with the \starlight\ and \superchic\ codes, and are found to be fully consistent with each other. The SF values are close to unity and each derived from CRs via tag-and-probe techniques as explained below.

\subsection{Exclusive diphoton final state}

The overall exclusive diphoton efficiency within the phase space defined by Table~\ref{tab:fiducialregion} derived from the LbL scattering MC simulation amounts to $\varepsilon^{\gaga} = (13.5 \pm 0.3)\%$. This result is mostly driven by the inefficiencies of the single-photon reconstruction and identification ($\varepsilon^{\PGg,\text{reco+ID}}\approx 40\%$), and of the trigger ($\varepsilon^{\gaga,\text{trig}}\approx 80\%$) in the $\et = 2\GeV$ regions closest to the kinematic threshold of our selection.
The quoted uncertainty here is statistical only, reflecting the finite size of the LbL scattering MC sample. This MC-based efficiency is cross-checked and corrected using CRs in data, as explained below.

The photon reconstruction efficiency $\varepsilon^{\PGg,\text{reco}}$ is extracted from data using the TnP approach by selecting $\gaga\to\EE$ events in which one of the electrons emits a hard-bremsstrahlung photon because of interaction with the material of the tracker. In such a case, the electron imparts a large fraction of its energy to the photon and thereby cannot reach the ECAL to be identified as an electron. However, it is reconstructed in the tracker as a charged particle.
{\tolerance=8000 In a first step, hard-bremsstrahlung events are selected among events passing a trigger requiring one L1 EG cluster with $\et>5\GeV$ that have exactly two oppositely charged particle tracks and exactly one electron reconstructed. Among the selected events, only those with exactly one photon compatible with a hard bremsstrahlung are kept. Such events are used to estimate the efficiency in a TnP procedure, via
\begin{equation}
\varepsilon^{\PGg,\text{reco, hard-brem}}_\text{data} = \frac{\text{N}^\text{reco,hard-brem}_\text{passing}}{\text{N}^\text{reco, hard-brem}_\text{probe}},
\end{equation}
where the denominator and numerator are defined as follows:
\begin{itemize}
\item $\text{N}^\text{reco, hard-brem}_\text{probe}$: Electrons are selected if (i) their direction matches with one of the two reconstructed tracks within a radius $\Delta R_\text{\Pe,track} = \sqrt{\smash[b]{(\eta^{\Pe}-\eta^\text{track})^{2} + (\phi^{\Pe}-\phi^\text{track})^{2}}} < 1.0$, (ii) they have $\et^{\Pe} > 5\GeV$, and (iii) their associated ECAL supercluster is matched within $\Delta R_\text{\Pe,L1\,EG}<0.1$ to an L1 EG cluster with $\et>5\GeV$. The $\pt^\text{unmatch}$ of the track that is not matched with the electron must be between 0.65 and 2\GeV, since it is assumed that the track is generated by the electron after bremsstrahlung emission. The $\pt^\text{unmatch} < 2\GeV$ requirement ensures that this low-\pt charged particle is sufficiently bent by the magnetic field, and thus the expected photon (extrapolated to the ECAL) and the second electron are sufficiently separated. Events entering the denominator are not required to have a reconstructed photon. 
\item $\text{N}^\text{reco, hard-brem}_\text{passing}$: Events from the denominator are also included in the numerator if a photon is found with $\et>2\GeV$ that passes the identification criteria.
\end{itemize}

The efficiency is extracted by selecting signal-like events through a fit to the acoplanarity distribution between the (tag) electron and the (probe) charged particle track and amounts to $\varepsilon^{\PGg,\text{reco}}_\text{data} = (76.7 \pm 2.4)\%$, to be compared with $\varepsilon_\text{MC}^{\PGg,\text{reco}} = (78.6 \pm 0.6)\%$ in the MC simulation, where uncertainties are systematic (as well as all other SF uncertainties quoted in this section). The ratio of these efficiencies is used to define the corresponding $\text{SF}^{\PGg,\text{reco}} = 0.98 \pm 0.03$ scale factor.

The third term of Eq.~(\ref{eq:corr_fac}) accounts for the $\varepsilon^{\PGg,\text{ID}}$ efficiency, which is estimated using the $\text{N}^\text{reco, hard-brem}_\text{passing}$ events passing the TnP method described above where, in addition, they are required to have $\Aco^\text{(tag,probe)}< 0.06$ to select mostly signal events. The efficiency is extracted by counting the number of events that pass all selections and amounts to $\varepsilon^{\PGg,\text{ID}}_\text{data} = (50.0 \pm 4.5)\%$ to be compared with $\varepsilon_\text{MC}^{\PGg,\text{ID}} = (51.5 \pm 3.5)\%$ in the MC simulation for this CR. The corresponding SF is estimated by taking the ratio of data and MC efficiencies and amounts to $\text{SF}^{\PGg,\text{ID}} = 0.95 \pm 0.05$.

The fourth term of Eq.~(\ref{eq:corr_fac}) accounts for the trigger selection efficiency. The analysis trigger comprises two main ingredients, the requirement of two L1 electron/photon objects above a given $\et$ threshold, and a veto on energy in the HF. Each component is studied independently. The efficiency for reconstructing an L1 EG cluster with $\et>2\GeV$ is verified using the TnP technique on $\gaga\to\EE$ events, where the dielectron acoplanarity is fit to extract the signal and measure the efficiency. Events are further selected using a supporting trigger requiring one L1 EG cluster with $\et>5\GeV$ with the same HF energy veto as the analysis trigger. The L1 EG cluster used in the trigger is matched (using the same criterion mentioned above) to one of the two electrons reconstructed offline, called the tag. The other electron in the event is the probe, and it qualifies as a passing probe if it is matched (within a $\Delta R_\text{\Pe,L1\,EG} < 0.5$ radius) to an L1 EG cluster with $\et>2\GeV$. The efficiency is defined as the fraction of probes that are also passing probes, and it is in the 45--100\% range with the lowest efficiency found close to the $\et = 2\GeV$ threshold. The efficiencies in data and MC simulation are found to be $(92.0 \pm 0.2)\%$ and $(91.2 \pm 0.1)\%$, respectively, corresponding to $\text{SF}^\text{trig,EG} = 1.01 \pm 0.01$. The same $\gaga\to\EE$ sample is used to assess the efficiency of the HF veto component of the trigger, using a complementary trigger with the same photon/electron $\et$ threshold but without any HF veto requirement. Events passing the reference trigger are required to have exactly two opposite-charge electrons passing all identification criteria, matched with L1 EG objects with $\et > 2\GeV$ within $\Delta R_\text{\Pe,L1\,EG} < 0.5$, and no additional charged particles in the event. The fraction of these events also passing the main trigger corresponds to the HF veto efficiency at L1. The efficiency was estimated by counting the number of events with acoplanarity ${<}0.01$ and found to be $87.0 \pm 8.0\%$ in data and fully efficient (within a $\pm 0.4\%$ uncertainty) in the simulation. The scale factor is estimated by taking the ratio of both efficiencies, and amounts to $\text{SF}^\text{trig,HF} = 0.87 \pm 0.05$. 

The last two terms of Eq.~(\ref{eq:corr_fac}) account for the efficiency of the charged and neutral exclusivity selections. These efficiencies are estimated from the fraction of events passing the dielectron selection criteria with the exception of the exclusivity criteria. Using the acoplanarity distribution to extract the signal, it is found that $(93.0 \pm 0.8)\%$ of the events feature no additional track in the event, to be compared with $(99.5 \pm 0.4)\%$ in the simulation. The associated scale factor is $\text{SF}^\text{ch.excl} = 0.93 \pm 0.01$. A similar strategy is used for the neutral exclusivity selection, this time in events passing the corresponding requirements. This efficiency is found to be $(80.8 \pm 0.8)\%$ in data, and $(95.1 \pm 0.4)\%$ in simulation. This scale factor is then $\text{SF}^\text{neut.excl} = 0.85 \pm 0.01$. 

\subsection{Exclusive dielectron final state}

The overall exclusive dielectron efficiency within the phase space defined by Table~\ref{tab:fiducialregion} derived from the simulation samples of the BW process amounts to $\varepsilon^{\Pe\Pe} = (7.2 \pm 0.1)\%$. The efficiency to reconstruct dielectrons is about half that for diphotons because of the comparatively softer nature of the $\EE$ produced in the BW process compared with the photons produced in LbL scattering. The single $\Pepm$ of the BW process have a steeper $\et$ spectrum than the photons from LbL scattering and, therefore, have a smaller probability to pass the different energy selection thresholds. Such losses are further enhanced because the overall probability for both electrons to pass the trigger requirements, or to be concurrently reconstructed above the chosen \pt and invariant mass thresholds, depends on the individual efficiencies squared.

Most of the SFs for the exclusive dielectron final state are common with those obtained for the diphoton case because they are computed using the same TnP samples, except for the reconstruction and identification efficiency. For the latter, the TnP technique is employed using a fit to the acoplanarity distribution in $\gaga\to\EE$ events, as done for the diphoton case, except that now the probe is a charged particle track that becomes a passing probe if it is matched to an electron passing the reconstruction and identification criteria. An efficiency of $(67.0\pm 0.6)\%$ is found in data, to be compared with $(71.1 \pm 0.1)\%$ in the MC simulation, corresponding to a scale factor of $\text{SF}^{\Pe,\text{reco+ID}} = 0.94 \pm 0.01$.

\subsection{Summary of the efficiencies}

Applying Eq.~(\ref{eq:corr_fac}), data-driven efficiency factors of $C^{\gaga} = (8.0 \pm 1.0)\%$ and  $C^{\Pe\Pe} = (4.4 \pm 0.3)\%$ are found for the exclusive diphoton and dielectron final states, respectively. The final data-driven efficiency factors have a relative uncertainty of 12.5\% and 6.6\% for the LbL and BW processes, respectively, with the uncertainties of each individual factor being propagated in quadrature. The MC-based efficiencies, the individual data-to-simulation SFs, and the final data-driven efficiency factors, are summarized in Table~\ref{tab:eff_summary} for the diphoton and dielectron channels. The overall $C^{\gaga,\Pe\Pe}$ factors are both about a factor of two smaller than those determined in our previous analysis using 2015 data~\cite{CMS:2018erd}. This is due to comparatively noisier calorimeters that worsen the low-\et photon and electron reconstruction, and reduce the neutral exclusivity efficiency.

\begin{table*}[htbp]
  \centering
 \topcaption{\label{tab:eff_summary}
  Summary of the overall efficiencies from simulation ($\varepsilon^{\gaga,\Pe\Pe}$), individual data-to-simulation SFs, and data-driven efficiency factors ($C^{\gaga,\Pe\Pe}$) obtained for the exclusive diphoton and dielectron analyses. The quoted uncertainties in $\varepsilon^{\gaga,\Pe\Pe}$, SF, and $C^{\gaga,\Pe\Pe}$  are statistical, systematic, and statistical and systematic added in quadrature, respectively.}
 \begin{tabular}{lrcl} \hline
  Diphoton efficiency from simulation & $\varepsilon^{\gaga}$ & $=$ & $(13.5 \pm 0.3)\%$ \\
  $\PGg$ reco. and ID data-to-simulation scale factor& $\text{SF}^{\PGg,\text{reco+ID}}$ & $=$ & $ 0.92 \pm 0.06$ \\
  Dielectron efficiency from simulation & $\varepsilon^{\Pe\Pe}$ & $=$ & $ (7.2 \pm 0.1)\%$ \\
  $\Pepm$ reco. and ID data-to-simulation scale factor& $\text{SF}^{\Pe,\text{reco+ID}}$ & $=$ & $ 0.94 \pm 0.01$ \\ [\cmsTabSkip]
  Trigger selection  data-to-simulation scale factor& $\text{SF}^{\gaga, \text{trig}}$ & $=$ & $0.88 \pm 0.05$ \\
  Charged exclusivity data-to-simulation scale factor & $\text{SF}^\text{ch.excl}$ & $=$ & $ 0.93 \pm 0.01$ \\
  Neutral exclusivity data-to-simulation scale factor & $\text{SF}^\text{neut.excl}$ & $=$ & $ 0.85 \pm 0.01$ \\ [\cmsTabSkip]
  Diphoton global efficiency, Eq.~(\ref{eq:corr_fac}) & $C^{\gaga}$ & $=$& $(8.0 \pm 1.1)\%$  \\
  Dielectron global efficiency, Eq.~(\ref{eq:corr_fac}) & $C^{\Pe\Pe}$ & $=$ & $(4.4 \pm 0.3)\%$  \\  \hline
 \end{tabular}

\end{table*}

\section{Measurement of the Breit--Wheeler process}
\label{sec:QEDee_results}

Following the reconstruction and selection criteria explained above, about 20\,000 exclusive dielectron pairs pass the fiducial criteria listed in Table~\ref{tab:fiducialregion}. Table~\ref{tab:qed_yields_inverted} indicates the number of events remaining after each analysis step in the data and in the simulations. The latter are shown for \superchic$\,+\,$FSR(\photos++) and \starlight\ MC simulations. The number of events in the first four rows do not match exactly for data and MC simulations because these selection requirements accept a fraction of nonexclusive backgrounds that are not included in the latter. However, the data-to-simulation agreement is much better when the full exclusivity is added to the selection criteria, reaching differences of a few percent once all requirements are applied.

\begin{table*}[htbp!]
  \centering
    \topcaption{Exclusive dielectron yields after applying each selection criteria in data and MC simulations. The MC simulation yields match the integrated luminosity of the measurement, $\sigma_\text{fid,MC}\mathcal{L}_\text{int}$, and are corrected by the SFs listed in Table~\ref{tab:eff_summary}. The (\%) column indicates the percentage of events remaining after applying the selection with respect to the previous row.\label{tab:qed_yields_inverted}}
  \cmsTable{
    \begin{tabular}{lrrrrrr}
    Selection criterion   & \multicolumn{2}{c}{Data}	  & \multicolumn{2}{c}{\superchic} & \multicolumn{2}{c}{\starlight}  \\
            & &	  & \multicolumn{2}{c}{$+\,$\photos++ } & &  \\
                          & $N_\text{events}$ & (\%) & $N_\text{events}$ & (\%) & $N_\text{events}$ & (\%) \\
    \hline
    Trigger						                               & 4\,600\,672 & \NA & 53\,500 & \NA  & 47\,700 & \NA \\
	Two reco+ID $\Pepm$ with $\pt>2\GeV$, $\abs{\eta}<2.2$ & 38\,053 & 1  & 28\,700 & 54  & 26\,400 & 55  \\
	Dielectron mass $m^{\Pe\Pe} > 5\GeV$		              & 35\,716 & 94 & 28\,700 & 100 & 26\,300 & 100 \\
	Charged exclusivity selection                         & 30\,198   & 85 & 26\,300 & 92  & 24\,000 & 91  \\
	Neutral exclusivity	selection			                    & 23\,464	  & 78 & 21\,100 & 80  & 19\,500 & 81  \\
	Dielectron $\pt^{\Pe\Pe} < 1\GeV$		                  & 20\,909   & 89 & 19\,500 & 92  & 18\,600 & 95  \\ 	
  Dielectron $\Aco^{\Pe\Pe} < 0.01$	                    & 20\,161   & 96 & 19\,200 & 98  & 18\,400 & 99  \\
  ZDC$^{-} < 7\TeV$ or ZDC$^{+} < 7\TeV$                & 19\,689   & 98 & 19\,200 & 100 & 18\,400 & 100 \\
  \end{tabular}
  }
\end{table*}

The kinematic distributions of the $\gaga\to \EE$ events passing all the analysis criteria are shown in Fig.~\ref{fig:qed_distributions}, together with the \superchic$\,+\,$FSR(\photos++) and \starlight\ MC simulations, at the detector-level (\ie\ without efficiency and bin-migration corrections). The absolute value of the cosine of the scattering angle with respect to the $z$-axis in the Collins--Soper frame~\cite{Collins:1977iv} is defined as
\begin{equation}
\abs{\cos{\theta^{*}}} = 2\,\left|\frac{E^{\Pe\Pe}p_z^{\Pe_1}-p_z^{\Pe\Pe}E^{\Pe_1}}{m^{\Pe\Pe} \mT^{\Pe\Pe}} \right|,
\end{equation}
where $E$, $p_{z}$, $m$, and $\mT$ are, respectively, the energy, the $z$-axis component of the momentum, the invariant mass, and the transverse mass ($\mT = \sqrt{E^2-p_z^2}$) of either the leading electron (superscript $\Pe_{1}$) or the dielectron (superscript $\Pe\Pe$).

A good data-to-simulation agreement is found for all variables within uncertainties, except for the \starlight\ predictions of the dielectron transverse momentum and acoplanarity. These quantities are better reproduced by \superchic$\,+\,$FSR(\photos++) because this latter simulation includes the emission of FSR photons that recoil against the BW pair, boosting it above $\pt^{\Pe\Pe}\approx 0.2\GeV$ and inducing larger azimuthal acoplanarities for $\Aco^{\Pe\Pe} >0.006$. Higher-order (NLO) QED corrections (currently missing in the MC event generators) have been shown to further increase the dilepton pair acoplanarity and transverse momentum~\cite{Shao:2024dmk}, and would improve the agreement with the data in the tails of the $\Aco^{\Pe\Pe}$ and $\pt^{\Pe\Pe}$ distributions.

\begin{figure}[hbtp!]
\centering
 \includegraphics[width=0.44\textwidth]{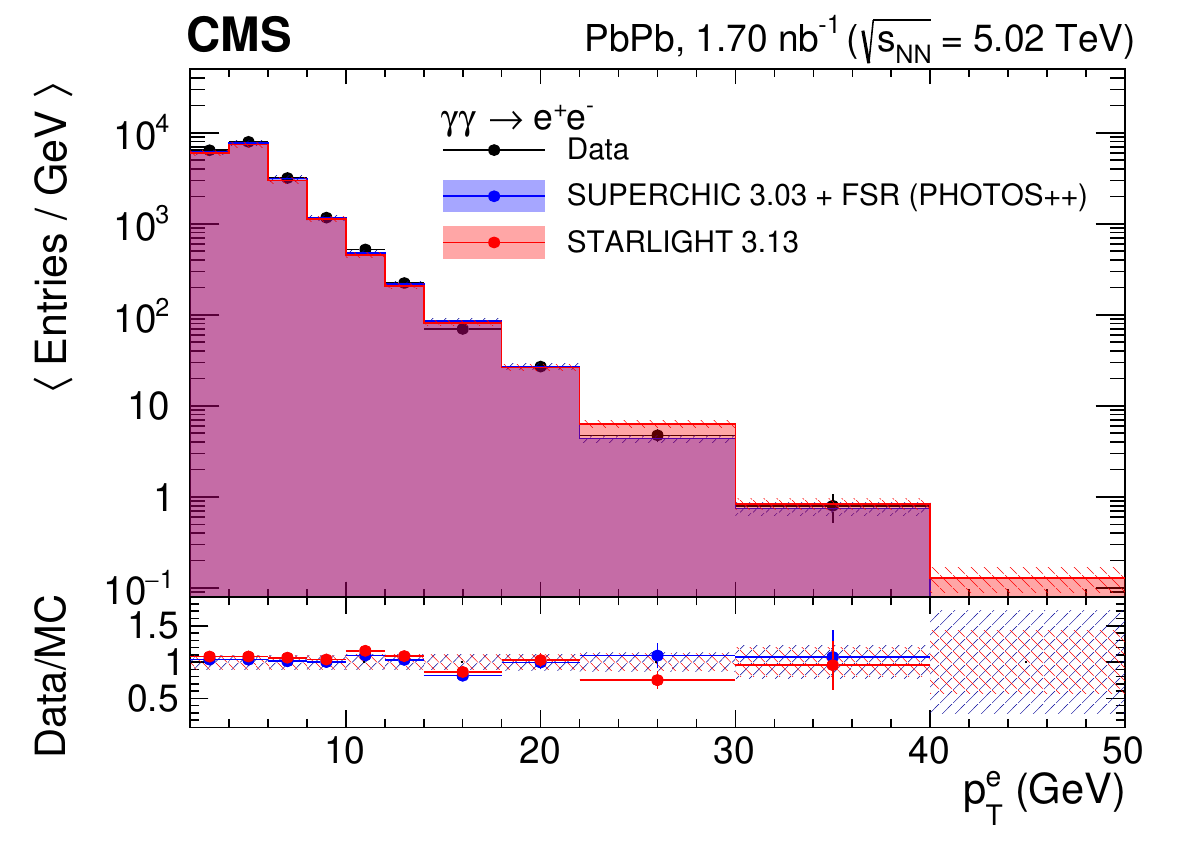}
 \includegraphics[width=0.44\textwidth]{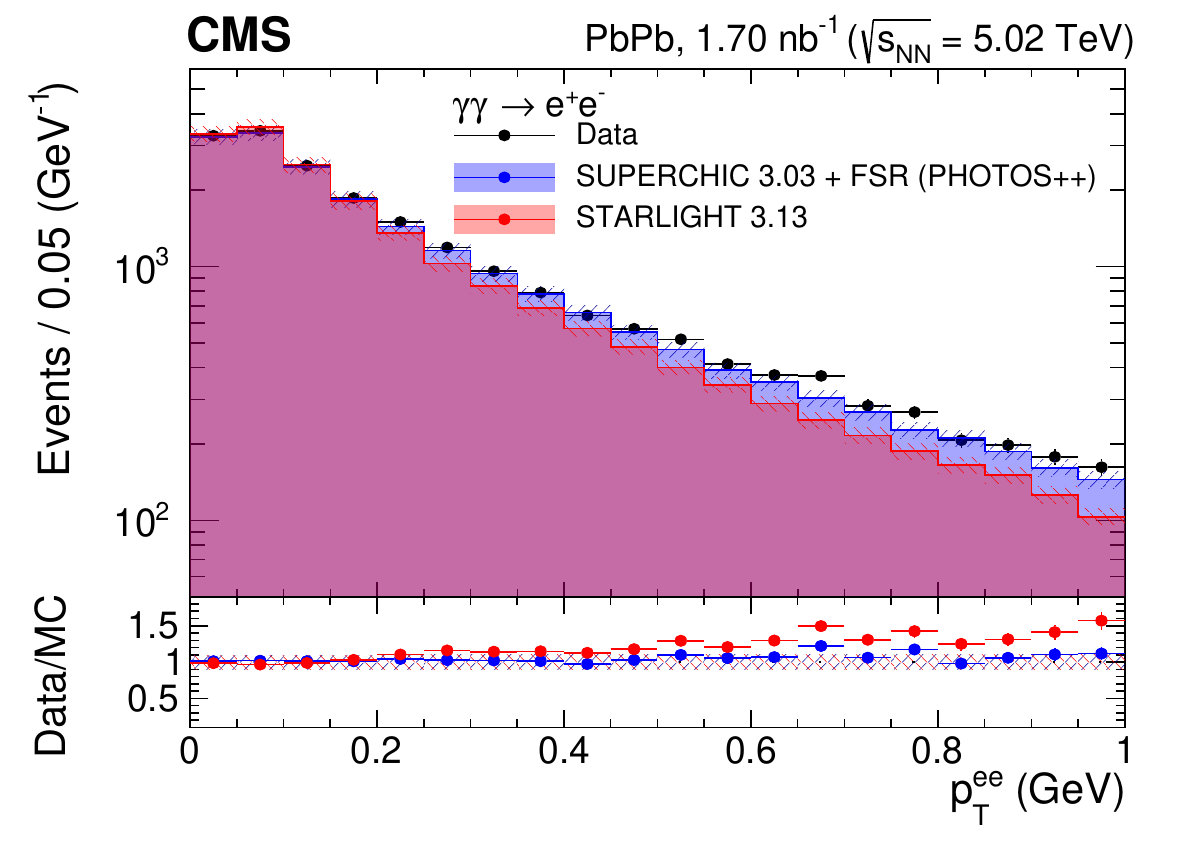}
 \includegraphics[width=0.44\textwidth]{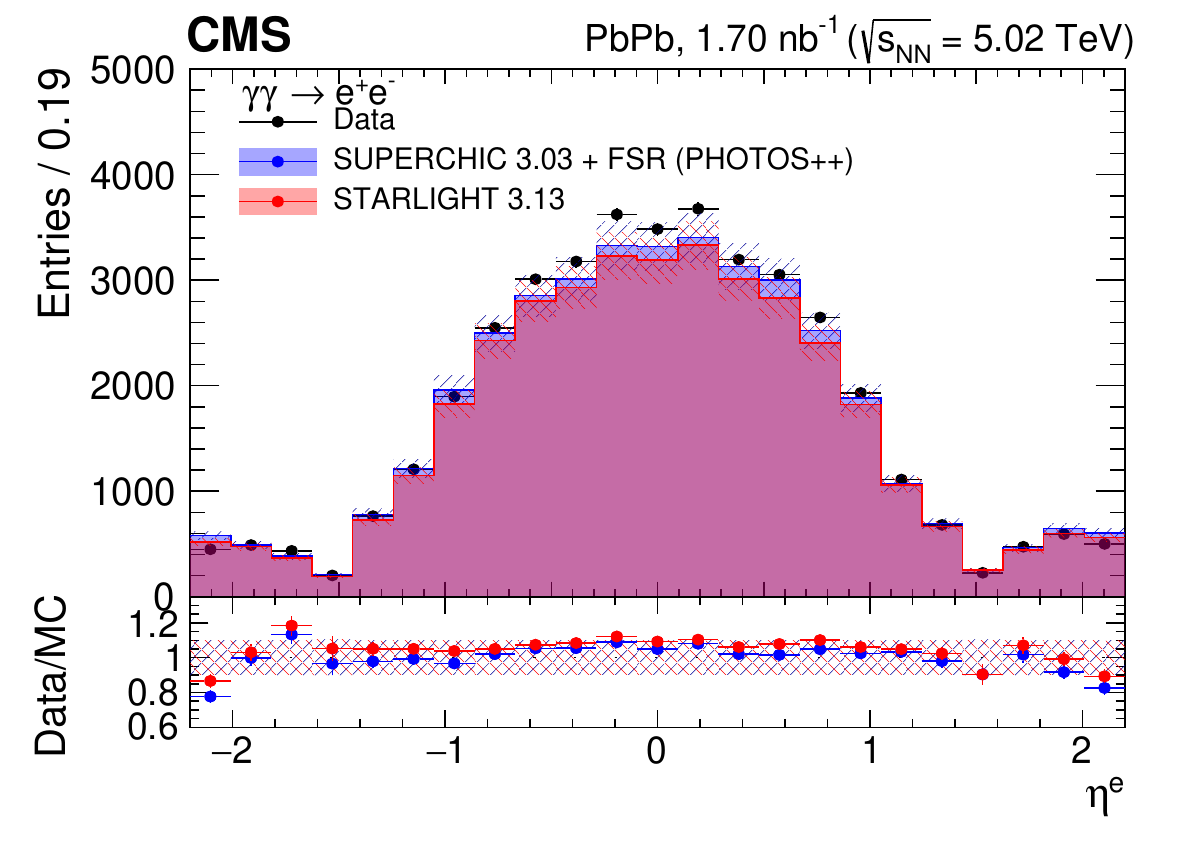}
 \includegraphics[width=0.44\textwidth]{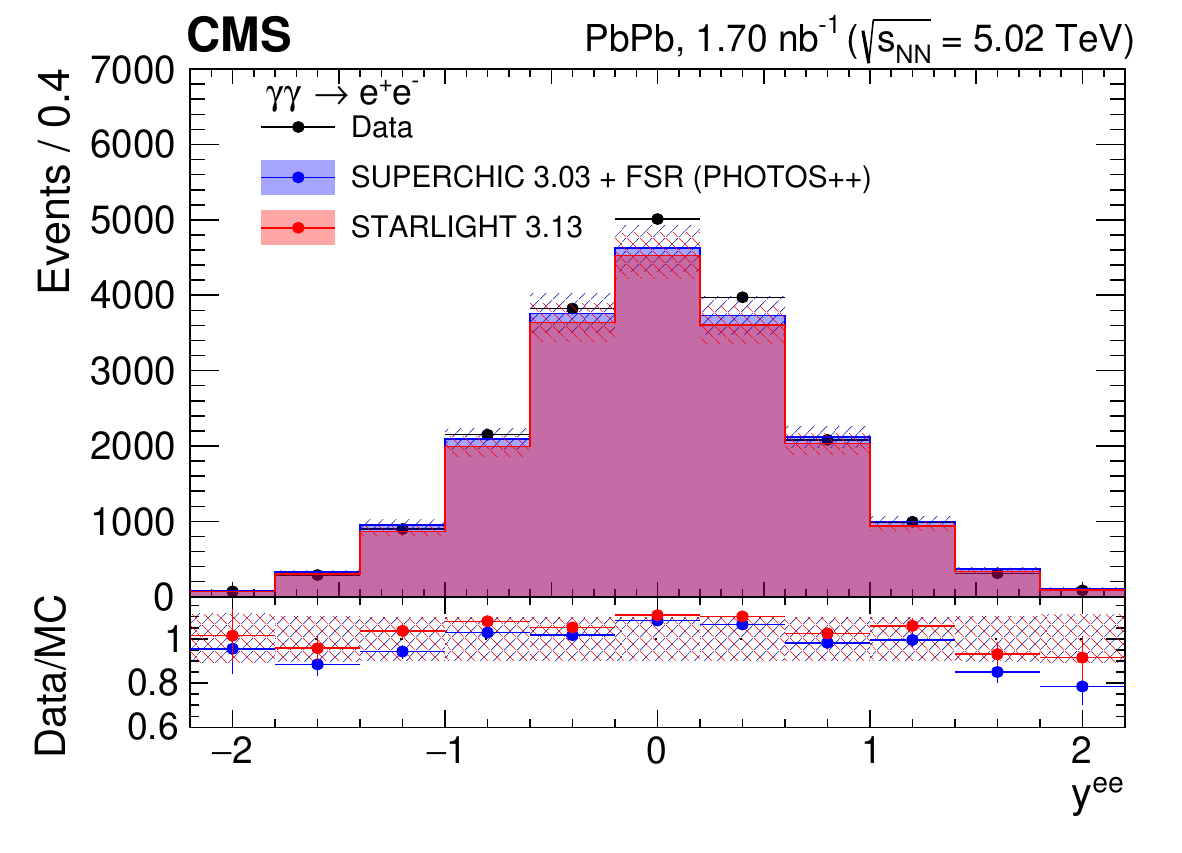}
 \includegraphics[width=0.44\textwidth]{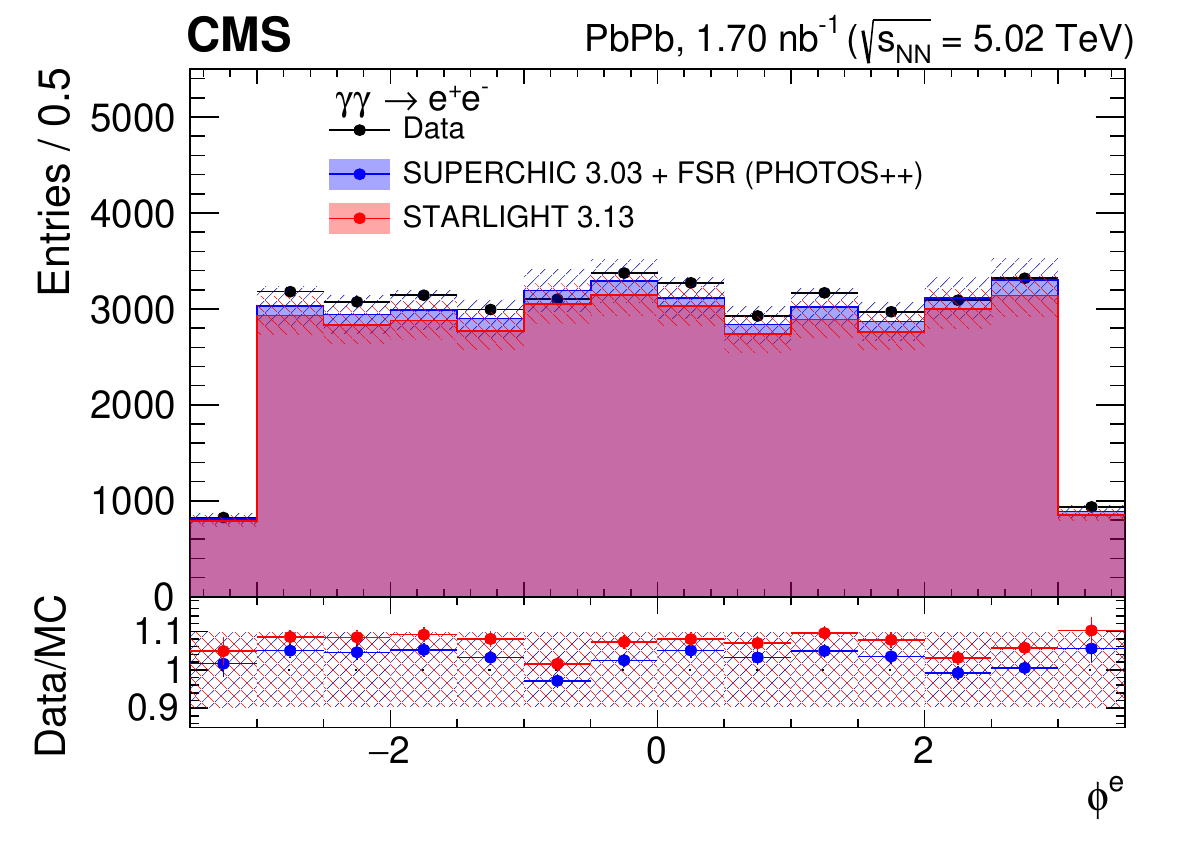}
 \includegraphics[width=0.44\textwidth]{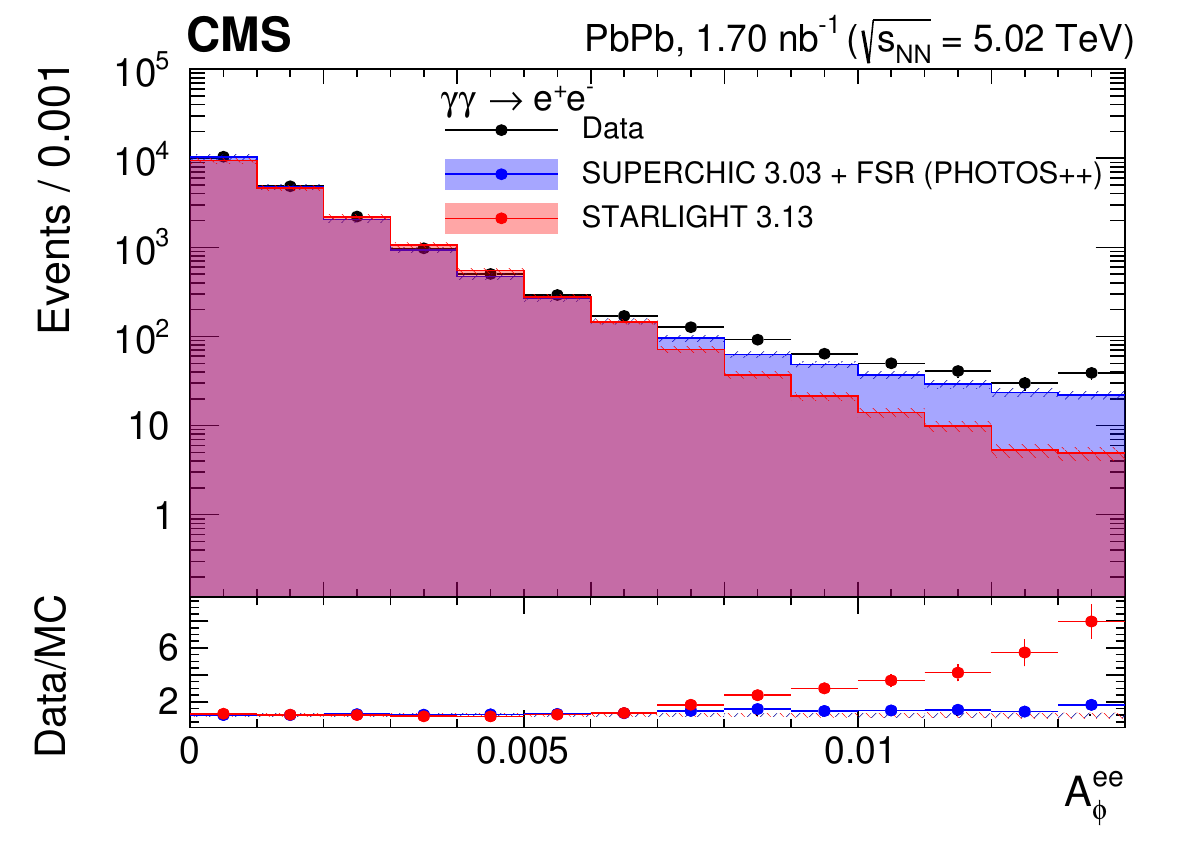}
 \includegraphics[width=0.44\textwidth]{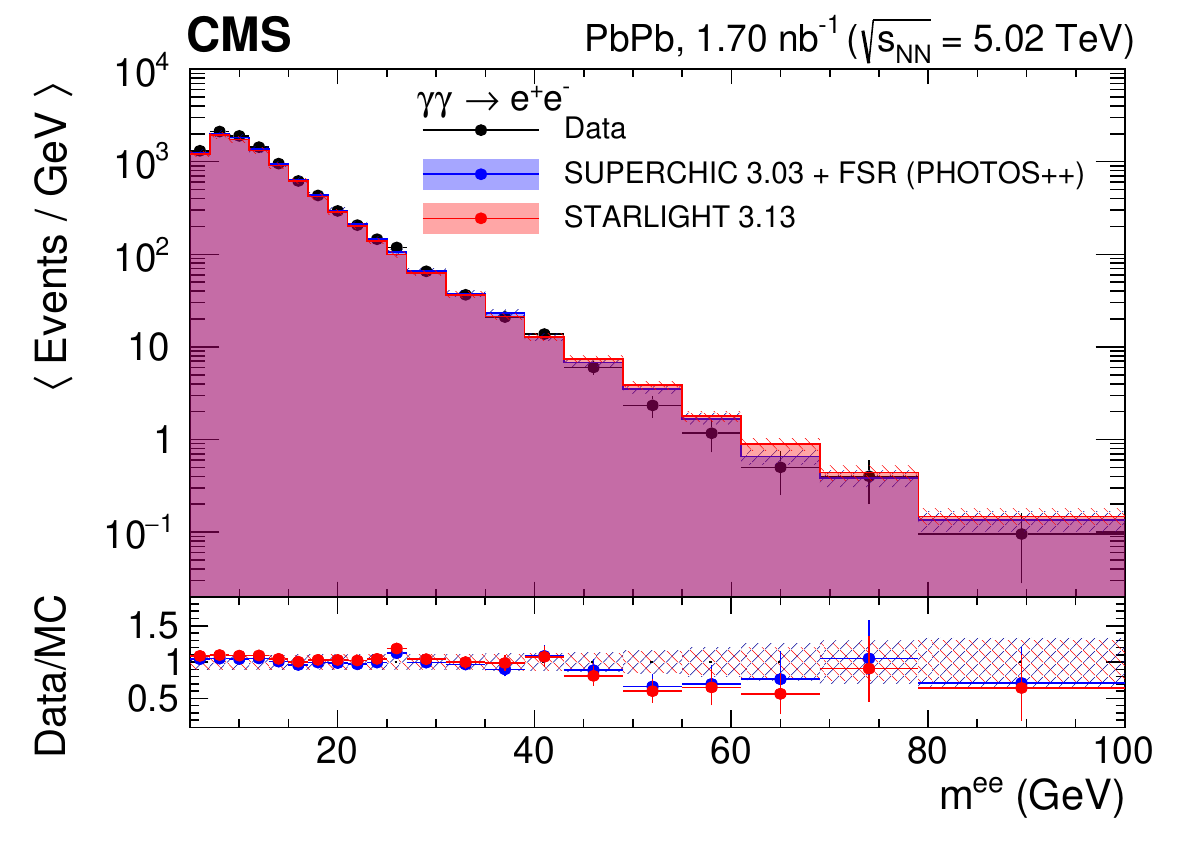}
 \includegraphics[width=0.44\textwidth]{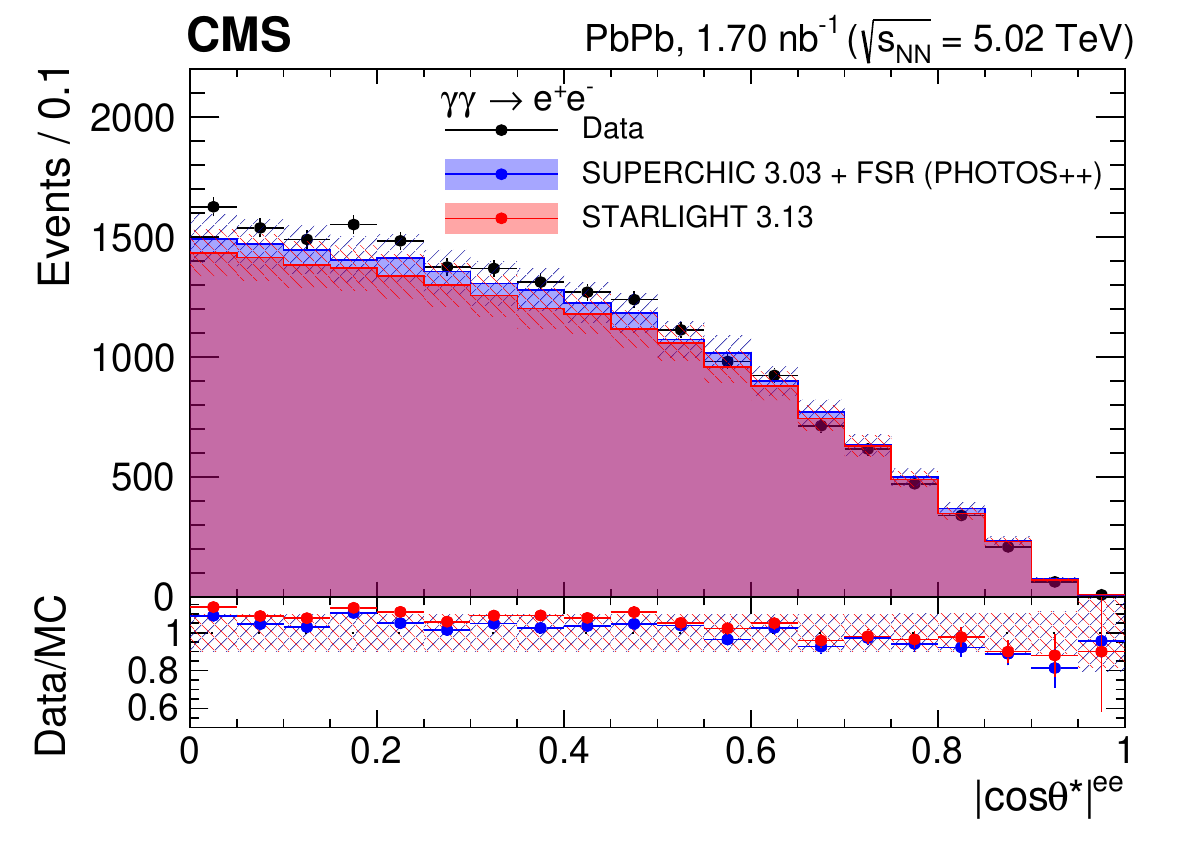}
 \caption{Detector-level kinematic distributions for exclusive $\EE$ events passing  the analysis requirements (Table~\ref{tab:fiducialregion}) in the data (black points), and in \superchic$\,+\,$FSR(\photos++) and \starlight\ simulations (histograms). The MC simulations are normalized to match $\sigma_\text{fid,MC}\mathcal{L}_\text{int}$, and corrected with the SFs listed in Table~\ref{tab:eff_summary}. The $\pt^\Pe$ and $m^{\Pe\Pe}$ distributions display the number of events per bin, divided by the bin width. Ratios of the data to MC expectation are shown in the lower panels. Error bars around the data points (hatched bands) indicate statistical (quadrature sum of MC statistical and systematic) uncertainties.}
\label{fig:qed_distributions}
\end{figure}

The systematic uncertainties for the exclusive dielectron measurement are summarized in Table~\ref{tab:syst_summary_qed}. The main source of uncertainty in the measured $\gaga\to\EE$ yields is that of the trigger efficiency, and amounts to $\pm6.2\%$. The integrated luminosity uncertainty is 1.7\%~\cite{CMS:2025rzq}. The sum in quadrature of all uncertainties propagates into a global 6.9\% uncertainty in the BW yields.

\begin{table}[htbp]
 \centering
 \topcaption{\label{tab:syst_summary_qed} Summary of relative systematic uncertainties in the measurement of exclusive dielectron cross sections.}
  \begin{tabular}{lc}  \hline
   Trigger SF & 6.2\% \\
   MC-based $\EE$ efficiency & 2\% \\
   Electron reconstruction and identification SF & $(2\times 0.5)\%$ \\
   Charged exclusivity SF & 1\% \\
   Neutral exclusivity SF & 1\% \\
   Integrated luminosity & 1.7\%\\  [\cmsTabSkip]
   Total & 6.9\% \\ \hline
  \end{tabular}
\end{table}

The BW scattering fiducial cross section, for electron pairs passing all selections listed in Table~\ref{tab:fiducialregion}, is
\begin{linenomath*}
\begin{equation}
\sigma_\text{fid}(\gaga\to\EE) = \frac{N^{\Pe\Pe,\text{data}}}{C^{\Pe\Pe} \mathcal{L}_\text{int}} = 263.5 \pm 1.8 \stat \pm 17.8 
\syst\mub,
\label{eq:qed_cross_section_noZDCcut}
\end{equation}
\end{linenomath*}
with $N^{\Pe\Pe,\text{data}} = 19\,689$, $C^{\Pe\Pe} = 4.4 \pm 0.3\%$, and $\mathcal{L}_\text{int} = 1.70 \pm 0.03\nbinv$. The measured cross section can be compared to the current state-of-the-art theoretical prediction given by the LO QED calculations complemented with photon FSR, which accounts for a good fraction of the NLO corrections~\cite{Shao:2024dmk}. To provide a consistent comparison across predictions and data, FSR emission has been added to the MC output of the three models and the fiducial phase space requirements applied to the generated events. The corresponding theoretical predictions are $\sigma_\text{fid}(\gaga\to\EE) = 225\mub$ for \starlight$\,+\,$FSR(\textsc{py}8), $\sigma_\text{fid}(\gaga\to\EE) = 261\mub$ for \superchic$\,+\,$FSR(\photos++), and $\sigma_\text{fid}(\gaga\to\EE) = 265\mub$ for \gammaUPC/\textsc{mg5}$\,+\,$FSR(\textsc{py}8). Theoretical uncertainties (not quoted) due to missing higher-order QED corrections are of the order of a few percent \cite{Shao:2024dmk}. The measured BW cross section is in very good agreement with the two latter predictions, but it is 15\% larger than the \starlight$\,+\,$FSR(\textsc{py}8) result. It is worth noting that the cross sections for the pure LO process, without FSR, amount to $\sigma_\text{fid}(\gaga\to\EE) = 251, 293$, and $297\mub$ for \starlight, \superchic, and \gammaUPC, respectively. The inclusion of photon FSR reduces the fiducial $\gaga\to\EE$ cross section because about 10\% of the events fail to satisfy the kinematic requirements: either the radiating $\Pepm$ falls below the $\et=2\GeV$ threshold at low pair masses, or the $\EE$ pair goes above the $\pt^{\Pe\Pe}<1\GeV$ criterion at high invariant masses. Numerical differences among $\sigma_\text{fid}$ predictions can be traced to different implementations of the Pb photon flux and the nuclear nonoverlap (\ie\ exclusivity) condition. Variations of the nonoverlap condition computed with a Glauber model for varying Pb radius and diffusivity parameters~\cite{Loizides:2017ack} propagate into a few percent differences in the cross sections. The major difference among the MC models comes from their implemented photon fluxes (electric-dipole form-factor in \starlight, and charged form-factor in \superchic\ and \gammaUPC). The present BW measurement favours the more realistic charged form-factor $\PGg$ flux from the Pb ions.

Finally, the number of forward neutrons emitted in the EMD of the two interacting ions is also measured in BW events passing the fiducial criteria of Table~\ref{tab:fiducialregion} with the exception of the maximal ZDC activity requirement. The number of neutrons is determined based on the energy deposition in the ZDC detectors, correcting for bin migrations, and EMD pileup events, as described in Refs.~\cite{CMS:2020skx,CMS:2023snh}. Table~\ref{tab:qed_ratios} lists the fraction of different neutron multiplicity classes (0n, 1n, and $X$n with $X \geq 1$) on each ZDC side measured in the BW process, and Fig.~\ref{fig:qed_ratios} shows them in graphical form, compared with the \superchic~4.2~\cite{Harland-Lang:2023ohq}, \starlight~3.13 (this generator does not compute all measured categories), and \gammaUPC~1.6~\cite{Crepet:2024} predictions. In general, a good agreement is found between data and EMD models, except for the 0n1n+1n0n and 1n$X$n+$X$n1n categories where differences $\pm20\%$, or larger, are found. The latest \gammaUPC~1.6 predictions show the best accord with the measured neutron category probabilities as this model includes fits to more differential photoexcitation cross section data not included in the other two event generators.

\begin{table*}[htb]
  \centering
    \topcaption{Probability of different neutron multiplicity classes (0n, 1n, and $X$n with $X \geq 1$) measured on each ZDC side for the exclusive $\EE$ events passing the fiducial criteria (first four rows of Table~\ref{tab:fiducialregion}), compared with the predictions of \superchic~4.2, \starlight~3.13, and \gammaUPC~1.6 for the deexcitation of the Pb ions in EMD processes. 
The experimental (MC model) uncertainties quoted are the square sum of statistical and systematic (MC statistical) sources.
    \label{tab:qed_ratios}}
    \begin{tabular}{c cccc}
    Neutron multiplicity     & \multicolumn{4}{c}{Probability (\%)} \\  
    category                 &  Data & \superchic~4.2  & \starlight~3.13 & \gammaUPC~1.6 \\ \hline
    0n0n & $74.0 \pm 0.7 $ & $76.6 \pm 1.0 $ & $74.5 \pm 1.0 $ & $74.9 \pm 2.4 $ \\
    0n$X$n + $X$n0n & $19.8 \pm 0.5 $ & $18.6 \pm 0.2 $ & $19.1 \pm 1.0 $ & $19.5 \pm 1.2 $ \\
    $X$n$X$n & $6.2 \pm 0.2 $ & $4.9 \pm 0.1 $ & $5.9 \pm 0.5 $ & $ 5.6 \pm 1.2 $\\
    0n1n + 1n0n & $4.5 \pm 0.2 $ & $6.4 \pm 0.1 $ & \NA & $ 5.8 \pm 0.2 $\\
    1n$X$n + $X$n1n & $3.7 \pm 0.1 $ & $3.0 \pm 0.0 $  & \NA & $ 3.0 \pm 0.2 $\\
    1n1n & $0.54 \pm 0.04 $ & $0.5 \pm 0.0 $ & $0.4 \pm 0.1 $ & $ 0.45 \pm 0.03 $\\ 
    \end{tabular}
\end{table*}

\begin{figure}[hbtp!]
\centering
 \includegraphics[width=0.7\textwidth]{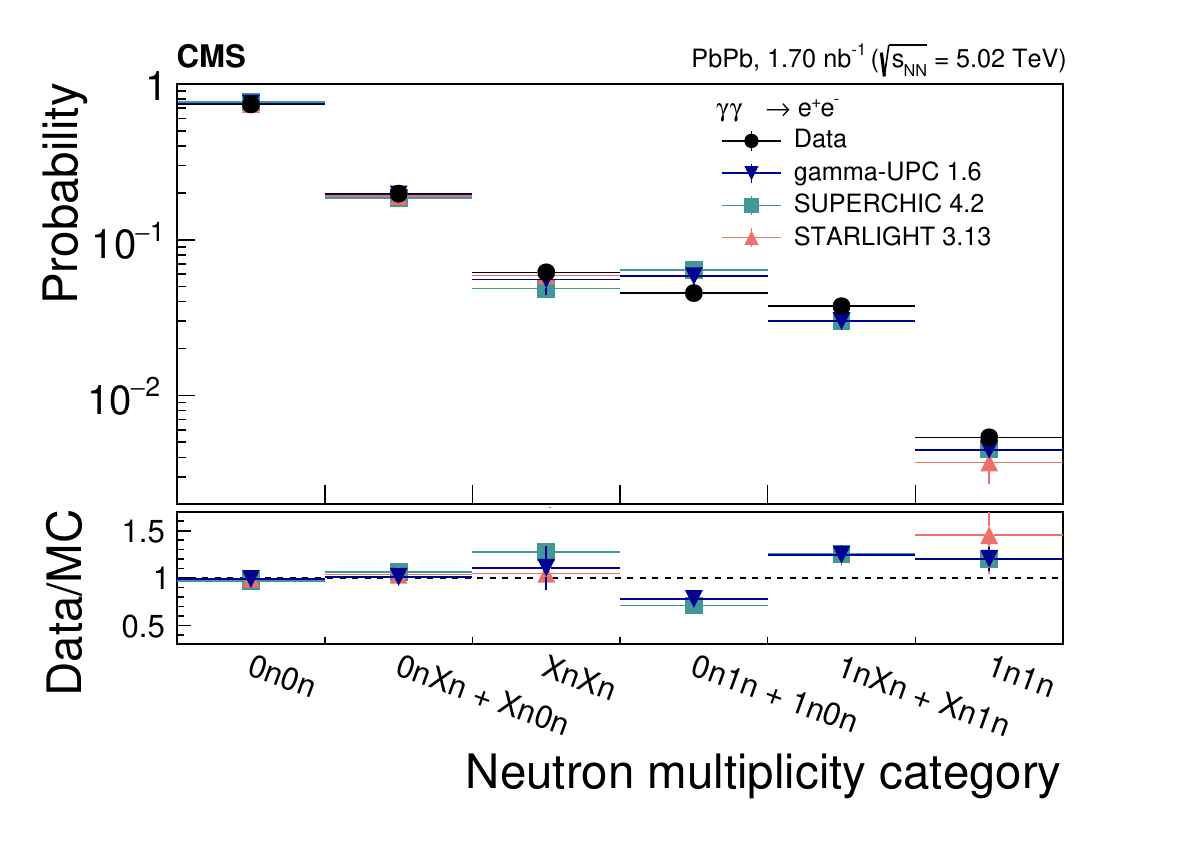}
 \caption{Probability for different neutron multiplicity classes (0n, 1n, and $X$n with $X \geq 1$) measured on each ZDC side for the exclusive $\EE$ events passing the fiducial phase space defined in Table~\ref{tab:fiducialregion}. The measured ratios are compared with \superchic~4.2, \starlight~3.13, and \gammaUPC~1.6 predictions.}
\label{fig:qed_ratios}
\end{figure}

The uncorrected kinematic distributions of the exclusive dielectron events obtained after all selection criteria (Fig.~\ref{fig:qed_distributions}) are unfolded to the particle level in the fiducial phase space defined in Table~\ref{tab:fiducialregion}. 
The only background to the BW process passing our fiducial criteria (Table 1) is exclusive $\PGU$ photoproduction, followed by the dielectron decay of the charmonium meson, which contributes to the $m^{\Pe\Pe} \approx 10$ GeV invariant mass bin. We have estimated this contribution to be below 1\% and neglected it in this study. 
The unfolding procedure corrects for bin migrations in the differential distributions. The default unfolding procedure is carried out with the \superchic$\,+\,$FSR(\photos++) MC simulation. In addition, unfolding based on the \starlight\ and \gammaUPC$\,+\,$FSR(\textsc{py}8) MC samples is performed to estimate the uncertainty due to the choice of the prior model. 
Each kinematic distribution is unfolded independently. The matrix inversion method is employed to unfold $m^{\Pe\Pe}$, $y^{\Pe\Pe}$ and $\abs{\cos\theta^{*}}^{\Pe\Pe}$ distributions with the \textsc{RooUnfold} package~\cite{Adye:2011gm}. For $\pt^{\Pe\Pe}$, unfolding with unregularized matrix inversion is not viable because of its sensitivity to statistical fluctuations, which lead to bin-to-bin oscillations in the final differential distributions. 
For this variable, the D'Agostini iterative unfolding with early stopping~\cite{DAgostini:1994fjx} has been used with three iterations (a value for which the ratio of unfolded and true distributions is very close to the ratio of reconstructed data and MC distributions). Prior to the unfolding, the response matrices are corrected for all SFs listed in Table~\ref{tab:eff_summary}. 
The unfolding procedure introduces an additional uncertainty in the final differential cross sections, beyond those listed in Table~\ref{tab:syst_summary_qed}, due to the MC-dependent shape of the input kinematic distributions. It amounts to $\pm 5\%$ on average, and $\pm 15\%$ in the tails of the distributions, and is added in quadrature bin-by-bin with the rest of the systematic uncertainties (hatched bands in Fig.~\ref{fig:QED_data_vs_theory}).

Figure~\ref{fig:QED_data_vs_theory} shows the comparison of the differential cross sections measured in data to the generator-level predictions from three MC considered: \superchic$\,+\,$FSR(\photos++), \starlight$\,+\,$FSR(\textsc{py}8), and \gammaUPC$\,+\,$FSR(\textsc{py}8). Within uncertainties, good agreement between data and predictions is found, except in the overall normalization of the \starlight$\,+\,$FSR(\textsc{py}8) distributions, which tend to consistently undershoot the measurements by about a factor of 15\%. The incorporation of the recoil due to the photon FSR improves the agreement between the \starlight\ prediction and the data in the tail of the pair $\pt^{\Pe\Pe}$ distribution in comparison with the default \starlight\ result (Fig.~\ref{fig:qed_distributions}, upper right).

\begin{figure}[hbtp!]
\centering
 \includegraphics[width=0.49\textwidth]{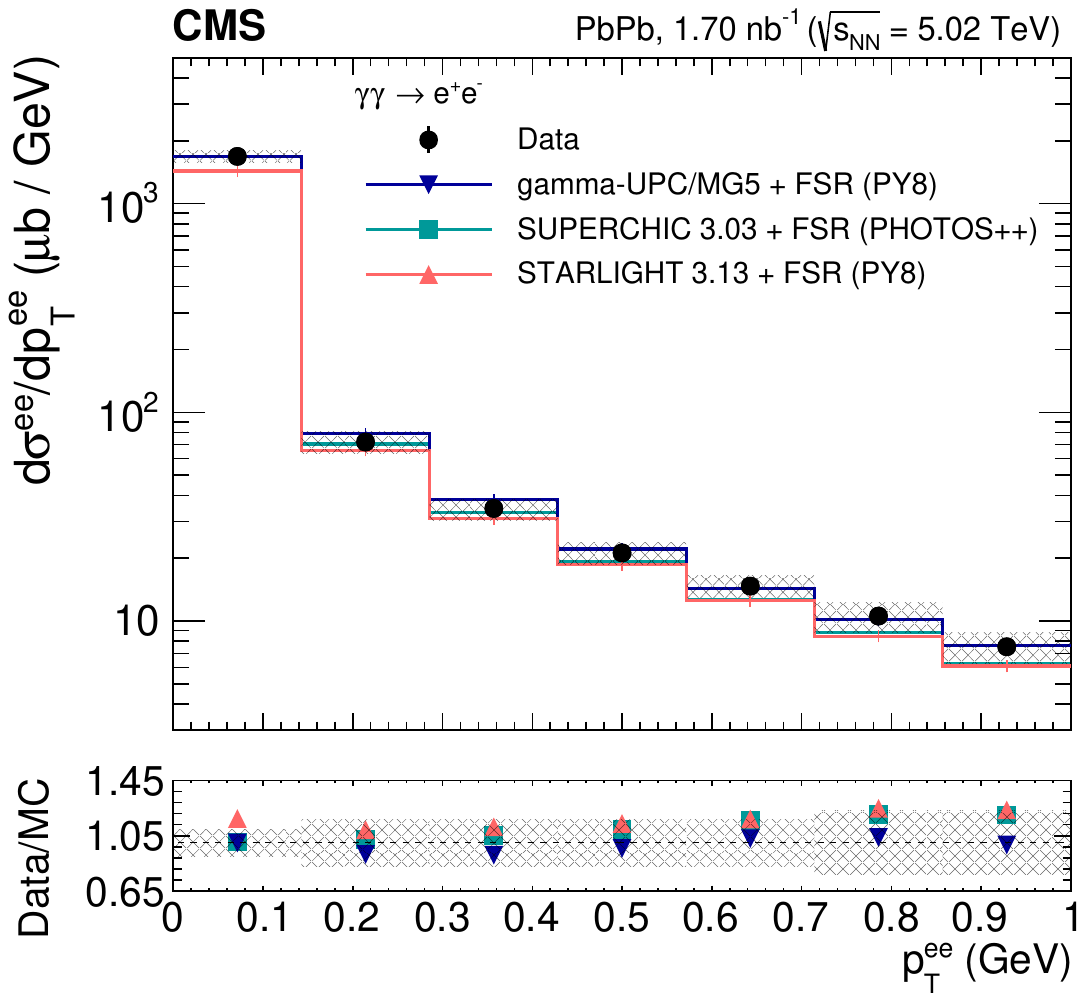}
 \includegraphics[width=0.49\textwidth]{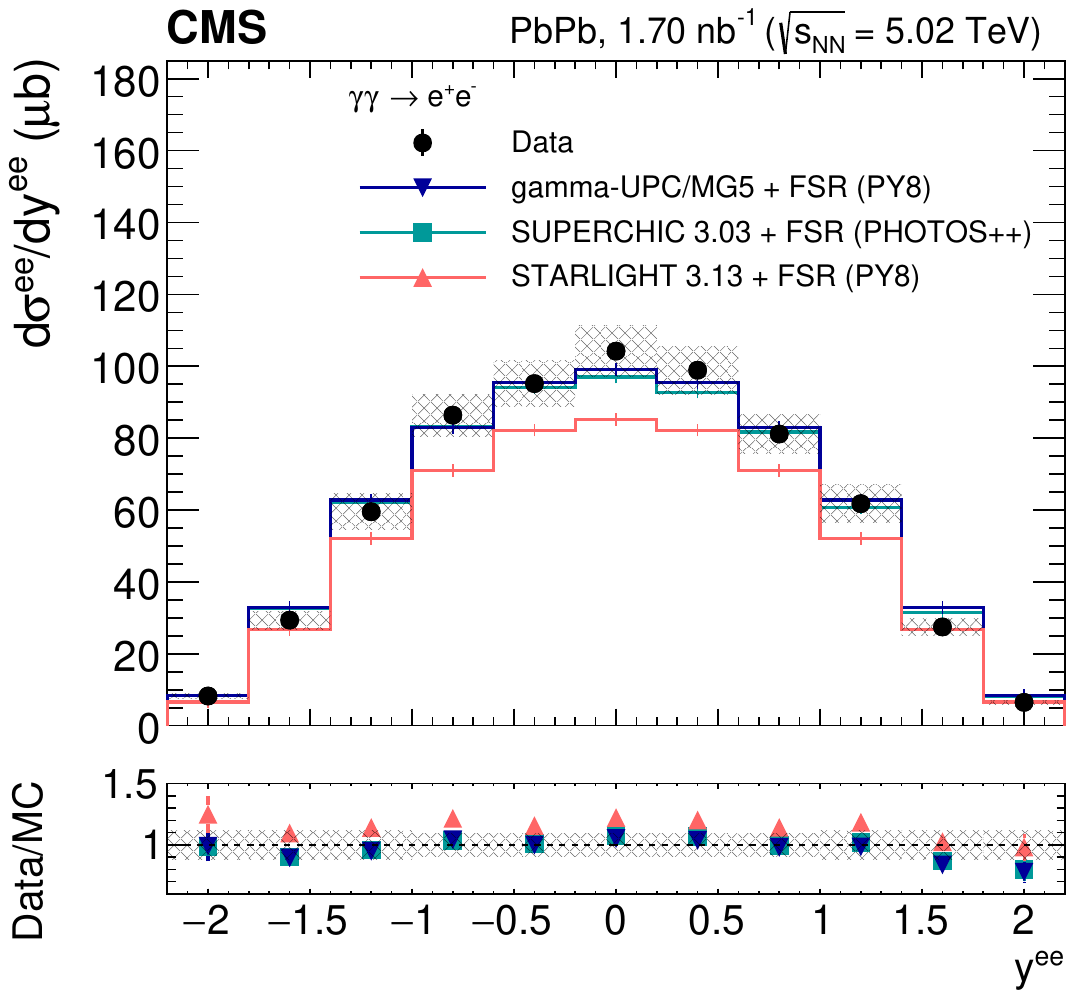}
 \includegraphics[width=0.49\textwidth]{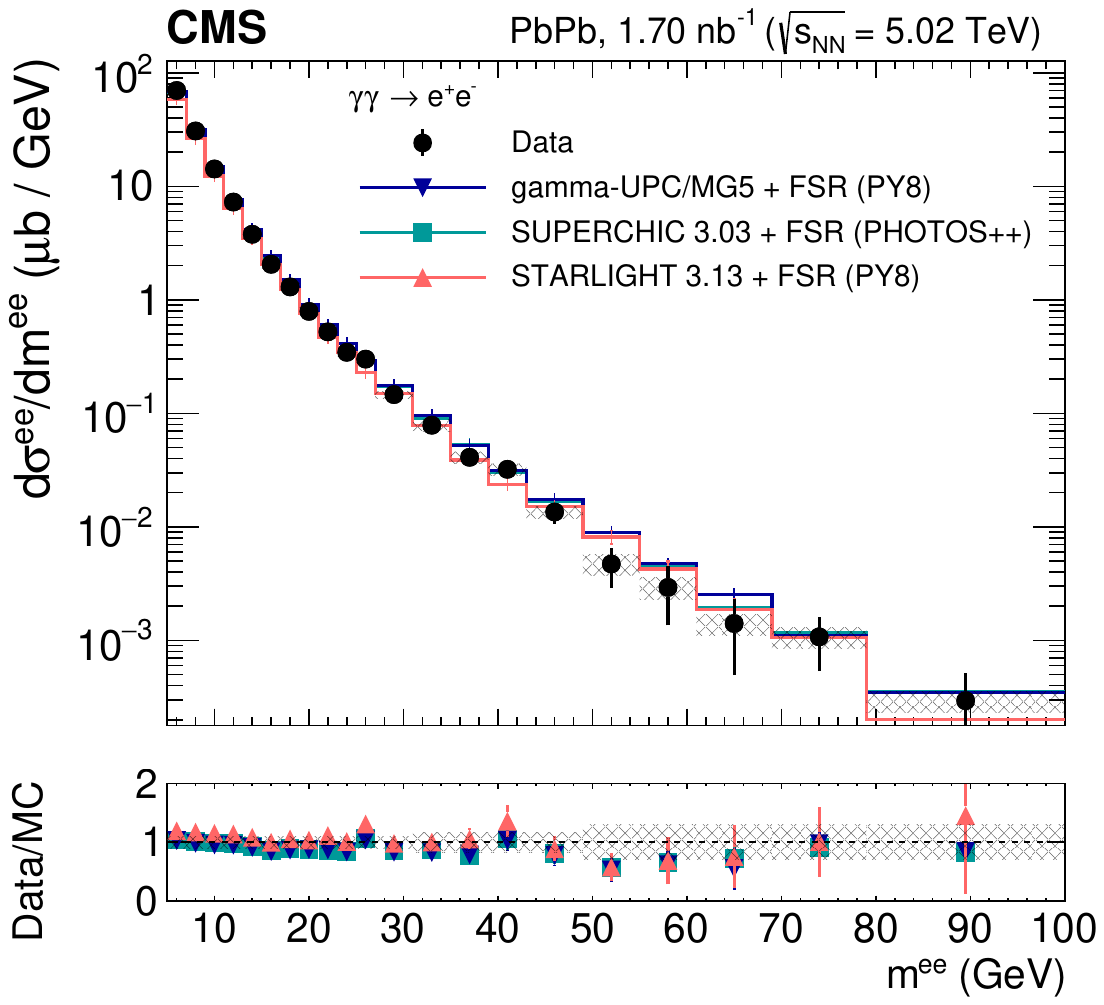}
 \includegraphics[width=0.49\textwidth]{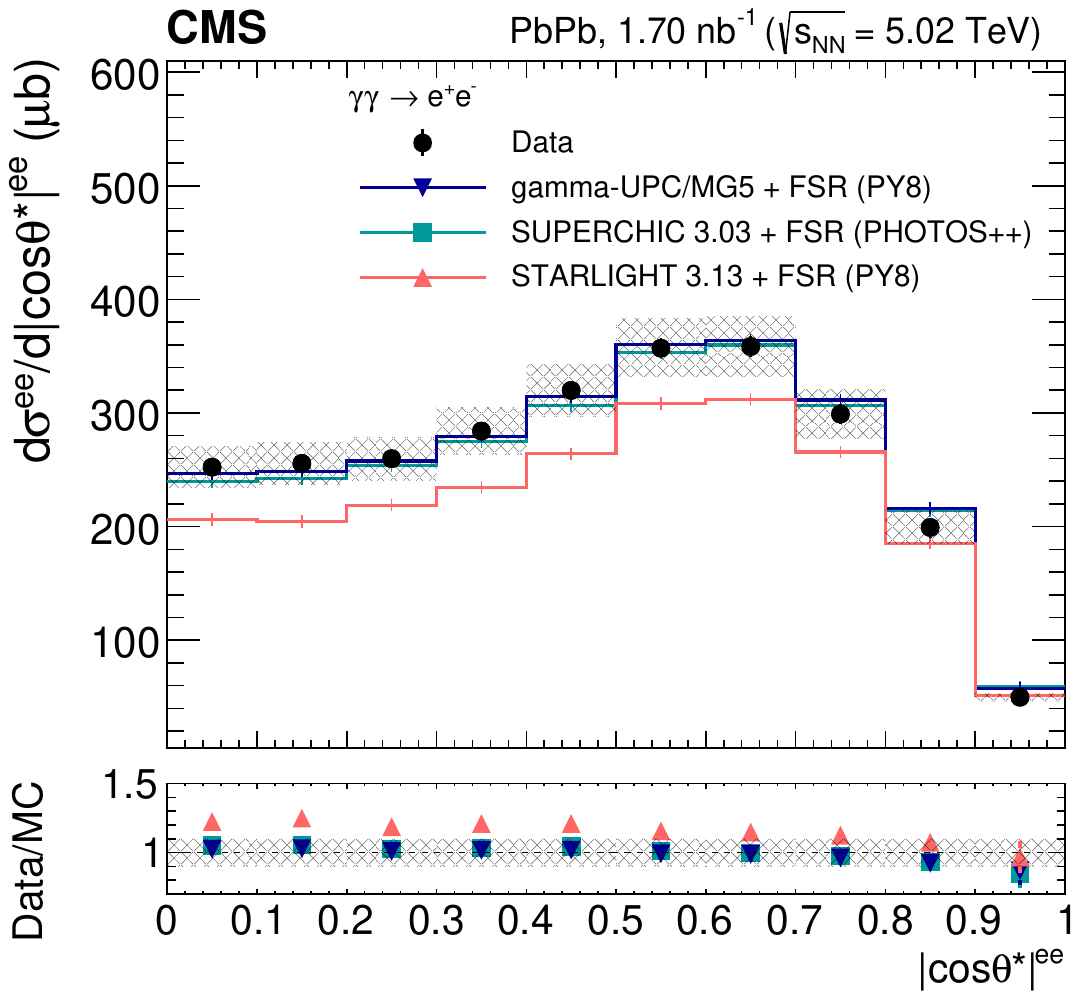}
 \caption{Differential cross sections for exclusive dielectron production, in the fiducial phase space defined in Table~\ref{tab:fiducialregion}, as functions of the pair $\pt$ (upper left), rapidity (upper right), invariant mass (lower left), and  $\abs{\cos\theta^{*}}$ (lower right). Data (black points) are compared with \superchic$\,+\,$FSR(\photos++), \starlight$\,+\,$FSR(\textsc{py}8), and \gammaUPC$\,+\,$FSR(\textsc{py}8) predictions. Vertical bars (hatched bands) show statistical (systematic) uncertainties.}
\label{fig:QED_data_vs_theory}
\end{figure}

\clearpage
\section{Measurement of light-by-light scattering}
\label{sec:LbL_results}

The exclusive diphoton signal is extracted after applying all selection criteria described in Section~\ref{sec:evt_sel}, and estimating and subtracting the residual backgrounds. The exclusive dielectron measurement has no significant physical background, but the much rarer LbL process receives contributions from BW $\gaga\to\EE$ (where the $\Pem$ and $\Pep$ both radiate a hard bremsstrahlung photon and/or are misidentified as photons) and from CEP $\Pg\Pg\to \gaga$ events. The number of events remaining in the data and MC simulations after applying each selection criteria (Section~\ref{sec:evt_sel}) is summarized in Table~\ref{tab:lbl_yields_inverted}. The numbers in the first rows, until applying exclusivity criteria, do not match between data and MC predictions, as the MC generation does not include nonexclusive backgrounds. It is also worth noting that the impact of the neutral exclusivity requirements is significant because they remove many events with noise in the electromagnetic and hadronic calorimeters. For the final selection, 26 events are observed in the signal region to be compared with the expected 12.8 LbL signal counts, and 10.1 CEP plus 1.9 BW FSR backgrounds events, determined as explained next. 

\begin{table*}[htb]
  \centering
  \topcaption{Exclusive diphoton yields after applying each selection criteria in data and MC simulations. The simulation yields are scaled by the integrated luminosity of the measurement and corrected by the SFs listed in Table~\ref{tab:eff_summary}. The (\%) column indicates the percentage of events remaining after applying the selection with respect to the previous row.
  \label{tab:lbl_yields_inverted}}
  \cmsTable{
    \begin{tabular}{lrrrrrrrrrr} 
    Selection criterion   & \multicolumn{2}{c}{Data}	  & \multicolumn{2}{c}{\superchic} & \multicolumn{2}{c}{\superchic$\,+\,$\photos++} & \multicolumn{2}{c}{\starlight} & \multicolumn{2}{c}{\superchic}  \\
    				       & \multicolumn{2}{c}{}		  & \multicolumn{2}{c}{LbL}  & \multicolumn{2}{c}{$\gaga\to\EE\,+\,$FSR}     & \multicolumn{2}{c}{$\gaga\to\EE$}     & \multicolumn{2}{c}{CEP}   \\ 
                          & $N_\text{events}$ & (\%) & $N_\text{events}$ & (\%) & $N_\text{events}$ & (\%) & $N_\text{events}$ & (\%) & $N_\text{events}$ & (\%) \\
    \hline
    Trigger						                   & 4\,600\,672  & \NA	&  47.1	    & \NA & 53\,400 & \NA & 47\,700 & \NA & 181.6 & \NA \\
    Two reco+ID $\PGg$ with & \multirow{2}{*}{1\,019\,569} & \multirow{2}{*}{22}  & \multirow{2}{*}{23.6} & \multirow{2}{*}{50} 
                                           & \multirow{2}{*}{24\,800}	  & \multirow{2}{*}{46} & \multirow{2}{*}{22\,600}	  & \multirow{2}{*}{47} & \multirow{2}{*}{88.9}	& \multirow{2}{*}{49} \\
     $\et>2\GeV$, $\abs{\eta}<2.2$           & 	      &      &   	 & & 	     & & 	  & \\
	Diphoton mass $m^{\gaga} > 5\GeV$		     & 582\,349 & 57  &  23.1 & 98  & 24\,200 & 97  & 22\,000 & 97 & 84.4 &	95  \\
	Charged exclusivity selection            & 436\,172 & 75  &  18.9 & 82  & 26.5    & 0.1 & 19.6 & 0.1 & 71.3 & 84  \\
	Neutral exclusivity	selection			       & 207	    & 0.05 &  15.0 & 79  & 20.6	  & 78  & 14.9 & 76  & 57.2 &  80  \\
	Diphoton $\pt < 1\GeV$		               & 83	      & 40   &  14.3 & 95  & 15.0	  & 73  & 11.5 & 77  & 47.3 &	83  \\
  ZDC$^{-} < 7\TeV$ or ZDC$^{+} < 7\TeV$   & 81       & 98   &  14.3 & 100 & 15.0   & 100 & 11.5 & 100 & 47.3 & 100 \\	
  Diphoton $\Aco^{\gaga} < 0.01$	 & 26  	 & 32       &  12.8 & 90   & 2.3 & 15     & 1.5 & 13  & 10.1 & 21  \\
  \end{tabular}
  }
\end{table*}

\subsection{Background subtraction}
\label{sec:backgd}

Because of its much higher rate than the LbL process, the exclusive production of electron pairs ($\gaga\to\EE$) can be a source of misidentified diphoton events in two circumstances. Misidentification of an electron as a photon can occur when the electron track is not reconstructed and/or when the electron pair emits one (or two) hard bremsstrahlung photon(s). The hard-bremsstrahlung emission can occur within the detector material, and such a contribution is already properly included in the \GEANTfour simulation, as well as in the efficiency SFs derived using CRs in data. The emission of physical FSR (prior to the electron/positron reaching the detector) is included in the \superchic$\,+\,$FSR(\photos++) and \gammaUPC$\,+\,$FSR(\textsc{py}8)~MC samples, where soft and collinear photons can be emitted by the $\EE$ pair. In addition, \gammaUPC/\MGvATNLO\ samples were generated for the $\gaga\to\EE\PGg$ and $\gaga\to\EE\gaga$ processes, to have an alternative simulation with harder photons. The study of the three simulated samples indicates that the only source of BW background to the LbL measurement is due to $\EE$ pairs that suffer hard bremsstrahlung in the material and get misidentified as photons. This conclusion was further confirmed from the data themselves by studying a sample of $\gaga\to\EE\PGg(\PGg)$ bremsstrahlung events where electrons were not identified, but their low-\et tracks could still be reconstructed. Since both \superchic\ and \starlight\ simulations of the BW background agree within uncertainties, their average is taken and any differences between them in the signal region are assigned as a systematic uncertainty, as explained below. The background from BW events, obtained by averaging out the \superchic\ and \starlight\ predictions, is thus estimated to be $1.9 \pm 0.4$ events in the signal region below $\Aco^{\gaga} = 0.01$.

\begin{figure}[hbtp!]
\centering
\includegraphics[width=0.8\textwidth]{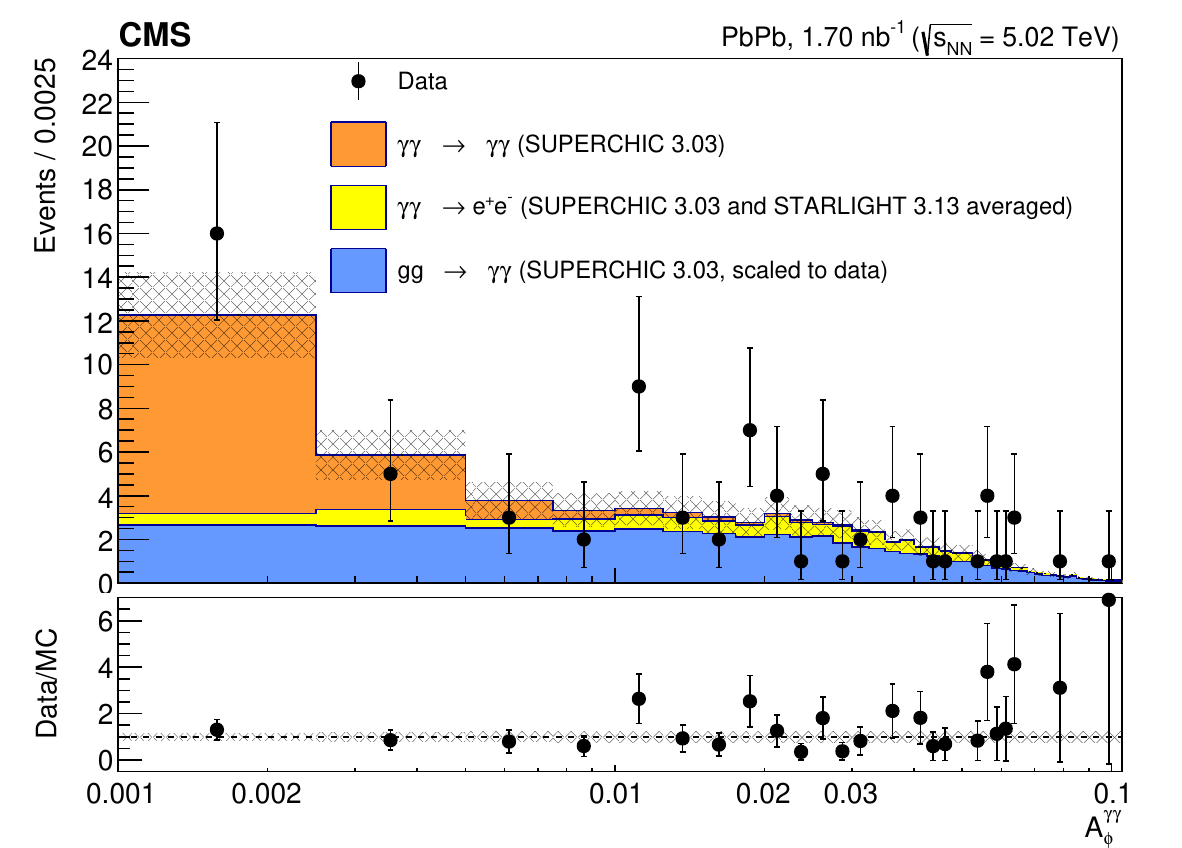}
\caption{Diphoton acoplanarity distribution over $\Aco^{\gaga} = 0$\nobreakdashes--0.1 in events passing the fiducial criteria of Table~\ref{tab:fiducialregion} (except the $\Aco^{\gaga}<0.01$ one) measured in data (black points) compared with the predictions for the LbL signal (orange histogram), the BW process (yellow histogram), and the CEP (blue histogram, normalized to data in the region $\Aco^{\gaga}>0.015$ as explained in the text) backgrounds. Error bars on the data points show statistical uncertainties, and dashed bands on the stacked histograms (and at unity in the lower data/MC ratio) represent systematic uncertainties.
\label{fig:kinematic_distributions_acoplanarity}}
\end{figure}

The theoretical CEP cross section (including coherent and incoherent contributions) in PbPb collisions is in principle expected to be much smaller than the LbL one~\cite{Khoze:2004ak,Harland-Lang:2018iur}, and the final-state diphoton \pt is harder than the LbL one~\cite{d'Enterria:2013yra}, such that less events pass the maximum \pt(pair) requirement in the LbL kinematic selection criteria. The CEP cross section has, however, very large uncertainties. The most striking feature of photons coming from LbL scattering is how nearly back-to-back they are, whereas any other diphoton background(s) typically features a larger azimuthal difference. Therefore, rather than relying fully on the MC predictions for the CEP background estimation, a method based on CRs in data is used instead that normalizes the sum of any remaining backgrounds to the tail of the diphoton acoplanarity distribution where no LbL signal is expected. Such a method also includes by definition any other potential exclusive processes remaining in the selected events (such as \eg\ scalar and tensor bottomonium resonances that can be produced via $\gaga\to \PGhb,\PGc_{\PQb}$ and decay back to diphotons, although their expected yields~\cite{Shao:2022cly} are much smaller than the LbL continuum signal). For this purpose, as done in the 2015 PbPb run analysis~\cite{CMS:2018erd}, the CEP MC contribution is scaled to the data in the acoplanarity tail $\Aco^{\gaga} > 0.015$ where no significant LbL signal is expected and, from there, it is extrapolated to the signal region: $\Aco^{\gaga} < 0.01$. The tail-based normalization factor is defined as:
\begin{equation}
f_\text{CEP,MC}^\text{norm,backgd} = \frac{N_\text{data}(\Aco^{\gaga} > 0.015) - N_\text{LbL,MC}(\Aco^{\gaga} > 0.015) - N_\text{BW,MC}(\Aco^{\gaga} > 0.015)} {N_\text{CEP,MC}(\Aco^{\gaga} > 0.015)},
\label{eq:CEPINcoh_norm}
\end{equation}
where $N(\Aco^{\gaga} > 0.015)$ indicates the number of exclusive diphoton candidate events observed in data and MC simulations in the region $\Aco^{\gaga}>0.015$.

Figure~\ref{fig:kinematic_distributions_acoplanarity} shows the diphoton acoplanarity distribution over $\Aco^{\gaga} = 0$--0.1 measured in data (black points) compared with MC predictions for the LbL process (orange histogram), the BW process (yellow histogram), and CEP (blue histogram, normalized to data as explained above). There are two sources of systematic uncertainty in the normalization of the CEP background in the signal region, which are added in quadrature: $\pm15\%$ to account for the finite number of measured events in the sideband acoplanarity region above $\Aco^{\gaga} = 0.015$ to which the simulation has been scaled, and $\pm14\%$ assigned to account for possible differences between the actual acoplanarity shape of the simulated samples and the observed events. This latter uncertainty is derived from the differences in the integral of the total background yield by varying the background acoplanarity distribution to be more similar in shape to the CEP or BW distributions for $\Aco^{\gaga}<0.015$.
The uncertainties of statistical (15\%) and nonstatistical (19\%) nature combined in quadrature amount to $\pm24\%$, as shown in Table~\ref{tab:syst_summary}. The background in the signal region below $\Aco^{\gaga} = 0.01$ from CEP plus any other acoplanar diphoton processes is thus estimated to be $10.1 \pm 2.4$ events. Adding the contribution from misidentified $\gaga\to\EE$ pairs, the total background in the LbL signal region amounts to $12.0 \pm 2.9$ events.

Figure~\ref{fig:kinematic_distributions_photons} shows the comparison of single photon and diphoton kinematic variables for data and MC simulations. All MC contributions are normalized to match $\sigma_\text{fid,MC}\mathcal{L}_\text{int}$ and are multiplied by the SFs listed in Table~\ref{tab:eff_summary}. The CEP MC has been scaled to match the data in the region of acoplanarity $\Aco^{\gaga}>0.015$ (Fig.~\ref{fig:kinematic_distributions_acoplanarity}), as explained above. Both the measured yields and kinematic distributions are in accord with the combination of the LbL scattering signal plus BW process and scaled-CEP background expectations.

\begin{figure}[hbtp!]
\centering
\includegraphics[width=0.45\textwidth]{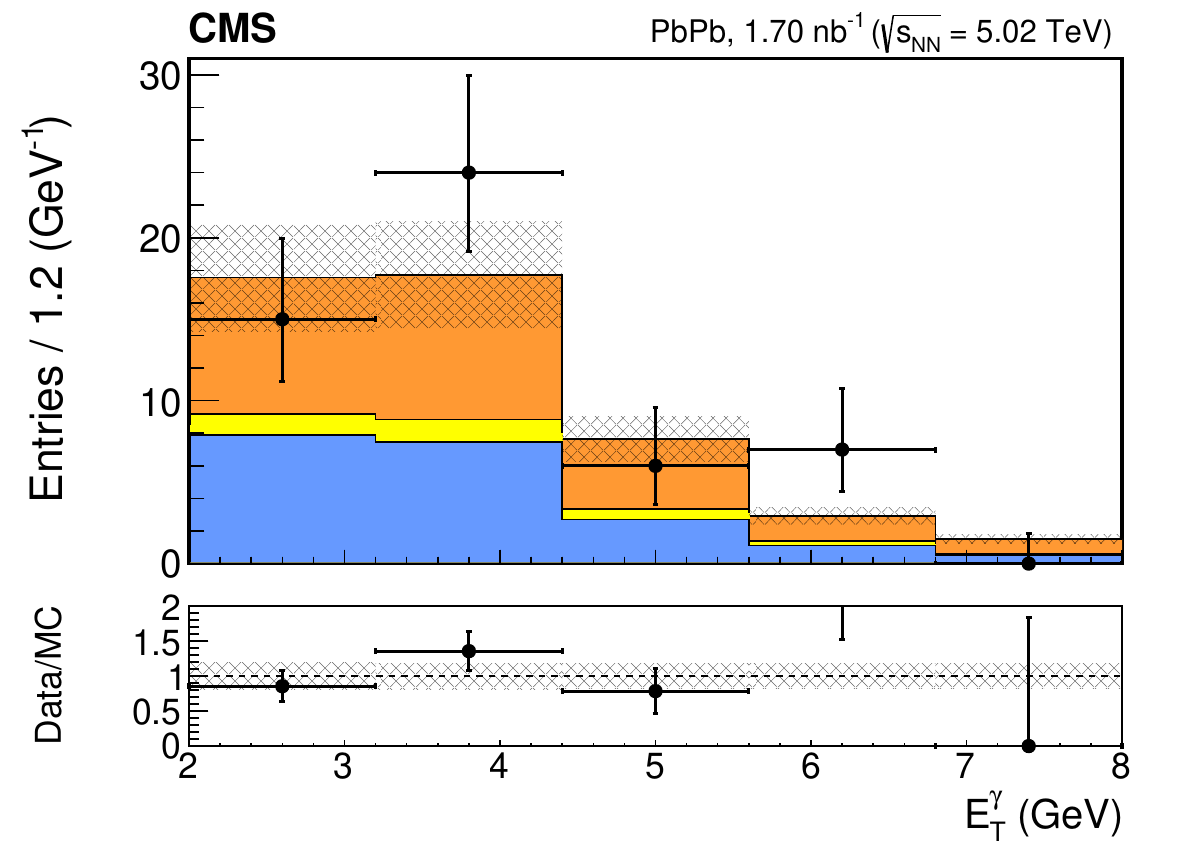}
\includegraphics[width=0.45\textwidth]{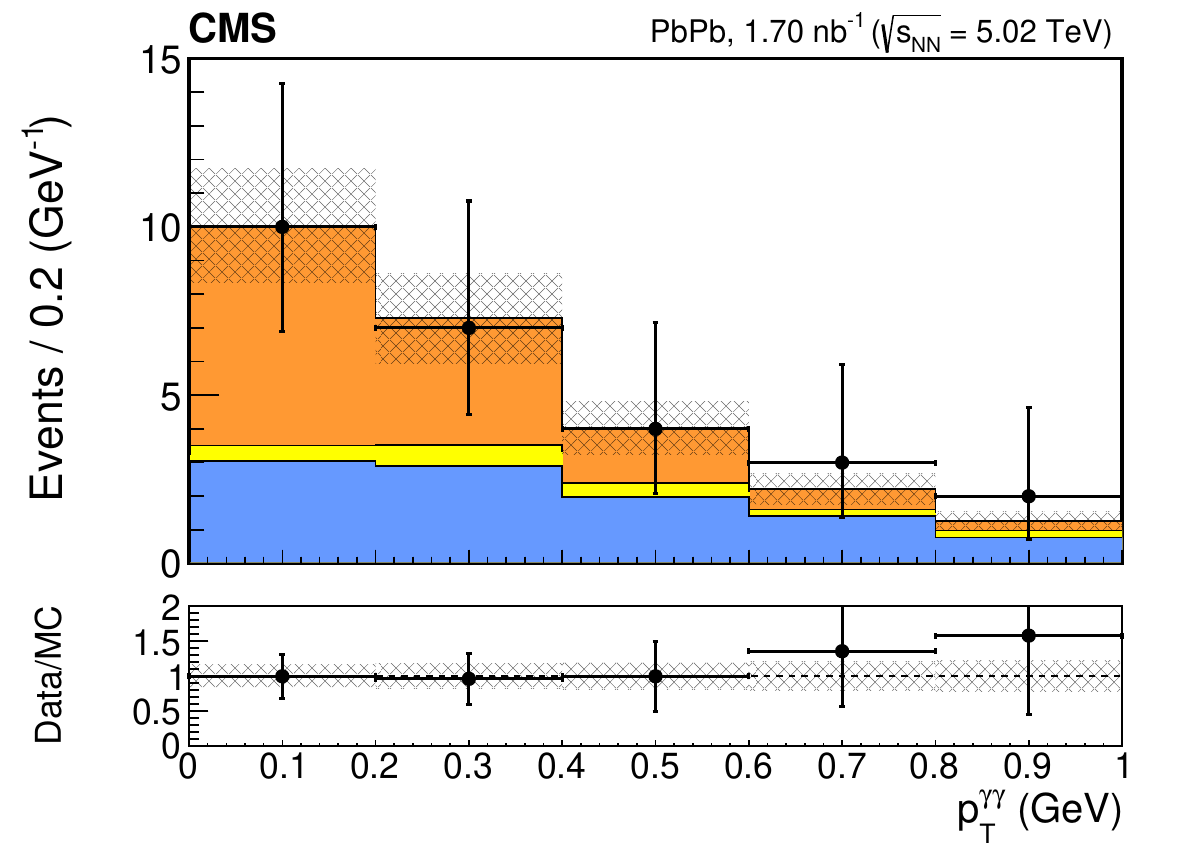}
\includegraphics[width=0.45\textwidth]{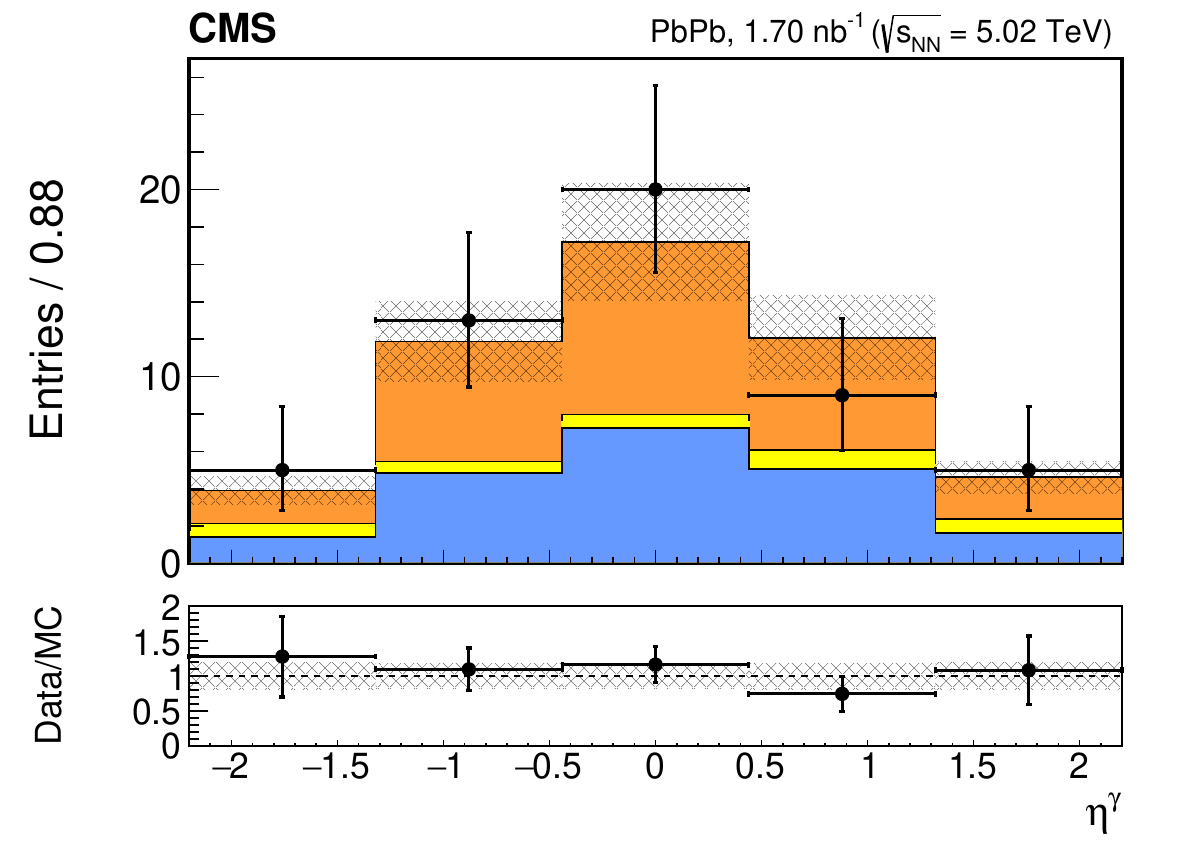}
\includegraphics[width=0.45\textwidth]{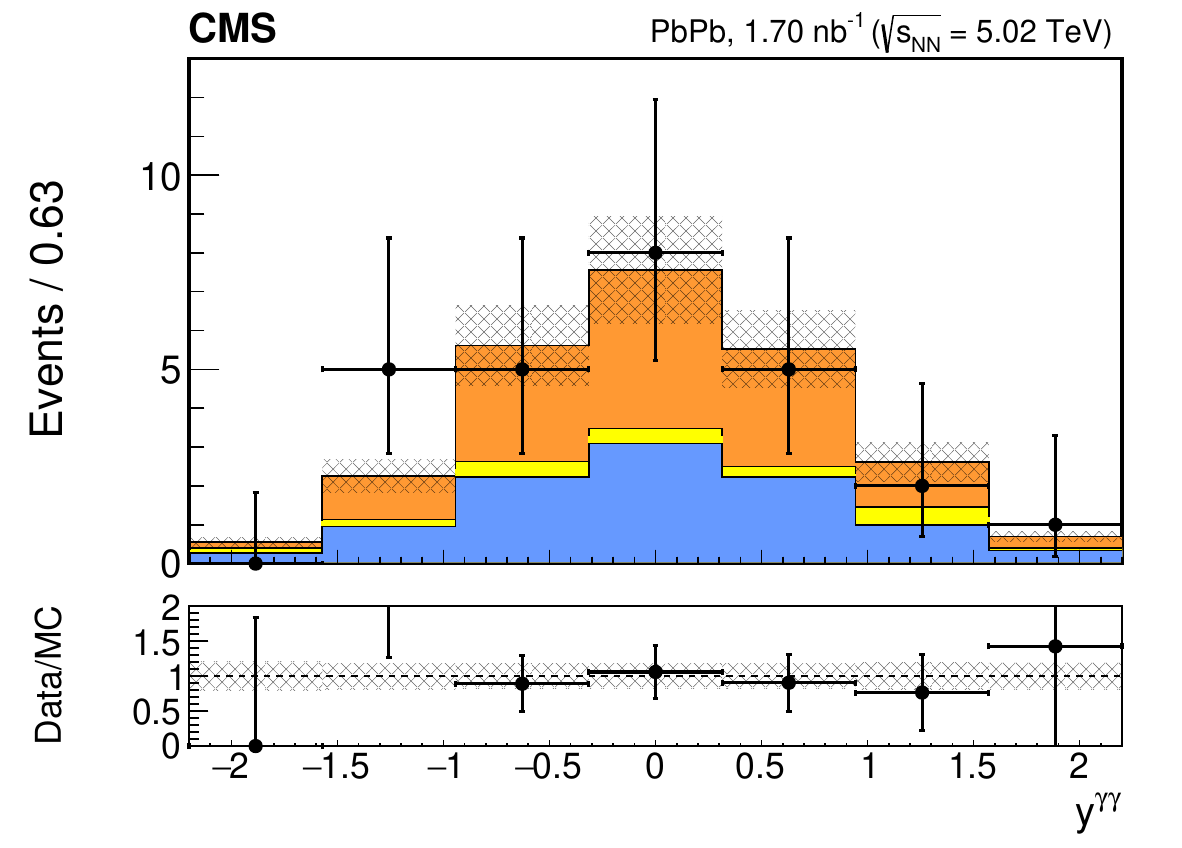}
\includegraphics[width=0.45\textwidth]{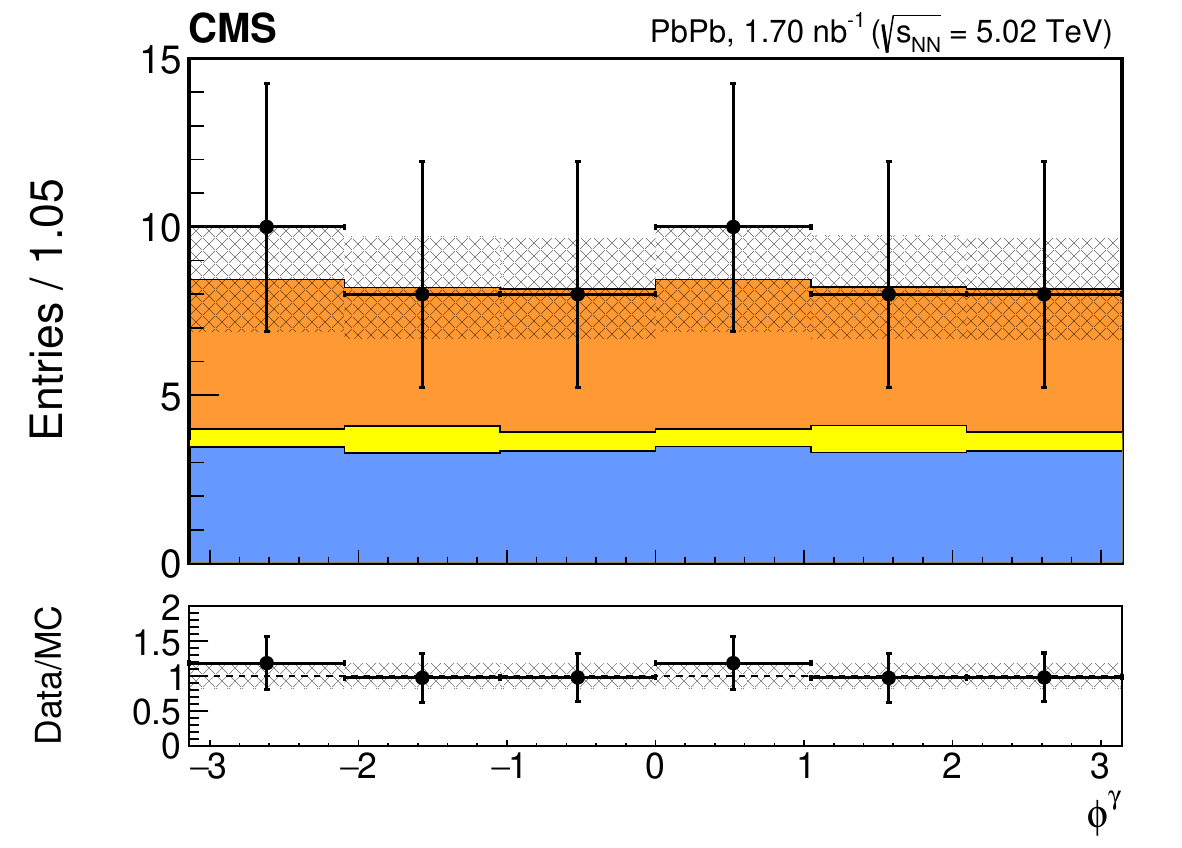}
\includegraphics[width=0.45\textwidth]{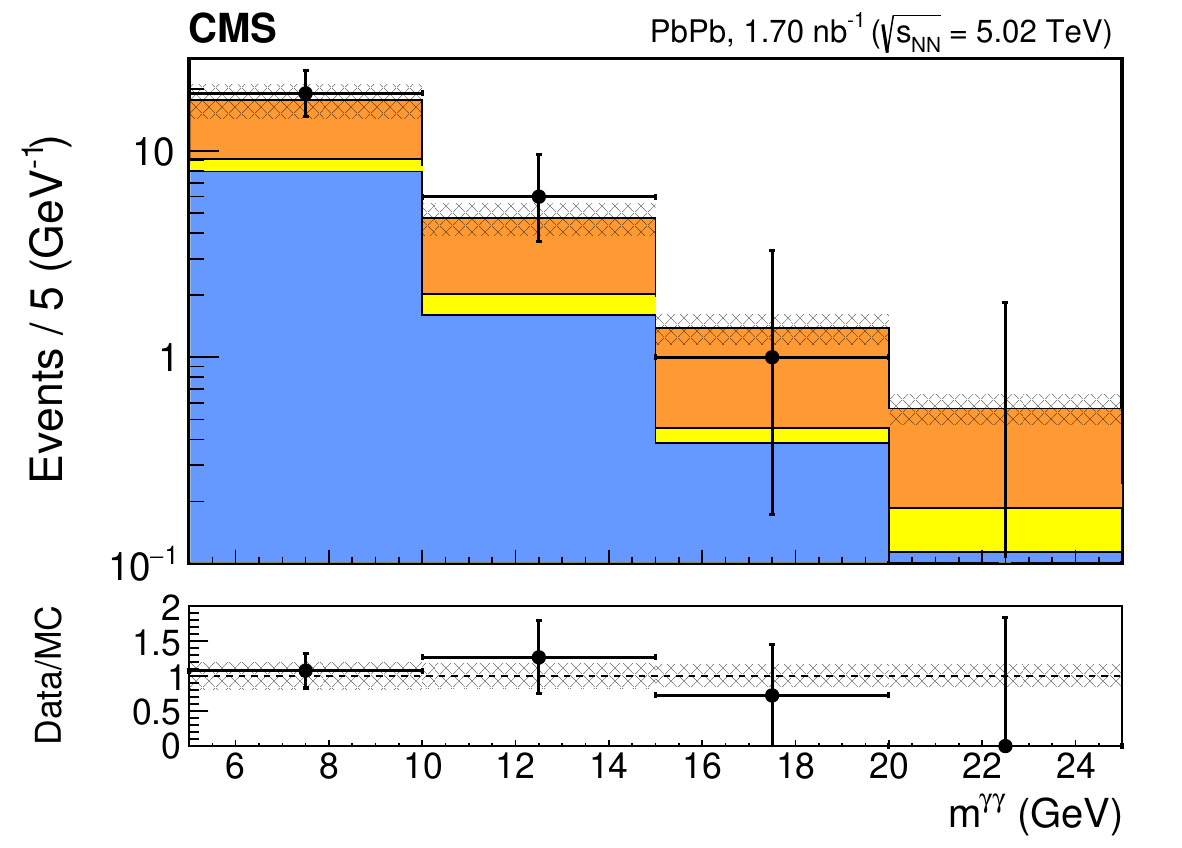}
\includegraphics[width=0.45\textwidth]{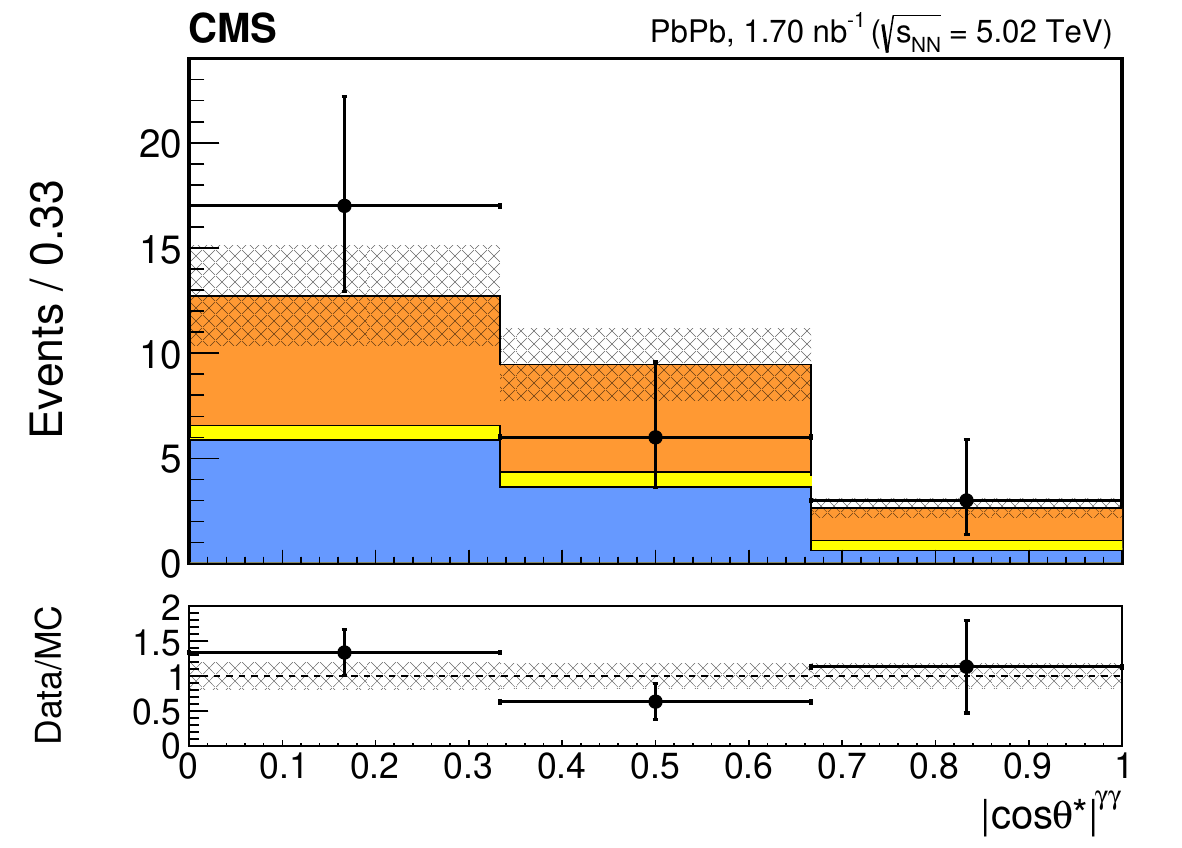}
\includegraphics[width=0.45\textwidth]{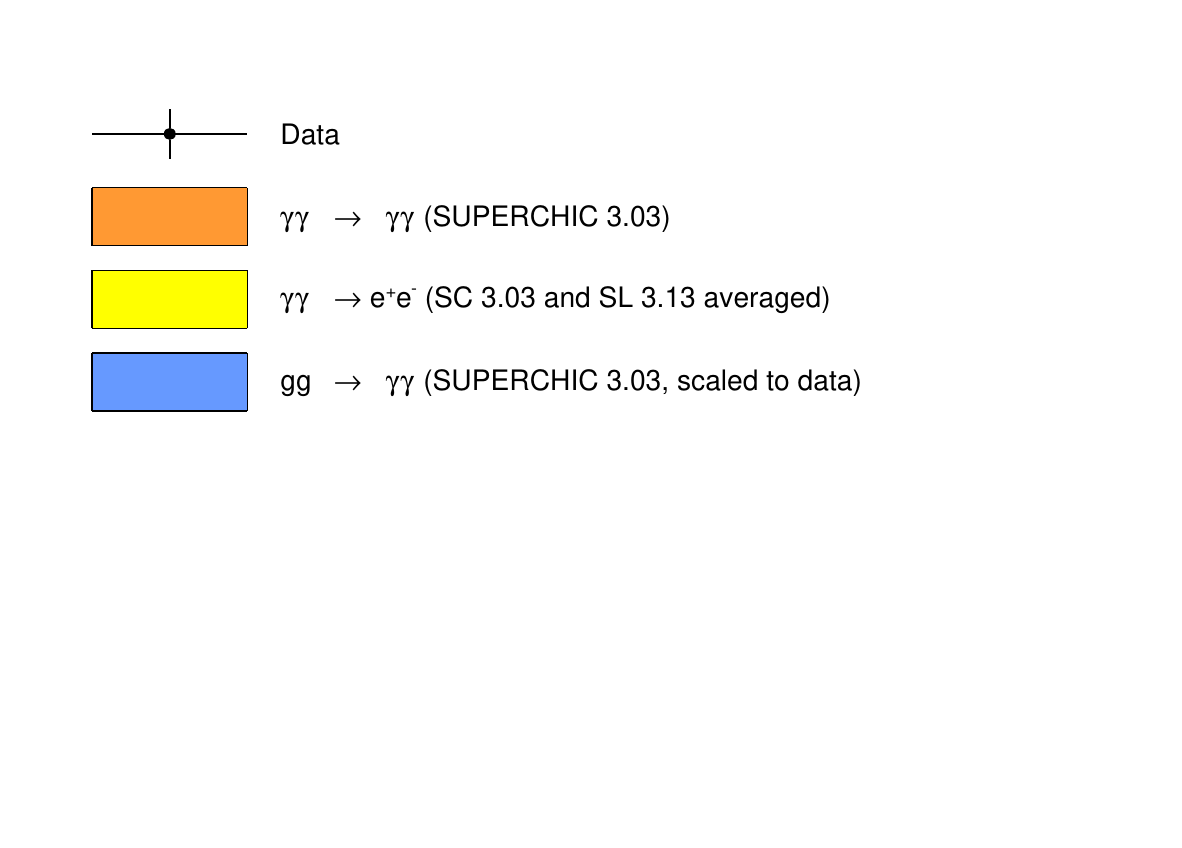}
\caption{Detector-level kinematic distributions for candidate exclusive diphoton events passing all selection criteria (Table~\ref{tab:fiducialregion}) in the data (black points) compared with the simulated LbL scattering signal (orange histogram) and backgrounds from the BW (yellow histogram) and CEP (blue histogram, scaled as described in the text) processes. The MC simulations are normalized to match $\sigma_\text{fid,MC}\mathcal{L}_\text{int}$, and corrected with the SFs listed in Table~\ref{tab:eff_summary}. Error bars on the data points show statistical uncertainties, and dashed bands on the stacked histograms (and at unity in the lower-panel data/MC ratios) represent systematic uncertainties.
\label{fig:kinematic_distributions_photons}
}
\end{figure}

\subsection{Systematic uncertainties and signal significance}

The systematic uncertainties in the LbL scattering measurement are summarized in Table~\ref{tab:syst_summary}.
The most important source of uncertainty in the LbL yields is that of the background normalization and shape combined, and amounts to $\pm21\%$. The data-driven efficiency factor, $C^{\gaga}$ given by Eq.~(\ref{eq:corr_fac}), has an uncertainty of 12.5\%, mostly dominated by the data-to-simulation SF of the trigger efficiency. The integrated luminosity uncertainty, relevant for the final cross section extraction, is 1.7\%~\cite{CMS:2025rzq}.

\begin{table}[htbp]
\centering
\topcaption{\label{tab:syst_summary}
 Summary of relative systematic uncertainties in the measurement of the LbL scattering cross section.}
\begin{tabular}{lc} \hline
Background normalization & 15\% \\
Background shape & 14\% \\
Exclusive diphoton SFs and efficiency & 12.5\%\\
Integrated luminosity & 1.7\%\\  [\cmsTabSkip] 
Total (statistical/nonstatistical) & 24\% (15\%/19\%) \\ \hline
\end{tabular}
\end{table}

The compatibility of the data with the background-only hypothesis is evaluated from the measured acoplanarity distribution (Fig.~\ref{fig:kinematic_distributions_acoplanarity}) using a profile-likelihood ratio test statistic modified for upper limits~\cite{Cowan:2010js}. Systematic uncertainties are included by introducing nuisance parameters that modulate the number of expected events following a log-normal probability density function. The uncertainty from the finite size of the MC samples is also included as an additional nuisance parameter for each bin of the histogram~\cite{Barlow:1993dm,Conway:2011in}.
The significance for the excess at low diphoton acoplanarity in data, estimated from the expected distribution of the test statistic for the background-only hypothesis obtained with the asymptotic formula~\cite{Cowan:2010js}, is 4.7 standard deviations (3.8 standard deviations expected).

In addition, the significance of the LbL signal has been recalculated by adding the number of measured counts in the analysis of the 2015 PbPb run data~\cite{CMS:2018erd}, which showed 14 events observed, compared with expectations of $9.0 \pm 0.9$ events for the LbL signal and $4.0 \pm 1.2 \stat$ for the background processes. The difference in signal and background efficiencies between 2015 and 2018 mainly results from tighter selections applied in the latter due to the ageing of the calorimeters that lead to increased noise levels. After correcting for a small fiducial acceptance difference (photons were previously measured over $\abs{\eta^\PGg}<2.4$, instead of $\abs{\eta^\PGg}<2.2$ now), both results were analyzed with the \textsc{Combine} tool~\cite{CMS:2024onh}. By exploiting the shape of the combined acoplanarity distribution, an asymptotic significance of 6.2 standard deviations (5.5 expected) is obtained.

\subsection{Fiducial LbL cross section}

The LbL scattering fiducial cross section for photon pairs passing all selections listed in Table~\ref{tab:fiducialregion}, is determined from a fit of the signal strength in the combined acoplanarity distributions of 2015 and 2018, using the \superchic\ cross section of $\sigma_\text{fid}^\text{LO} = 93$\unit{nb} as a reference, and yields
\begin{linenomath*}
\begin{equation}
\sigma_\text{fid}(\gaga\to\gaga) = 107 \pm 24 \stat \pm 13 \syst~\unit{nb}.
\label{eq:Req}
\end{equation}
\end{linenomath*}
The experimental measurement agrees within uncertainties with the theoretical predictions at LO, $\sigma_\text{fid}^\text{LO} = 93$\unit{nb}, and NLO: $\sigma_\text{fid}^\text{NLO} = 95.5^{+2.0}_{-1.0}~\text{(scale)}^{+1.0}_{-1.5}~\text{(param)}$\unit{nb}, obtained with \gammaUPC@NLO. The predicted NLO cross section has a total relative uncertainty of about 2\%, shared in about equal parts between the missing higher-order uncertainties~\cite{H:2023wfg,H:2023znv} and parametric uncertainties from the Pb survival probability. The latter are derived by varying the nuclear radius and diffusivity of the Pb nucleus with a Glauber MC model~\cite{Loizides:2017ack}.

The uncorrected kinematic distributions of the selected exclusive diphoton events obtained after background subtraction (Fig.~\ref{fig:kinematic_distributions_photons}) are unfolded to the particle level in the fiducial phase space defined in Table~\ref{tab:fiducialregion} using the same \textsc{RooUnfold} procedure applied to the BW distributions (Section~\ref{sec:QEDee_results}). Prior to the unfolding, the response matrix is corrected for all SFs listed in Table~\ref{tab:eff_summary}, determined previously. The matrix inversion method has been used to obtain the diphoton rapidity and invariant mass differential cross sections. The results are presented in Fig.~\ref{fig:unfold_lbl_ratios}. The upper panels present the unfolded distributions measured in the data, compared with the corresponding \superchic\ and \gammaUPC@NLO predictions, and the lower panels their ratio. Good agreement between both predictions and the unfolded data for diphoton rapidity and invariant mass can be seen within experimental uncertainties.

\begin{figure}[hbtp!]
\centering
\includegraphics[width=0.49\textwidth]{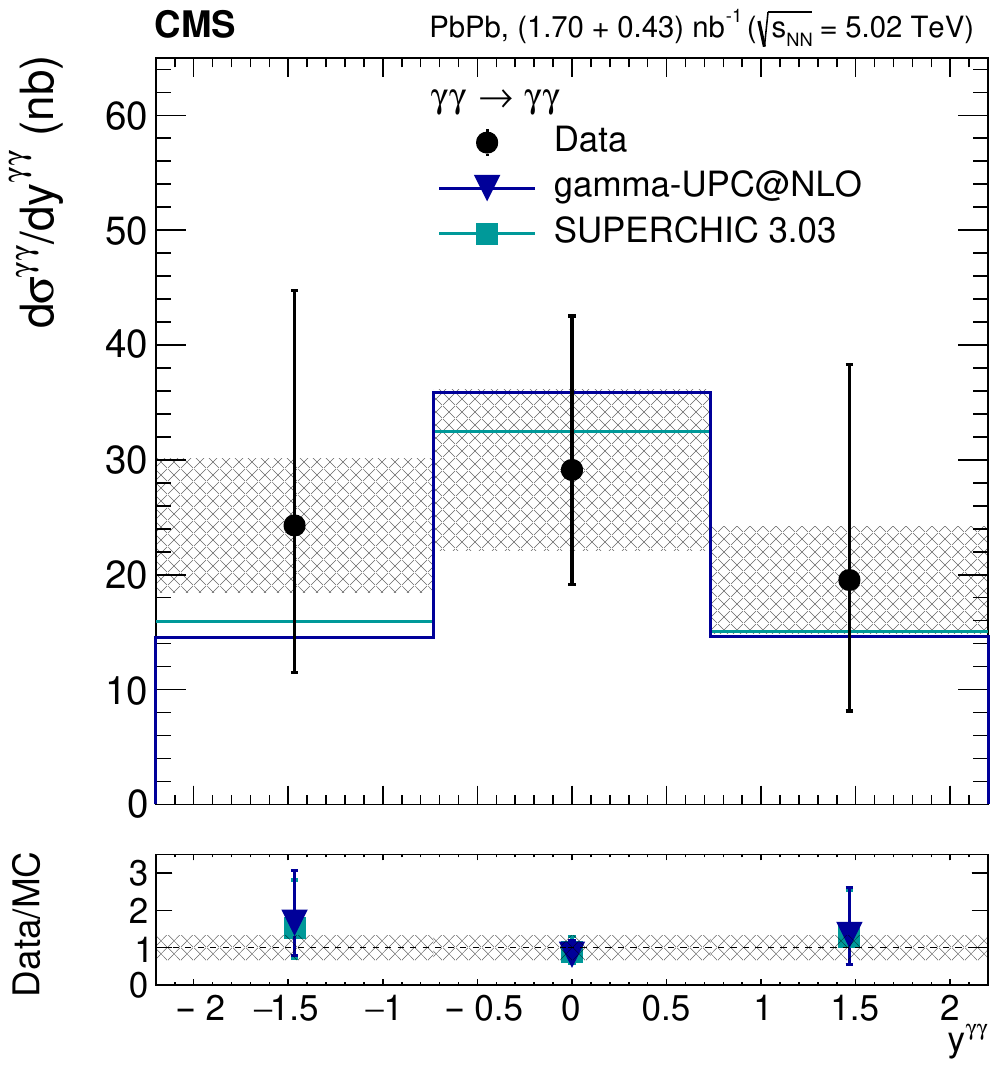}
\includegraphics[width=0.49\textwidth]{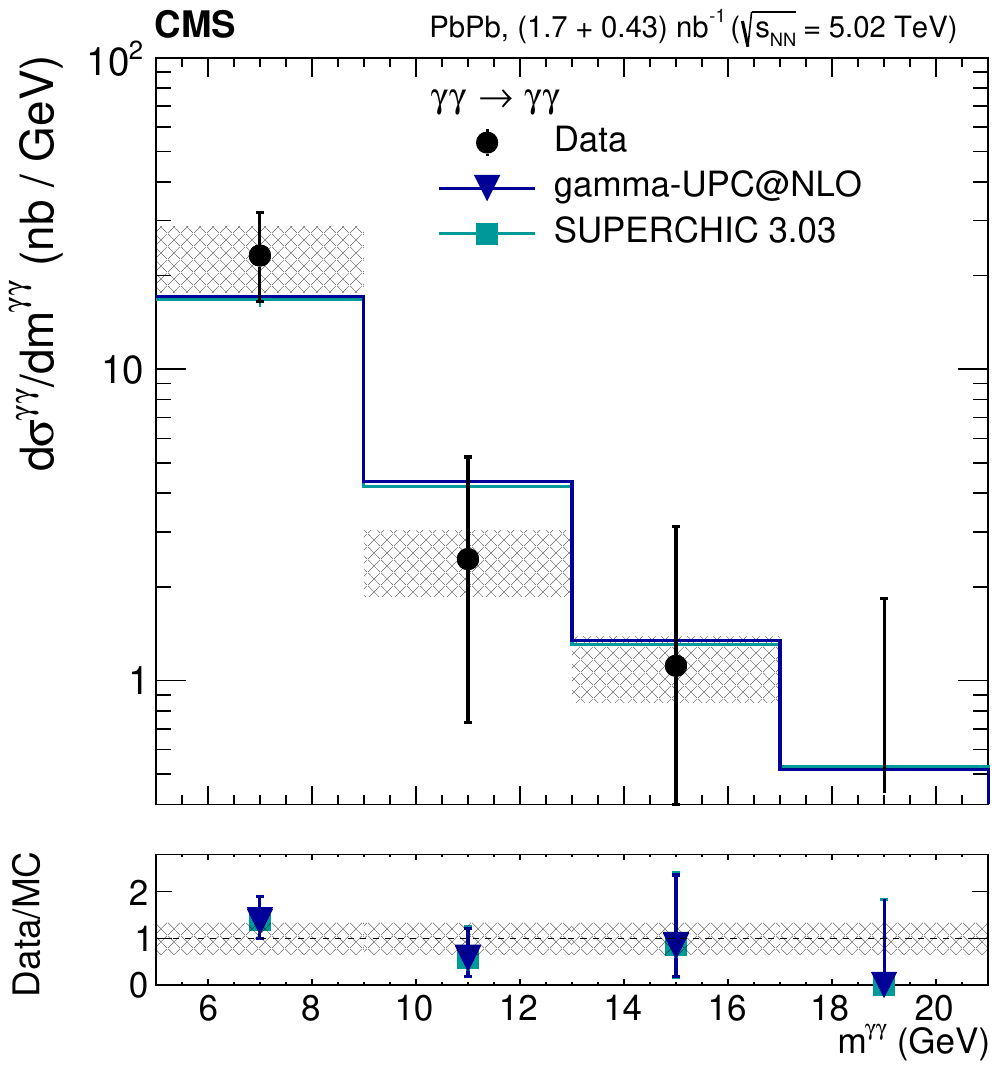}
\caption{Differential exclusive diphoton cross sections in the fiducial phase space defined in Table~\ref{tab:fiducialregion} as a function of the diphoton rapidity (left) and invariant mass (right) measured in data (black points) compared with  \superchic\ and \gammaUPC@NLO predictions. The lower panels show the corresponding data/MC ratios. Vertical bars (hatched bands) indicate statistical (systematic) uncertainties.}
\label{fig:unfold_lbl_ratios}
\end{figure}

\section{Limits on axion-like particles}
\label{sec:ALPs}

Searching for an excess with respect to the expected $\gaga\to \gaga$ continuum has been proposed to identify the production of ALPs in UPCs~\cite{Knapen:2016moh}. The measured invariant mass distribution of candidate diphoton events (Fig.~\ref{fig:kinematic_distributions_photons}, third row, right) is used to search for possible narrow resonances. The LbL, BW, and CEP continuum processes are considered as backgrounds in this search. Different \superchic\ MC samples are generated for ALP masses, $m_{\Pa}$, ranging from 5 to 100\GeV. The calculation of the ALP photon-fusion cross sections is based on the Lagrangian density
\begin{equation}
\mathcal{L} \supset \frac{1}{2}\partial_\mu a \partial^\mu a-\frac{m_{\Pa}^2}{2}a^2-\frac{g_{\Pa\PGg}}{4}a F^{\mu \nu}\widetilde{F}_{\mu\nu},\;\text{with } g_{\Pa\PGg} \equiv \mathcal{C}_{\gaga}/\Lambda, 
\label{eq:Laxion}
\end{equation}
where $a$ is the ALP field, $\widetilde{F}_{\mu\nu}$ is the photon field strength dual tensor, and the dimensionful ALP-$\PGg$ coupling strength $g_{\Pa\PGg}$ is inversely proportional to the high-energy scale $\Lambda$ associated with the spontaneous breaking of an approximate Peccei--Quinn global U$(1)$ symmetry~\cite{Peccei:1977hh}. The effective dimensionless coefficient $\mathcal{C}_{\gaga}$ can be used to rescale the ALP-$\PGg$ coupling whenever the ALP also interacts with, and consequently also decays to, other SM particles. The photon-dominant, or photophilic, $\mathcal{C}_{\gaga} = 1$ case considered here is the most common one found in the literature~\cite{Agrawal:2021dbo,dEnterria:2021ljz}.

Examples of two simulated ALP signals with masses $m_{\Pa} = 14, 30\GeV$ and $g_{\Pa\PGg}=0.25\TeV^{-1}$ are shown in the left plot of Fig.~\ref{fig:alp_limits_xsec} (gray and red histograms, respectively).
The simulated ALP samples are reconstructed and processed as done for the LbL final state to estimate the acceptance $\mathcal{A}$ and efficiency $\varepsilon$, as well as the expected reconstructed diphoton mass template distributions. Corrections to the efficiency estimated in the MC simulation are derived based on data, and applied in the same way as for the LbL final state. The diphoton mass templates used in the limit-setting procedure are built using the fully simulated and reconstructed ALP samples, after the final selection. Their cross sections are scaled by $\sigma_{\gaga\to\Pa} \mathcal{L}_\text{int} \mathcal{A} \varepsilon$, with $\sigma_{\gaga\to\Pa}$ arbitrarily fixed to 10\unit{nb} as reference point. 

Upper limits at 95\% confidence level (\CL) are then determined using the \textsc{Combine} tool with asymptotic formulae. A binned maximum likelihood fit is performed in the same way as done in the LbL analysis. The systematic uncertainty is 100\% correlated between the LbL background and the ALP signal. An additional uncorrelated uncertainty of $\pm3\%$ is added to the ALP signal cross section corresponding to the MC statistical uncertainty propagated to its reconstruction efficiency. The results of this analysis are combined with our previous limits~\cite{CMS:2018erd}, assuming fully uncorrelated uncertainties. 

The \CLs criterion~\cite{CLS2,CLS1}, with a profile-likelihood ratio as test statistic~\cite{ATLAS:2011tau}, is used to extract exclusion limits in the $\sigma(\gaga\to\Pa\to\gaga)$ cross section at 95\% \CL, as well as the 68 and 95\% bands around the expected limits. Limits on $\sigma(\gaga\to\Pa\to\gaga)$ cross section for ALPs with masses 5--100\GeV are set in the 5--200\unit{nb} range, as shown in Fig.~\ref{fig:alp_limits_xsec} (right). The red curves in this plot indicate the expected ALP cross sections as a function of $m_{\Pa}$ for decreasing values of the coupling ($g_{\Pa\PGg}=0.3, 0.1,$ and $0.05\TeV^{-1}$).

The limits on the $\sigma(\gaga\to\Pa\to\gaga)$ cross sections shown in Fig.~\ref{fig:alp_limits_xsec} (right) are used to determine exclusion regions in the $g_{\Pa\PGg}$ versus $m_{\Pa}$ plane. Constraints on the ALP mass and its coupling to photons derived from accelerator and collider searches~\cite{CMS:2018erd,ATLAS:2020hii,limits_lep,limits_opal,PrimEx:2010fvg,Aloni:2019ruo,Belle-II:2020jti,BESIII:2022rzz,limits_atlas_2gamma,limits_atlas_3gamma,Knapen:2016moh,TOTEM:2021zxa,CMS:2023jgd,ATLAS:2023zfc}, beam dumps~\cite{CHARM:1985anb,Riordan:1987aw,Dolan:2017osp,Dobrich:2019dxc,NA64:2020qwq,Capozzi:2023ffu}, and supernova particle decays~\cite{Caputo:2022mah} are compared with those obtained from the current PbPb data in Fig.~\ref{fig:alp_limits_coupling}. All results assume $\mathcal{C}_{\gaga} = 1$, \ie\ a $\mathcal{B}(\Pa\to\gaga)=100\%$ branching fraction. The exclusion limits extracted here are the most stringent to date over the $m_{\Pa} = 5$\nobreakdashes--10\GeV range.

\begin{figure}[hbtp]
\centering
\includegraphics[width=0.49\textwidth]{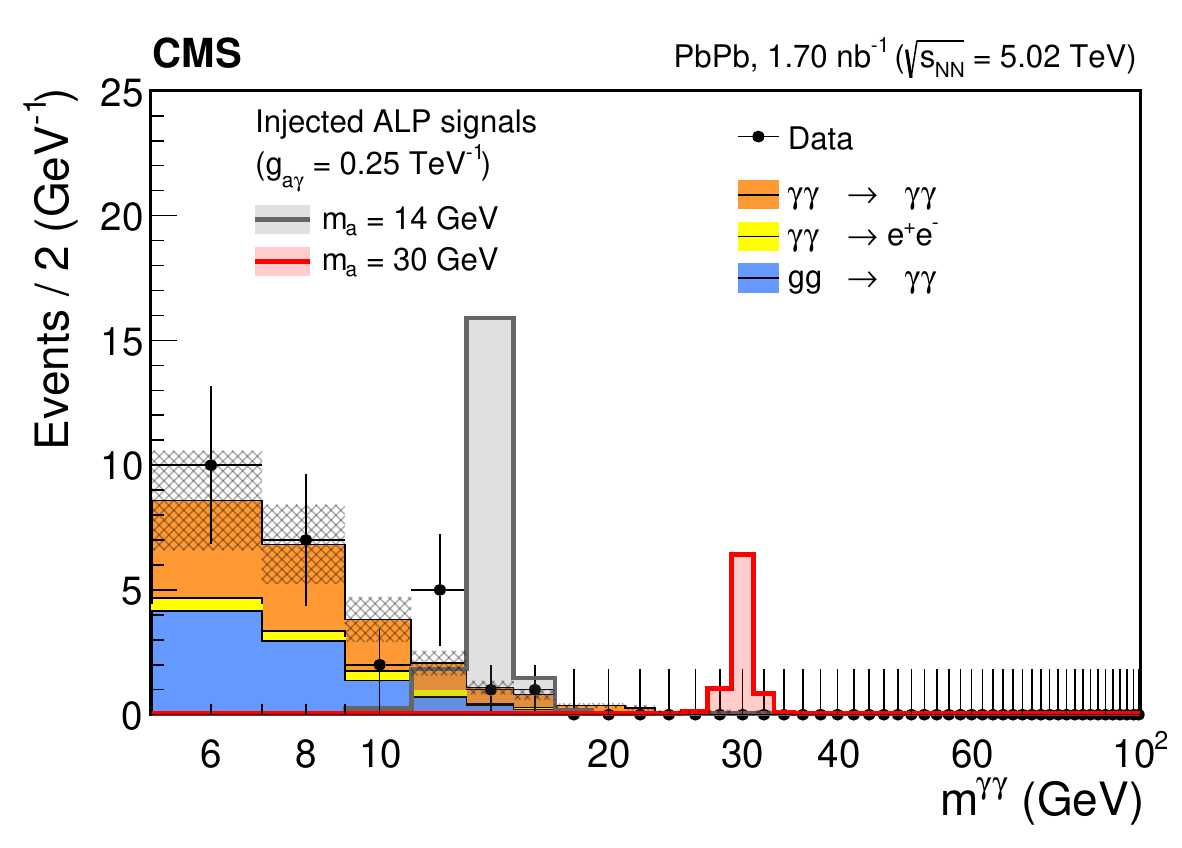}
\includegraphics[width=0.49\textwidth]{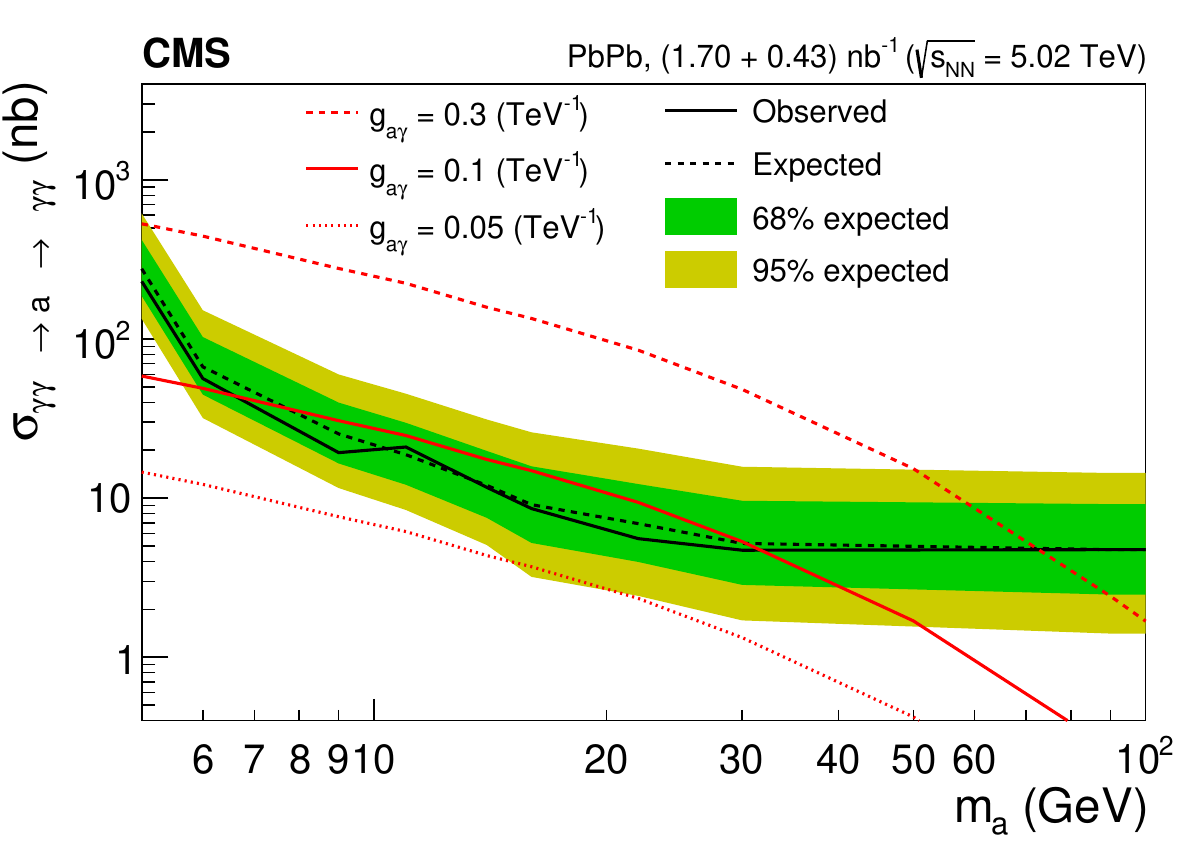}
\caption{Left: Exclusive diphoton invariant mass distribution measured in data (black points) with the expected LbL, BW, and CEP backgrounds (orange, yellow, and blue histograms, respectively), and two arbitrary ALP signals injected at masses $m_{\Pa} = 14$ and 30\GeV with $g_{\Pa\PGg}=0.25\TeV^{-1}$ (gray and red histograms, respectively).
Right: Observed (solid black line) and expected (dotted black line) 95\% \CL limits on the ALP production cross section $\sigma(\gaga\to\Pa\to\gaga)$ as a function of mass $m_{\Pa}$. The inner (green) and outer (yellow) bands indicate the regions containing 68 and 95\%, respectively, of the distribution of limits expected under the background-only hypothesis. The red curves indicate the expected ALP cross sections as a function of $m_{\Pa}$ for decreasing photon couplings ($g_{\Pa\PGg} = 0.3, 0.1, 0.05\TeV^{-1}$, upper to lower).}
\label{fig:alp_limits_xsec}
\end{figure}

\begin{figure}[hbtp]
\centering
\includegraphics[width=0.7\textwidth]{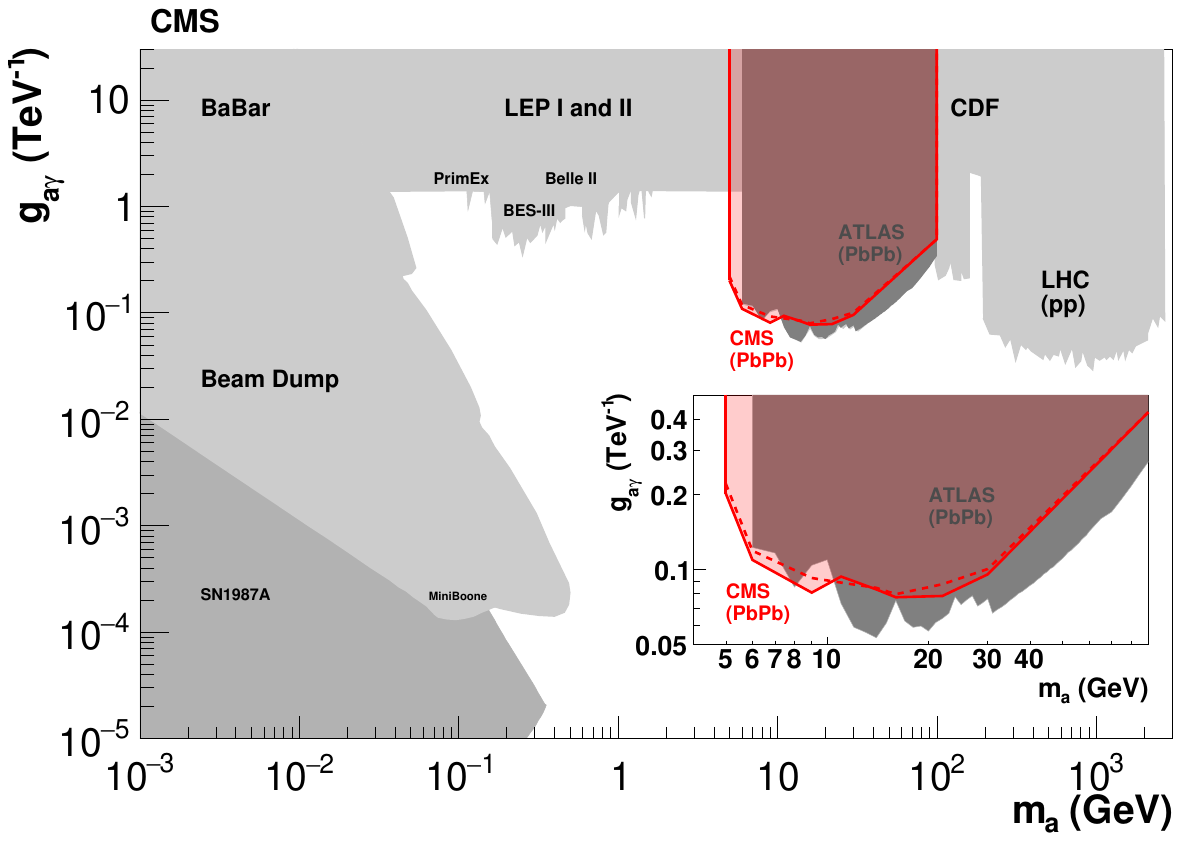}
\caption{Exclusion limits at 95\% \CL in the axion-photon coupling $g_{\Pa\PGg}$ versus axion mass $m_{\Pa}$ plane, for the operator $\frac{1}{4\Lambda}aF\widetilde{F}$ (assuming ALPs coupled only to photons) derived from multiple measurements (gray areas) compared with the  limits extracted in this analysis (red area, the corresponding expected limits are indicated with a dashed line). Previous limits have been obtained from data from LHC PbPb~\cite{CMS:2018erd,ATLAS:2020hii}, LEP~\cite{limits_lep,limits_opal,Knapen:2016moh}, PrimEx~\cite{PrimEx:2010fvg,Aloni:2019ruo}, BELLE~II~\cite{Belle-II:2020jti}, BES-III~\cite{BESIII:2022rzz}, LHC (pp)~\cite{limits_atlas_2gamma,limits_atlas_3gamma,TOTEM:2021zxa,CMS:2023jgd,ATLAS:2023zfc}, and beam dumps~\cite{CHARM:1985anb,Riordan:1987aw,Dolan:2017osp,Dobrich:2019dxc,NA64:2020qwq,Capozzi:2023ffu}, as well as from SN1987A supernova constraints~\cite{Caputo:2022mah}.
\label{fig:alp_limits_coupling}}
\end{figure}

\section{Summary}
\label{sec:summ}

Measurements of light-by-light scattering (LbL, $\gaga\to\gaga$) and the Breit--Wheeler process (BW, $\gaga\to\EE$) are reported in ultraperipheral collisions of lead ions at the LHC. The data, corresponding to an integrated luminosity of 1.70\nbinv, were collected in 2018 by the CMS experiment at a centre-of-mass energy per nucleon pair of 5.02\TeV. The LbL and BW processes are studied in events with exclusively produced $\gaga$ and $\EE$ pairs, respectively. Each reconstructed particle is required to have a transverse energy of $\et^{\PGg,\Pe}>2\GeV$, a pseudorapidity of $\abs{\eta^{\PGg,\Pe}}<2.2$, and the pairs to have an invariant mass of $m^{\gaga,\Pe\Pe}> 5\GeV$, a transverse momentum of $\pt^{\gaga,\Pe\Pe}<1\GeV$, and an azimuthal acoplanarity of $(1-\Delta \phi^{\gaga,\Pe\Pe}/\pi) <0.01$. The selected events are required to have no additional neutral particles with $\et>1\GeV$ over $\abs{\eta}<5.2$, as well as no charged particles with $\pt>0.3\GeV$ over $\abs{\eta}<2.4$.

About 20\,000 events pass the selection criteria for the BW process, and their detector-level kinematic distributions are consistent with simulated events generated with the \superchic~3.03 and \starlight~3.13 Monte Carlo (MC) codes based on quantum electrodynamics calculations at leading order. A fiducial cross section of $\sigma_\text{fid}(\gaga\to\Pe^+\Pe^-)= 263.5 \pm 1.8 \stat \pm 17.8 \syst\mub$ is measured. The BW fiducial cross section and unfolded $\EE$ transverse momentum, rapidity, and invariant mass distributions are compared with the predictions of the \starlight, \superchic, and \gammaUPC/\MGvATNLO MC event generators, including photon final-state radiation (FSR) simulated with the \photos++ or \PYTHIA8 codes. The addition of photon FSR leads to better agreement of the calculations with the measured dielectron differential distributions. The \superchic\ and \gammaUPC\ predictions, both based on the charged form factor photon flux of the lead ion, are in better agreement with the data than the \starlight\ calculations, which are based on an electric-dipole form factor. The probabilities of different multiplicities of forward neutrons emitted due to the electromagnetic excitation of the ions in the BW process are also measured, showing best agreement with the \gammaUPC\ model expectations.

In the LbL final state, 26 exclusive diphoton candidate events are observed after applying all selection criteria, compared with an expectation of 12.8 events predicted for the signal and 12.0 for the background, the latter dominated by contributions from central exclusive (gluon mediated) production scaled to the data (10.1 events) with some remaining counts from the BW process (1.9 events). Combined with previous results, the significance of the LbL signal with respect to the background-only hypothesis is above five standard deviations. The measured fiducial LbL scattering cross section, $\sigma_\text{fid}(\gaga\to\gaga) = 107 \pm 24 \stat \pm 13 \syst \unit{nb}$, is consistent with theoretical predictions at next-to-leading order accuracy. The unfolded diphoton rapidity and invariant mass differential cross sections show good agreement with the theoretical expectations.

Exploiting the measured invariant mass distribution of exclusive diphoton events, new limits on the resonant production of axion-like particles coupled to photons are set in the mass vs.\ axion-photon coupling plane. Couplings larger than $g_{\Pa\PGg} \approx 0.1\text{\nobreakdashes--}0.4\TeV^{-1}$ can be excluded over $m_{\Pa} = 5$\nobreakdashes--100\GeV, including the most stringent constraints to date in the 5--10\GeV range.

\begin{acknowledgments}
  We congratulate our colleagues in the CERN accelerator departments for the excellent performance of the LHC and thank the technical and administrative staffs at CERN and at other CMS institutes for their contributions to the success of the CMS effort. In addition, we gratefully acknowledge the computing centres and personnel of the Worldwide LHC Computing Grid and other centres for delivering so effectively the computing infrastructure essential to our analyses. Finally, we acknowledge the enduring support for the construction and operation of the LHC, the CMS detector, and the supporting computing infrastructure provided by the following funding agencies: SC (Armenia), BMBWF and FWF (Austria); FNRS and FWO (Belgium); CNPq, CAPES, FAPERJ, FAPERGS, and FAPESP (Brazil); MES and BNSF (Bulgaria); CERN; CAS, MoST, and NSFC (China); MINCIENCIAS (Colombia); MSES and CSF (Croatia); RIF (Cyprus); SENESCYT (Ecuador); ERC PRG, RVTT3 and MoER TK202 (Estonia); Academy of Finland, MEC, and HIP (Finland); CEA and CNRS/IN2P3 (France); SRNSF (Georgia); BMBF, DFG, and HGF (Germany); GSRI (Greece); NKFIH (Hungary); DAE and DST (India); IPM (Iran); SFI (Ireland); INFN (Italy); MSIP and NRF (Republic of Korea); MES (Latvia); LMTLT (Lithuania); MOE and UM (Malaysia); BUAP, CINVESTAV, CONACYT, LNS, SEP, and UASLP-FAI (Mexico); MOS (Montenegro); MBIE (New Zealand); PAEC (Pakistan); MES and NSC (Poland); FCT (Portugal); MESTD (Serbia); MCIN/AEI and PCTI (Spain); MOSTR (Sri Lanka); Swiss Funding Agencies (Switzerland); MST (Taipei); MHESI and NSTDA (Thailand); TUBITAK and TENMAK (Turkey); NASU (Ukraine); STFC (United Kingdom); DOE and NSF (USA).
  
  \hyphenation{Rachada-pisek} Individuals have received support from the Marie-Curie programme and the European Research Council and Horizon 2020 Grant, contract Nos.\ 675440, 724704, 752730, 758316, 765710, 824093, 101115353, 101002207, and COST Action CA16108 (European Union); the Leventis Foundation; the Alfred P.\ Sloan Foundation; the Alexander von Humboldt Foundation; the Science Committee, project no. 22rl-037 (Armenia); the Belgian Federal Science Policy Office; the Fonds pour la Formation \`a la Recherche dans l'Industrie et dans l'Agriculture (FRIA-Belgium); the F.R.S.-FNRS and FWO (Belgium) under the ``Excellence of Science -- EOS" -- be.h project n.\ 30820817; the Beijing Municipal Science \& Technology Commission, No. Z191100007219010 and Fundamental Research Funds for the Central Universities (China); the Ministry of Education, Youth and Sports (MEYS) of the Czech Republic; the Shota Rustaveli National Science Foundation, grant FR-22-985 (Georgia); the Deutsche Forschungsgemeinschaft (DFG), among others, under Germany's Excellence Strategy -- EXC 2121 ``Quantum Universe" -- 390833306, and under project number 400140256 - GRK2497; the Hellenic Foundation for Research and Innovation (HFRI), Project Number 2288 (Greece); the Hungarian Academy of Sciences, the New National Excellence Program - \'UNKP, the NKFIH research grants K 131991, K 133046, K 138136, K 143460, K 143477, K 146913, K 146914, K 147048, 2020-2.2.1-ED-2021-00181, TKP2021-NKTA-64, and 2021-4.1.2-NEMZ\_KI-2024-00036 (Hungary); the Council of Science and Industrial Research, India; ICSC -- National Research Centre for High Performance Computing, Big Data and Quantum Computing and FAIR -- Future Artificial Intelligence Research, funded by the NextGenerationEU program (Italy); the Latvian Council of Science; the Ministry of Education and Science, project no. 2022/WK/14, and the National Science Center, contracts Opus 2021/41/B/ST2/01369 and 2021/43/B/ST2/01552 (Poland); the Funda\c{c}\~ao para a Ci\^encia e a Tecnologia, grant CEECIND/01334/2018 (Portugal); the National Priorities Research Program by Qatar National Research Fund; MCIN/AEI/10.13039/501100011033, ERDF ``a way of making Europe", and the Programa Estatal de Fomento de la Investigaci{\'o}n Cient{\'i}fica y T{\'e}cnica de Excelencia Mar\'{\i}a de Maeztu, grant MDM-2017-0765 and Programa Severo Ochoa del Principado de Asturias (Spain); the Chulalongkorn Academic into Its 2nd Century Project Advancement Project, and the National Science, Research and Innovation Fund via the Program Management Unit for Human Resources \& Institutional Development, Research and Innovation, grant B39G670016 (Thailand); the Kavli Foundation; the Nvidia Corporation; the SuperMicro Corporation; the Welch Foundation, contract C-1845; and the Weston Havens Foundation (USA).  
\end{acknowledgments}

\bibliography{auto_generated}

\providecommand{\href}[2]{#2}\begingroup\raggedright\begin{thebibliography}{100}%
\makeatletter
\providecommand{\hrefCMSnoop }[0]{\@secondoftwo}%
\makeatother
\providecommand{\doi}{\texttt{doi:}\begingroup \urlstyle{tt}\Url}

\bibitem{vonWeizsacker:1934nji}
\hrefCMSnoop {}{C.~F. von Weizsacker, ``{Radiation emitted in collisions of
  very fast electrons}'',} \textit{ Z. Phys.} \textbf{ 88} (1934) 612,
\href{http://dx.doi.org/10.1007/BF01333110}{\doi{10.1007/BF01333110}}.

\bibitem{Williams:1934ad}
\hrefCMSnoop {}{E.~J. Williams, ``{Nature of the high-energy particles of
  penetrating radiation and status of ionization and radiation formulae}'',}
  \textit{ Phys. Rev.} \textbf{ 45} (1934) 729,
\href{http://dx.doi.org/10.1103/PhysRev.45.729}{\doi{10.1103/PhysRev.45.729}}.

\bibitem{Fermi:1925fq}
\hrefCMSnoop {}{E.~Fermi, ``{On the theory of collisions between atoms and
  electrically charged particles}'',} \textit{ Nuovo Cim.} \textbf{ 2} (1925)
  143, \href{http://dx.doi.org/10.1007/BF02961914}{\doi{10.1007/BF02961914}},
\href{http://www.arXiv.org/abs/hep-th/0205086}{\texttt{arXiv:hep-th/0205086}}.

\bibitem{Brodsky:1971ud}
\hrefCMSnoop {}{S.~J. Brodsky, T.~Kinoshita, and H.~Terazawa, ``Two photon
  mechanism of particle production by high-energy colliding beams'',} \textit{
  Phys. Rev. D} \textbf{ 4} (1971) 1532,
\href{http://dx.doi.org/10.1103/PhysRevD.4.1532}{\doi{10.1103/PhysRevD.4.1532}}.

\bibitem{Budnev:1975poe}
\hrefCMSnoop {}{V.~M. Budnev, I.~F. Ginzburg, G.~V. Meledin, and V.~G. Serbo,
  ``{The two photon particle production mechanism. Physical problems.
  Applications. Equivalent photon approximation}'',} \textit{ Phys. Rept.}
  \textbf{ 15} (1975) 181,
  \href{http://dx.doi.org/10.1016/0370-1573(75)90009-5}{\doi{10.1016/0370-1573(75)90009-5}}.

\bibitem{Vermaseren:1982cz}
\hrefCMSnoop {}{J.~A.~M. Vermaseren, ``Two photon processes at very
  high-energies'',} \textit{ Nucl. Phys. B} \textbf{ 229} (1983) 347,
  \href{http://dx.doi.org/10.1016/0550-3213(83)90336-X}{\doi{10.1016/0550-3213(83)90336-X}}.

\bibitem{Morgan:1994ip}
\hrefCMSnoop {}{D.~Morgan, M.~R. Pennington, and M.~R. Whalley, ``{A
  compilation of data on two photon reactions leading to hadron final
  states}'',} \textit{ J. Phys. G} \textbf{ 20 Suppl. 8A} (1994) A1,
  \href{http://dx.doi.org/10.1088/0954-3899/20/8A/001}{\doi{10.1088/0954-3899/20/8A/001}}.

\bibitem{Whalley:2001mk}
\hrefCMSnoop {}{M.~R. Whalley, ``A compilation of data on two photon
  reactions'',} \textit{ J. Phys. G} \textbf{ 27} (2001) A1,
  \href{http://dx.doi.org/10.1088/0954-3899/27/12A/301}{\doi{10.1088/0954-3899/27/12A/301}}.

\bibitem{Krauss:1997vr}
\hrefCMSnoop {}{F.~Krauss, M.~Greiner, and G.~Soff, ``{Photon and gluon induced
  processes in relativistic heavy ion collisions}'',} \textit{ Prog. Part.
  Nucl. Phys.} \textbf{ 39} (1997) 503,
  \href{http://dx.doi.org/10.1016/S0146-6410(97)00049-5}{\doi{10.1016/S0146-6410(97)00049-5}}.

\bibitem{Baltz:2007kq}
\hrefCMSnoop {}{A.~J. Baltz, ``The physics of ultraperipheral collisions at the
  {LHC}'',} \textit{ Phys. Rept.} \textbf{ 458} (2008) 1,
  \href{http://dx.doi.org/10.1016/j.physrep.2007.12.001}{\doi{10.1016/j.physrep.2007.12.001}},
\href{http://www.arXiv.org/abs/0706.3356}{\texttt{arXiv:0706.3356}}.

\bibitem{dEnterria:2008puz}
\hrefCMSnoop {}{D.~d'Enterria, M.~Klasen, and K.~Piotrzkowski, ``High-energy
  photon collisions at the {LHC}'',} \textit{ Nucl. Phys. B Proc. Suppl.}
  \textbf{ 179} (2008) 1,
  \href{http://dx.doi.org/10.1016/S0920-5632(08)00090-X}{\doi{10.1016/S0920-5632(08)00090-X}}.

\bibitem{deFavereaudeJeneret:2009db}
J.~de~Favereau~de Jeneret\hrefCMSnoop {}{ { et~al.}, ``{High energy photon
  interactions at the LHC}'',} 2009.
  \href{http://www.arXiv.org/abs/0908.2020}{\texttt{arXiv:0908.2020}}.

\bibitem{CMS:2024krd}
\hrefCMSnoop {}{{CMS Collaboration}, ``{Overview of high-density QCD studies
  with the CMS experiment at the LHC}'',} 2024.
  \href{http://www.arXiv.org/abs/2405.10785}{\texttt{arXiv:2405.10785}}.
  Submitted to \textit{Phys.\ Rept.}

\bibitem{d'Enterria:2013yra}
\hrefCMSnoop {}{D.~d'Enterria and G.~G. da~Silveira, ``{Observing
  light-by-light scattering at the Large Hadron Collider}'',} \textit{ Phys.
  Rev. Lett.} \textbf{ 111} (2013) 080405,
  \href{http://dx.doi.org/10.1103/PhysRevLett.111.080405}{\doi{10.1103/PhysRevLett.111.080405}},
  \href{http://www.arXiv.org/abs/1305.7142}{\texttt{arXiv:1305.7142}}.
[Erratum: \DOI{10.1103/PhysRevLett.116.129901}].

\bibitem{Bruce:2018yzs}
\hrefCMSnoop {}{R.~Bruce { et~al.}, ``{New physics searches with heavy-ion
  collisions at the CERN Large Hadron Collider}'',} \textit{ J. Phys. G}
  \textbf{ 47} (2020) 060501,
  \href{http://dx.doi.org/10.1088/1361-6471/ab7ff7}{\doi{10.1088/1361-6471/ab7ff7}},
  \href{http://www.arXiv.org/abs/1812.07688}{\texttt{arXiv:1812.07688}}.

\bibitem{dEnterria:2022sut}
\hrefCMSnoop {}{D.~d'Enterria { et~al.}, ``{Opportunities for new physics
  searches with heavy ions at colliders}'',} \textit{ J. Phys. G} \textbf{ 50}
  (2023) 050501,
  \href{http://dx.doi.org/10.1088/1361-6471/acc197}{\doi{10.1088/1361-6471/acc197}},
  \href{http://www.arXiv.org/abs/2203.05939}{\texttt{arXiv:2203.05939}}.

\bibitem{ATLAS:2017fur}
\hrefCMSnoop {}{{ATLAS Collaboration}, ``{Evidence for light-by-light
  scattering in heavy-ion collisions with the ATLAS detector at the LHC}'',}
  \textit{ Nature Phys.} \textbf{ 13} (2017) 852,
  \href{http://dx.doi.org/10.1038/nphys4208}{\doi{10.1038/nphys4208}},
  \href{http://www.arXiv.org/abs/1702.01625}{\texttt{arXiv:1702.01625}}.

\bibitem{CMS:2018erd}
\hrefCMSnoop {}{{CMS Collaboration}, ``{Evidence for light-by-light scattering
  and searches for axion-like particles in ultraperipheral PbPb collisions at
  $\sqrtsNN = 5.02$~TeV}'',} \textit{ Phys. Lett. B} \textbf{ 797} (2019)
  134826,
  \href{http://dx.doi.org/10.1016/j.physletb.2019.134826}{\doi{10.1016/j.physletb.2019.134826}},
  \href{http://www.arXiv.org/abs/1810.04602}{\texttt{arXiv:1810.04602}}.

\bibitem{Aad:2019ock}
\hrefCMSnoop {}{{ATLAS Collaboration}, ``{Observation of light-by-light
  scattering in ultraperipheral Pb+Pb collisions with the ATLAS detector}'',}
  \textit{ Phys. Rev. Lett.} \textbf{ 123} (2019) 052001,
  \href{http://dx.doi.org/10.1103/PhysRevLett.123.052001}{\doi{10.1103/PhysRevLett.123.052001}},
  \href{http://www.arXiv.org/abs/1904.03536}{\texttt{arXiv:1904.03536}}.

\bibitem{H:2023wfg}
\hrefCMSnoop {}{A.~A. H., E.~Chaubey, and H.-S. Shao, ``{Two-loop massive QCD
  and QED helicity amplitudes for light-by-light scattering}'',} \textit{ JHEP}
  \textbf{ 03} (2024) 121,
  \href{http://dx.doi.org/10.1007/JHEP03(2024)121}{\doi{10.1007/JHEP03(2024)121}},
  \href{http://www.arXiv.org/abs/2312.16966}{\texttt{arXiv:2312.16966}}.

\bibitem{H:2023znv}
A.~A. H\hrefCMSnoop {}{ { et~al.}, ``{Light-by-light scattering at
  next-to-leading order in QCD and QED}'',} \textit{ Phys. Lett. B} \textbf{
  851} (2024) 138555,
  \href{http://dx.doi.org/10.1016/j.physletb.2024.138555}{\doi{10.1016/j.physletb.2024.138555}},
  \href{http://www.arXiv.org/abs/2312.16956}{\texttt{arXiv:2312.16956}}.

\bibitem{Knapen:2016moh}
\hrefCMSnoop {}{S.~Knapen, T.~Lin, H.~K. Lou, and T.~Melia, ``Searching for
  axionlike particles with ultraperipheral heavy-ion collisions'',} \textit{
  Phys. Rev. Lett.} \textbf{ 118} (2017) 171801,
  \href{http://dx.doi.org/10.1103/PhysRevLett.118.171801}{\doi{10.1103/PhysRevLett.118.171801}},
\href{http://www.arXiv.org/abs/1607.06083}{\texttt{arXiv:1607.06083}}.

\bibitem{dEnterria:2021ljz}
\hrefCMSnoop {}{D.~d'Enterria, ``{Collider constraints on axion-like
  particles}'',} in \textit{ {Workshop on Feebly Interacting Particles}}.
\newblock 2, 2021.
\newblock
  \href{http://www.arXiv.org/abs/2102.08971}{\texttt{arXiv:2102.08971}}.

\bibitem{Sun:2014qba}
\hrefCMSnoop {}{H.~Sun, ``Large extra dimension effects through light-by-light
  scattering at the {CERN LHC}'',} \textit{ Eur. Phys. J. C} \textbf{ 74}
  (2014) 2977,
  \href{http://dx.doi.org/10.1140/epjc/s10052-014-2977-1}{\doi{10.1140/epjc/s10052-014-2977-1}},
\href{http://www.arXiv.org/abs/1406.3897}{\texttt{arXiv:1406.3897}}.

\bibitem{dEnterria:2023npy}
D.~d'Enterria\hrefCMSnoop {}{ { et~al.}, ``{Collider constraints on massive
  gravitons coupling to photons}'',} \textit{ Phys. Lett. B} \textbf{ 846}
  (2023) 138237,
  \href{http://dx.doi.org/10.1016/j.physletb.2023.138237}{\doi{10.1016/j.physletb.2023.138237}},
  \href{http://www.arXiv.org/abs/2306.15558}{\texttt{arXiv:2306.15558}}.

\bibitem{Ellis:2017edi}
\hrefCMSnoop {}{J.~Ellis, N.~E. Mavromatos, and T.~You, ``Light-by-light
  scattering constraint on {Born--Infeld} theory'',} \textit{ Phys. Rev. Lett.}
  \textbf{ 118} (2017) 261802,
  \href{http://dx.doi.org/10.1103/PhysRevLett.118.261802}{\doi{10.1103/PhysRevLett.118.261802}},
\href{http://www.arXiv.org/abs/1703.08450}{\texttt{arXiv:1703.08450}}.

\bibitem{Fichet:2014uka}
S.~Fichet\hrefCMSnoop {}{ { et~al.}, ``Light-by-light scattering with intact
  protons at the {LHC}: from standard model to new physics'',} \textit{ JHEP}
  \textbf{ 02} (2015) 165,
  \href{http://dx.doi.org/10.1007/JHEP02(2015)165}{\doi{10.1007/JHEP02(2015)165}},
\href{http://www.arXiv.org/abs/1411.6629}{\texttt{arXiv:1411.6629}}.

\bibitem{Breit:1934zz}
\hrefCMSnoop {}{G.~Breit and J.~A. Wheeler, ``{Collision of two light
  quanta}'',} \textit{ Phys. Rev.} \textbf{ 46} (1934) 1087,
  \href{http://dx.doi.org/10.1103/PhysRev.46.1087}{\doi{10.1103/PhysRev.46.1087}}.

\bibitem{Vane:1992ms}
C.~R. Vane\hrefCMSnoop {}{ { et~al.}, ``Electron positron pair production in
  {Coulomb} collisions of ultrarelativistic sulphur ions with fixed targets'',}
  \textit{ Phys. Rev. Lett.} \textbf{ 69} (1992) 1911,
  \href{http://dx.doi.org/10.1103/PhysRevLett.69.1911}{\doi{10.1103/PhysRevLett.69.1911}}.

\bibitem{CERESNA45:1994cpb}
\hrefCMSnoop {}{{CERES/NA45} Collaboration, ``{Measurement of
  electromagnetically produced {\EE} pairs in distant S-Pt collisions}'',}
  \textit{ Phys. Lett. B} \textbf{ 332} (1994) 471,
  \href{http://dx.doi.org/10.1016/0370-2693(94)91283-1}{\doi{10.1016/0370-2693(94)91283-1}}.

\bibitem{STAR:2004bzo}
\hrefCMSnoop {}{{STAR} Collaboration, ``{Production of {\EE} pairs accompanied
  by nuclear dissociation in ultra-peripheral heavy ion collision}'',} \textit{
  Phys. Rev. C} \textbf{ 70} (2004) 031902,
  \href{http://dx.doi.org/10.1103/PhysRevC.70.031902}{\doi{10.1103/PhysRevC.70.031902}},
  \href{http://www.arXiv.org/abs/nucl-ex/0404012}{\texttt{arXiv:nucl-ex/0404012}}.

\bibitem{STAR:2019wlg}
\hrefCMSnoop {}{{STAR} Collaboration, ``Measurement of {\EE} momentum and
  angular distributions from linearly polarized photon collisions'',} \textit{
  Phys. Rev. Lett.} \textbf{ 127} (2021) 052302,
  \href{http://dx.doi.org/10.1103/PhysRevLett.127.052302}{\doi{10.1103/PhysRevLett.127.052302}},
  \href{http://www.arXiv.org/abs/1910.12400}{\texttt{arXiv:1910.12400}}.

\bibitem{PHENIX:2009xtn}
\hrefCMSnoop {}{{PHENIX} Collaboration, ``{Photoproduction of $\JPsi$ and of
  high mass {\EE} in ultra-peripheral Au+Au collisions at $\sqrt{s} =
  200$~GeV}'',} \textit{ Phys. Lett. B} \textbf{ 679} (2009) 321,
  \href{http://dx.doi.org/10.1016/j.physletb.2009.07.061}{\doi{10.1016/j.physletb.2009.07.061}},
  \href{http://www.arXiv.org/abs/0903.2041}{\texttt{arXiv:0903.2041}}.

\bibitem{CDF:2006apx}
\hrefCMSnoop {}{{CDF} Collaboration, ``Observation of exclusive
  electron-positron production in hadron-hadron collisions'',} \textit{ Phys.
  Rev. Lett.} \textbf{ 98} (2007) 112001,
  \href{http://dx.doi.org/10.1103/PhysRevLett.98.112001}{\doi{10.1103/PhysRevLett.98.112001}},
  \href{http://www.arXiv.org/abs/hep-ex/0611040}{\texttt{arXiv:hep-ex/0611040}}.

\bibitem{CDF:2009xwv}
\hrefCMSnoop {}{{CDF} Collaboration, ``{Search for exclusive Z boson production
  and observation of high mass $\Pp\PAp\to\gamma\gamma\to \Pp+\ell\ell+\PAp$
  events in $\Pp\PAp$ collisions at $\sqrt{s}$ = 1.96 TeV}'',} \textit{ Phys.
  Rev. Lett.} \textbf{ 102} (2009) 222002,
  \href{http://dx.doi.org/10.1103/PhysRevLett.102.222002}{\doi{10.1103/PhysRevLett.102.222002}},
  \href{http://www.arXiv.org/abs/0902.2816}{\texttt{arXiv:0902.2816}}.

\bibitem{ALICE:2013wjo}
\hrefCMSnoop {}{{ALICE Collaboration}, ``{Charmonium and {\EE} pair
  photoproduction at mid-rapidity in ultra-peripheral Pb-Pb collisions at
  $\sqrtsNN = 2.76$~TeV}'',} \textit{ Eur. Phys. J. C} \textbf{ 73} (2013)
  2617,
  \href{http://dx.doi.org/10.1140/epjc/s10052-013-2617-1}{\doi{10.1140/epjc/s10052-013-2617-1}},
  \href{http://www.arXiv.org/abs/1305.1467}{\texttt{arXiv:1305.1467}}.

\bibitem{CMS:2012cve}
\hrefCMSnoop {}{{CMS Collaboration}, ``{Search for exclusive or semi-exclusive
  photon pair production and observation of exclusive and semi-exclusive
  electron pair production in pp collisions at $\sqrt{s}=7$~TeV}'',} \textit{
  JHEP} \textbf{ 11} (2012) 080,
  \href{http://dx.doi.org/10.1007/JHEP11(2012)080}{\doi{10.1007/JHEP11(2012)080}},
  \href{http://www.arXiv.org/abs/1209.1666}{\texttt{arXiv:1209.1666}}.

\bibitem{CMS:2018uvs}
\hrefCMSnoop {}{{CMS and TOTEM Collaborations}, ``{Observation of
  proton-tagged, central (semi)exclusive production of high-mass lepton pairs
  in pp collisions at 13 TeV with the CMS-TOTEM precision proton
  spectrometer}'',} \textit{ JHEP} \textbf{ 07} (2018) 153,
  \href{http://dx.doi.org/10.1007/JHEP07(2018)153}{\doi{10.1007/JHEP07(2018)153}},
  \href{http://www.arXiv.org/abs/1803.04496}{\texttt{arXiv:1803.04496}}.

\bibitem{ATLAS:2015wnx}
\hrefCMSnoop {}{{ATLAS Collaboration}, ``{Measurement of exclusive
  $\gamma\gamma\to \ell^+\ell^-$ production in proton-proton collisions at
  $\sqrt{s} = 7$ TeV with the ATLAS detector}'',} \textit{ Phys. Lett. B}
  \textbf{ 749} (2015) 242,
  \href{http://dx.doi.org/10.1016/j.physletb.2015.07.069}{\doi{10.1016/j.physletb.2015.07.069}},
  \href{http://www.arXiv.org/abs/1506.07098}{\texttt{arXiv:1506.07098}}.

\bibitem{ATLAS:2020mve}
\hrefCMSnoop {}{{ATLAS Collaboration}, ``Observation and measurement of forward
  proton scattering in association with lepton pairs produced via the photon
  fusion mechanism at {ATLAS}'',} \textit{ Phys. Rev. Lett.} \textbf{ 125}
  (2020) 261801,
  \href{http://dx.doi.org/10.1103/PhysRevLett.125.261801}{\doi{10.1103/PhysRevLett.125.261801}},
  \href{http://www.arXiv.org/abs/2009.14537}{\texttt{arXiv:2009.14537}}.

\bibitem{Khoze:2004ak}
\hrefCMSnoop {}{V.~A. Khoze, A.~D. Martin, M.~G. Ryskin, and W.~J. Stirling,
  ``{Diffractive $\gaga$ production at hadron colliders}'',} \textit{ Eur.
  Phys. J. C} \textbf{ 38} (2005) 475,
  \href{http://dx.doi.org/10.1140/epjc/s2004-02059-0}{\doi{10.1140/epjc/s2004-02059-0}},
\href{http://www.arXiv.org/abs/hep-ph/0409037}{\texttt{arXiv:hep-ph/0409037}}.

\bibitem{CMS:2025rzq}
\hrefCMSnoop {}{{CMS Collaboration}, ``{Luminosity measurement for lead-lead
  collisions at $\sqrtsNN = 5.02$~TeV in 2015 and 2018 at CMS}'',} 2025.
  \href{http://www.arXiv.org/abs/2503.03946}{\texttt{arXiv:2503.03946}}.

\bibitem{Baur:2003gv}
\hrefCMSnoop {}{G.~Baur, K.~Hencken, and D.~Trautmann, ``{Electromagnetic
  dissociation as a tool for nuclear structure and astrophysics}'',} \textit{
  Prog. Part. Nucl. Phys.} \textbf{ 51} (2003) 487,
  \href{http://dx.doi.org/10.1016/S0146-6410(03)90006-8}{\doi{10.1016/S0146-6410(03)90006-8}},
  \href{http://www.arXiv.org/abs/nucl-th/0304041}{\texttt{arXiv:nucl-th/0304041}}.

\bibitem{hepdata}
\hrefCMSnoop {}{}{HEPData} record for this analysis, 2024.
\newblock
  \href{http://dx.doi.org/10.17182/hepdata.155674}{\doi{10.17182/hepdata.155674}}.

\bibitem{CMS:2014pgm}
\hrefCMSnoop {}{{CMS Collaboration}, ``Description and performance of track and
  primary-vertex reconstruction with the {CMS} tracker'',} \textit{ JINST}
  \textbf{ 9} (2014) P10009,
  \href{http://dx.doi.org/10.1088/1748-0221/9/10/P10009}{\doi{10.1088/1748-0221/9/10/P10009}},
  \href{http://www.arXiv.org/abs/1405.6569}{\texttt{arXiv:1405.6569}}.

\bibitem{Grachov:2006ke}
\hrefCMSnoop {}{{CMS Collaboration}, ``{Status of zero degree calorimeter for
  CMS experiment}'',} \textit{ AIP Conf. Proc.} \textbf{ 867} (2006) 258,
  \href{http://dx.doi.org/10.1063/1.2396962}{\doi{10.1063/1.2396962}},
\href{http://www.arXiv.org/abs/nucl-ex/0608052}{\texttt{arXiv:nucl-ex/0608052}}.

\bibitem{CMS:2020cmk}
\hrefCMSnoop {}{{CMS Collaboration}, ``{Performance of the CMS Level-1 trigger
  in proton-proton collisions at $\sqrt{s} = 13$\,TeV}'',} \textit{ JINST}
  \textbf{ 15} (2020) P10017,
  \href{http://dx.doi.org/10.1088/1748-0221/15/10/P10017}{\doi{10.1088/1748-0221/15/10/P10017}},
  \href{http://www.arXiv.org/abs/2006.10165}{\texttt{arXiv:2006.10165}}.

\bibitem{CMS:2016ngn}
\hrefCMSnoop {}{{CMS Collaboration}, ``{The CMS trigger system}'',} \textit{
  JINST} \textbf{ 12} (2017) P01020,
  \href{http://dx.doi.org/10.1088/1748-0221/12/01/P01020}{\doi{10.1088/1748-0221/12/01/P01020}},
\href{http://www.arXiv.org/abs/1609.02366}{\texttt{arXiv:1609.02366}}.

\bibitem{CMS:2008xjf}
\hrefCMSnoop {}{{CMS Collaboration}, ``The {CMS} experiment at the {CERN}
  {LHC}'',} \textit{ JINST} \textbf{ 3} (2008) S08004,
  \href{http://dx.doi.org/10.1088/1748-0221/3/08/S08004}{\doi{10.1088/1748-0221/3/08/S08004}}.

\bibitem{CMS:2023gfb}
\hrefCMSnoop {}{{CMS Collaboration}, ``{Development of the CMS detector for the
  CERN LHC Run 3}'',} \textit{ JINST} \textbf{ 19} (2024) P05064,
  \href{http://dx.doi.org/10.1088/1748-0221/19/05/P05064}{\doi{10.1088/1748-0221/19/05/P05064}},
  \href{http://www.arXiv.org/abs/2309.05466}{\texttt{arXiv:2309.05466}}.

\bibitem{Klein:2016yzr}
S.~R. Klein\hrefCMSnoop {}{ { et~al.}, ``{\starlight: A Monte Carlo simulation
  program for ultra-peripheral collisions of relativistic ions}'',} \textit{
  Comput. Phys. Commun.} \textbf{ 212} (2017) 258,
  \href{http://dx.doi.org/10.1016/j.cpc.2016.10.016}{\doi{10.1016/j.cpc.2016.10.016}},
\href{http://www.arXiv.org/abs/1607.03838}{\texttt{arXiv:1607.03838}}.

\bibitem{Harland-Lang:2018iur}
\hrefCMSnoop {}{L.~A. Harland-Lang, V.~A. Khoze, and M.~G. Ryskin, ``{Exclusive
  LHC physics with heavy ions: \superchic~3}'',} \textit{ Eur. Phys. J. C}
  \textbf{ 79} (2019) 39,
  \href{http://dx.doi.org/10.1140/epjc/s10052-018-6530-5}{\doi{10.1140/epjc/s10052-018-6530-5}},
  \href{http://www.arXiv.org/abs/1810.06567}{\texttt{arXiv:1810.06567}}.

\bibitem{Davidson:2010ew}
\hrefCMSnoop {}{N.~Davidson, T.~Przedzinski, and Z.~Was, ``{PHOTOS interface in
  C++: Technical and physics documentation}'',} \textit{ Comput. Phys. Commun.}
  \textbf{ 199} (2016) 86,
  \href{http://dx.doi.org/10.1016/j.cpc.2015.09.013}{\doi{10.1016/j.cpc.2015.09.013}},
  \href{http://www.arXiv.org/abs/1011.0937}{\texttt{arXiv:1011.0937}}.

\bibitem{Shao:2022cly}
\hrefCMSnoop {}{H.-S. Shao and D.~d'Enterria, ``{gamma-UPC: automated
  generation of exclusive photon-photon processes in ultraperipheral proton and
  nuclear collisions with varying form factors}'',} \textit{ JHEP} \textbf{ 09}
  (2022) 248,
  \href{http://dx.doi.org/10.1007/JHEP09(2022)248}{\doi{10.1007/JHEP09(2022)248}},
  \href{http://www.arXiv.org/abs/2207.03012}{\texttt{arXiv:2207.03012}}.

\bibitem{Shao:2024dmk}
\hrefCMSnoop {}{H.-S. Shao and D.~d'Enterria, ``Dimuon and ditau production in
  photon-photon collisions at next-to-leading order in {QED}'',} \textit{ JHEP}
  \textbf{ 02} (2025) 023,
  \href{http://dx.doi.org/10.1007/JHEP02(2025)023}{\doi{10.1007/JHEP02(2025)023}},
  \href{http://www.arXiv.org/abs/2407.13610}{\texttt{arXiv:2407.13610}}.

\bibitem{MadGraph5}
J.~Alwall\hrefCMSnoop {}{ { et~al.}, ``The automated computation of tree-level
  and next-to-leading order differential cross sections, and their matching to
  parton shower simulations'',} \textit{ JHEP} \textbf{ 07} (2014) 079,
  \href{http://dx.doi.org/10.1007/JHEP07(2014)079}{\doi{10.1007/JHEP07(2014)079}},
  \href{http://www.arXiv.org/abs/1405.0301}{\texttt{arXiv:1405.0301}}.

\bibitem{Sjostrand:2014zea}
T.~Sj{\"o}strand\hrefCMSnoop {}{ { et~al.}, ``{An introduction to PYTHIA
  8.2}'',} \textit{ Comput. Phys. Commun.} \textbf{ 191} (2015) 159,
  \href{http://dx.doi.org/10.1016/j.cpc.2015.01.024}{\doi{10.1016/j.cpc.2015.01.024}},
  \href{http://www.arXiv.org/abs/1410.3012}{\texttt{arXiv:1410.3012}}.

\bibitem{Loizides:2017ack}
\hrefCMSnoop {}{C.~Loizides, J.~Kamin, and D.~d'Enterria, ``{Improved Monte
  Carlo Glauber predictions at present and future nuclear colliders}'',}
  \textit{ Phys. Rev. C} \textbf{ 97} (2018) 054910,
  \href{http://dx.doi.org/10.1103/PhysRevC.97.054910}{\doi{10.1103/PhysRevC.97.054910}},
  \href{http://www.arXiv.org/abs/1710.07098}{\texttt{arXiv:1710.07098}}.
  [Erratum: \DOI{10.1103/PhysRevC.99.019901}].

\bibitem{dEnterria:2020dwq}
\hrefCMSnoop {}{D.~d'Enterria and C.~Loizides, ``{Progress in the Glauber model
  at collider energies}'',} \textit{ Ann. Rev. Nucl. Part. Sci.} \textbf{ 71}
  (2021) 315,
  \href{http://dx.doi.org/10.1146/annurev-nucl-102419-060007}{\doi{10.1146/annurev-nucl-102419-060007}},
  \href{http://www.arXiv.org/abs/2011.14909}{\texttt{arXiv:2011.14909}}.

\bibitem{Agostinelli:2002hh}
\hrefCMSnoop {}{{GEANT4} Collaboration, ``{\GEANTfour}---a simulation
  toolkit'',} \textit{ Nucl. Instrum. Meth. A} \textbf{ 506} (2003) 250,
\href{http://dx.doi.org/10.1016/S0168-9002(03)01368-8}{\doi{10.1016/S0168-9002(03)01368-8}}.

\bibitem{CMS:2018wqs}
\hrefCMSnoop {}{{CMS Collaboration}, ``Precision measurement of the structure
  of the {CMS} inner tracking system using nuclear interactions'',} \textit{
  JINST} \textbf{ 13} (2018) P10034,
  \href{http://dx.doi.org/10.1088/1748-0221/13/10/P10034}{\doi{10.1088/1748-0221/13/10/P10034}},
  \href{http://www.arXiv.org/abs/1807.03289}{\texttt{arXiv:1807.03289}}.

\bibitem{Sirunyan:2017ulk}
\hrefCMSnoop {}{{CMS Collaboration}, ``{Particle-flow reconstruction and global
  event description with the CMS detector}'',} \textit{ JINST} \textbf{ 12}
  (2017) P10003,
  \href{http://dx.doi.org/10.1088/1748-0221/12/10/P10003}{\doi{10.1088/1748-0221/12/10/P10003}},
\href{http://www.arXiv.org/abs/1706.04965}{\texttt{arXiv:1706.04965}}.

\bibitem{CMS:2020uim}
\hrefCMSnoop {}{{CMS Collaboration}, ``Electron and photon reconstruction and
  identification with the {CMS} experiment at the {CERN} {LHC}'',} \textit{
  JINST} \textbf{ 16} (2021) P05014,
  \href{http://dx.doi.org/10.1088/1748-0221/16/05/P05014}{\doi{10.1088/1748-0221/16/05/P05014}},
  \href{http://www.arXiv.org/abs/2012.06888}{\texttt{arXiv:2012.06888}}.

\bibitem{CMS-DP-2022-006}
\href {https://cds.cern.ch/record/2806501}{{CMS Collaboration}, ``{Performance
  of low-$E_{\text{T}}$ electrons and photons using 2018 ultraperipheral
  PbPb}'',} CMS Detector Performance Note CMS-DP-2022-006, 2022.

\bibitem{CMS-DP-2020-021}
\href {https://cds.cern.ch/record/2717925}{{CMS Collaboration}, ``{ECAL} 2016
  refined calibration and {Run 2} summary plots'',} CMS Detector Performance
  Note CMS-DP-2020-021, 2020.

\bibitem{CMS:2010svw}
\hrefCMSnoop {}{{CMS Collaboration}, ``Measurements of inclusive {W} and {Z}
  cross sections in pp collisions at $\sqrt{s}=7$~{TeV}'',} \textit{ JHEP}
  \textbf{ 01} (2011) 080,
  \href{http://dx.doi.org/10.1007/JHEP01(2011)080}{\doi{10.1007/JHEP01(2011)080}},
  \href{http://www.arXiv.org/abs/1012.2466}{\texttt{arXiv:1012.2466}}.

\bibitem{CMS:2020skx}
\hrefCMSnoop {}{{CMS Collaboration}, ``{Observation of forward neutron
  multiplicity dependence of dimuon acoplanarity in ultraperipheral PbPb
  collisions at $\sqrtsNN = 5.02$~TeV}'',} \textit{ Phys. Rev. Lett.} \textbf{
  127} (2021) 122001,
  \href{http://dx.doi.org/10.1103/PhysRevLett.127.122001}{\doi{10.1103/PhysRevLett.127.122001}},
  \href{http://www.arXiv.org/abs/2011.05239}{\texttt{arXiv:2011.05239}}.

\bibitem{Crepet:2024}
\hrefCMSnoop {}{N.~Crepet, D.~d'Enterria, and H.-S. Shao, ``Improved modeling
  of $\gaga$ processes in ultraperipheral collisions at hadron colliders'',} in
  \textit{ {DIS2024: XXXI International Workshop on Deep Inelastic Scattering
  and Related Subjects, Grenoble, France, April 8-12, 2024}}.
\newblock 9, 2024.
\newblock
  \href{http://www.arXiv.org/abs/2409.18485}{\texttt{arXiv:2409.18485}}.

\bibitem{Collins:1977iv}
\hrefCMSnoop {}{J.~C. Collins and D.~E. Soper, ``Angular distribution of
  dileptons in high-energy hadron collisions'',} \textit{ Phys. Rev. D}
  \textbf{ 16} (1977) 2219,
  \href{http://dx.doi.org/10.1103/PhysRevD.16.2219}{\doi{10.1103/PhysRevD.16.2219}}.

\bibitem{CMS:2023snh}
\hrefCMSnoop {}{{CMS Collaboration}, ``{Probing small Bjorken-$x$ nuclear
  gluonic structure via coherent J/$\psi$ photoproduction in ultraperipheral
  PbPb collisions at $\sqrtsNN = 5.02$~TeV}'',} \textit{ Phys. Rev. Lett.}
  \textbf{ 131} (2023) 262301,
  \href{http://dx.doi.org/10.1103/PhysRevLett.131.262301}{\doi{10.1103/PhysRevLett.131.262301}},
  \href{http://www.arXiv.org/abs/2303.16984}{\texttt{arXiv:2303.16984}}.

\bibitem{Harland-Lang:2023ohq}
\hrefCMSnoop {}{L.~A. Harland-Lang, ``{Exciting ions: A systematic treatment of
  ultraperipheral heavy ion collisions with nuclear breakup}'',} \textit{ Phys.
  Rev. D} \textbf{ 107} (2023) 093004,
  \href{http://dx.doi.org/10.1103/PhysRevD.107.093004}{\doi{10.1103/PhysRevD.107.093004}},
  \href{http://www.arXiv.org/abs/2303.04826}{\texttt{arXiv:2303.04826}}.

\bibitem{Adye:2011gm}
\hrefCMSnoop {}{T.~Adye, ``Unfolding algorithms and tests using {RooUnfold}'',}
  in \textit{ {PHYSTAT 2011}}, p.~313.
\newblock CERN, Geneva, 2011.
\newblock \href{http://www.arXiv.org/abs/1105.1160}{\texttt{arXiv:1105.1160}}.
\newblock
  \href{http://dx.doi.org/10.5170/CERN-2011-006.313}{\doi{10.5170/CERN-2011-006.313}}.

\bibitem{DAgostini:1994fjx}
\hrefCMSnoop {}{G.~D'Agostini, ``{A multidimensional unfolding method based on
  Bayes' theorem}'',} \textit{ Nucl. Instrum. Meth. A} \textbf{ 362} (1995)
  487,
  \href{http://dx.doi.org/10.1016/0168-9002(95)00274-X}{\doi{10.1016/0168-9002(95)00274-X}}.

\bibitem{Cowan:2010js}
\hrefCMSnoop {}{G.~Cowan, K.~Cranmer, E.~Gross, and O.~Vitells, ``Asymptotic
  formulae for likelihood-based tests of new physics'',} \textit{ Eur. Phys. J.
  C} \textbf{ 71} (2011) 1554,
  \href{http://dx.doi.org/10.1140/epjc/s10052-011-1554-0}{\doi{10.1140/epjc/s10052-011-1554-0}},
  \href{http://www.arXiv.org/abs/1007.1727}{\texttt{arXiv:1007.1727}}.
[Erratum: \DOI{10.1140/epjc/s10052-013-2501-z}].

\bibitem{Barlow:1993dm}
\hrefCMSnoop {}{R.~J. Barlow and C.~Beeston, ``{Fitting using finite Monte
  Carlo samples}'',} \textit{ Comput. Phys. Commun.} \textbf{ 77} (1993) 219,
\href{http://dx.doi.org/10.1016/0010-4655(93)90005-W}{\doi{10.1016/0010-4655(93)90005-W}}.

\bibitem{Conway:2011in}
\hrefCMSnoop {}{J.~S. Conway, ``Incorporating nuisance parameters in
  likelihoods for multisource spectra'',} in \textit{ {Proceedings, PHYSTAT
  2011 Workshop on Statistical Issues Related to Discovery Claims in Search
  Experiments and Unfolding, CERN,Geneva, Switzerland 17-20 January 2011}},
  p.~115.
\newblock 2011.
\newblock \href{http://www.arXiv.org/abs/1103.0354}{\texttt{arXiv:1103.0354}}.
\newblock
\href{http://dx.doi.org/10.5170/CERN-2011-006.115}{\doi{10.5170/CERN-2011-006.115}}.

\bibitem{CMS:2024onh}
\hrefCMSnoop {}{{CMS Collaboration}, ``The {CMS} statistical analysis and
  combination tool: {\textsc{Combine}}'',} \textit{ Comput. Softw. Big Sci.}
  \textbf{ 8} (2024) 19,
  \href{http://dx.doi.org/10.1007/s41781-024-00121-4}{\doi{10.1007/s41781-024-00121-4}},
  \href{http://www.arXiv.org/abs/2404.06614}{\texttt{arXiv:2404.06614}}.

\bibitem{Peccei:1977hh}
\hrefCMSnoop {}{R.~D. Peccei and H.~R. Quinn, ``{CP conservation in the
  presence of instantons}'',} \textit{ Phys. Rev. Lett.} \textbf{ 38} (1977)
  1440,
  \href{http://dx.doi.org/10.1103/PhysRevLett.38.1440}{\doi{10.1103/PhysRevLett.38.1440}}.

\bibitem{Agrawal:2021dbo}
\hrefCMSnoop {}{P.~Agrawal { et~al.}, ``{Feebly-interacting particles: FIPs
  2020 workshop report}'',} \textit{ Eur. Phys. J. C} \textbf{ 81} (2021) 1015,
  \href{http://dx.doi.org/10.1140/epjc/s10052-021-09703-7}{\doi{10.1140/epjc/s10052-021-09703-7}},
  \href{http://www.arXiv.org/abs/2102.12143}{\texttt{arXiv:2102.12143}}.

\bibitem{CLS2}
\hrefCMSnoop {}{T.~Junk, ``Confidence level computation for combining searches
  with small statistics'',} \textit{ Nucl. Instrum. Meth. A} \textbf{ 434}
  (1999) 435,
  \href{http://dx.doi.org/10.1016/S0168-9002(99)00498-2}{\doi{10.1016/S0168-9002(99)00498-2}},
\href{http://www.arXiv.org/abs/hep-ex/9902006}{\texttt{arXiv:hep-ex/9902006}}.

\bibitem{CLS1}
\hrefCMSnoop {}{A.~L. Read, ``Presentation of search results: The
  {CL$_{\text{s}}$} technique'',} \textit{ J. Phys. G} \textbf{ 28} (2002)
  2693,
\href{http://dx.doi.org/10.1088/0954-3899/28/10/313}{\doi{10.1088/0954-3899/28/10/313}}.

\bibitem{ATLAS:2011tau}
\href {http://cdsweb.cern.ch/record/1379837}{{ATLAS and CMS Collaborations, LHC
  Higgs Combination Group}, ``Procedure for the {LHC Higgs} boson search
  combination in {S}ummer 2011'',} {CMS-NOTE-2011-005; ATL-PHYS-PUB-2011-11},
  2011.

\bibitem{ATLAS:2020hii}
\hrefCMSnoop {}{{ATLAS Collaboration}, ``{Measurement of light-by-light
  scattering and search for axion-like particles with 2.2 nb$^{-1}$ of Pb+Pb
  data with the ATLAS detector}'',} \textit{ JHEP} \textbf{ 11} (2021) 050,
  \href{http://dx.doi.org/10.1007/JHEP11(2021)050}{\doi{10.1007/JHEP11(2021)050}},
  \href{http://www.arXiv.org/abs/2008.05355}{\texttt{arXiv:2008.05355}}.

\bibitem{limits_lep}
\hrefCMSnoop {}{J.~Jaeckel and M.~Spannowsky, ``Probing {MeV} to 90 {GeV}
  axion-like particles with {LEP} and {LHC}'',} \textit{ Phys. Lett. B}
  \textbf{ 753} (2016) 482,
  \href{http://dx.doi.org/10.1016/j.physletb.2015.12.037}{\doi{10.1016/j.physletb.2015.12.037}},
\href{http://www.arXiv.org/abs/1509.00476}{\texttt{arXiv:1509.00476}}.

\bibitem{limits_opal}
\hrefCMSnoop {}{{OPAL} Collaboration, ``Multiphoton production in {\EE}
  collisions at $\sqrt{s}$ = 181 {GeV} to 209 {GeV}'',} \textit{ Eur. Phys. J.
  C} \textbf{ 26} (2003) 331,
  \href{http://dx.doi.org/10.1140/epjc/s2002-01074-5}{\doi{10.1140/epjc/s2002-01074-5}},
\href{http://www.arXiv.org/abs/hep-ex/0210016}{\texttt{arXiv:hep-ex/0210016}}.

\bibitem{PrimEx:2010fvg}
\hrefCMSnoop {}{{PrimEx} Collaboration, ``A new measurement of the $\pi^0$
  radiative decay width'',} \textit{ Phys. Rev. Lett.} \textbf{ 106} (2011)
  162303,
  \href{http://dx.doi.org/10.1103/PhysRevLett.106.162303}{\doi{10.1103/PhysRevLett.106.162303}},
  \href{http://www.arXiv.org/abs/1009.1681}{\texttt{arXiv:1009.1681}}.

\bibitem{Aloni:2019ruo}
\hrefCMSnoop {}{D.~Aloni, C.~Fanelli, Y.~Soreq, and M.~Williams,
  ``Photoproduction of axionlike particles'',} \textit{ Phys. Rev. Lett.}
  \textbf{ 123} (2019) 071801,
  \href{http://dx.doi.org/10.1103/PhysRevLett.123.071801}{\doi{10.1103/PhysRevLett.123.071801}},
  \href{http://www.arXiv.org/abs/1903.03586}{\texttt{arXiv:1903.03586}}.

\bibitem{Belle-II:2020jti}
\hrefCMSnoop {}{{Belle-II} Collaboration, ``{Search for axion-like particles
  produced in {\EE} collisions at Belle~II}'',} \textit{ Phys. Rev. Lett.}
  \textbf{ 125} (2020) 161806,
  \href{http://dx.doi.org/10.1103/PhysRevLett.125.161806}{\doi{10.1103/PhysRevLett.125.161806}},
  \href{http://www.arXiv.org/abs/2007.13071}{\texttt{arXiv:2007.13071}}.

\bibitem{BESIII:2022rzz}
\hrefCMSnoop {}{{BESIII} Collaboration, ``{Search for an axion-like particle in
  radiative J/\ensuremath{\psi} decays}'',} \textit{ Phys. Lett. B} \textbf{
  838} (2023) 137698,
  \href{http://dx.doi.org/10.1016/j.physletb.2023.137698}{\doi{10.1016/j.physletb.2023.137698}},
  \href{http://www.arXiv.org/abs/2211.12699}{\texttt{arXiv:2211.12699}}.

\bibitem{limits_atlas_2gamma}
\hrefCMSnoop {}{{ATLAS Collaboration}, ``Search for scalar diphoton resonances
  in the mass range 65--600 {GeV} with the {ATLAS} detector in pp collision
  data at $\sqrt{s}$ = 8 {TeV}'',} \textit{ Phys. Rev. Lett.} \textbf{ 113}
  (2014) 171801,
  \href{http://dx.doi.org/10.1103/PhysRevLett.113.171801}{\doi{10.1103/PhysRevLett.113.171801}},
\href{http://www.arXiv.org/abs/1407.6583}{\texttt{arXiv:1407.6583}}.

\bibitem{limits_atlas_3gamma}
\hrefCMSnoop {}{{ATLAS Collaboration}, ``Search for new phenomena in events
  with at least three photons collected in pp collisions at $\sqrt{s}$ = 8
  {TeV} with the {ATLAS} detector'',} \textit{ Eur. Phys. J. C} \textbf{ 76}
  (2016) 210,
  \href{http://dx.doi.org/10.1140/epjc/s10052-016-4034-8}{\doi{10.1140/epjc/s10052-016-4034-8}},
\href{http://www.arXiv.org/abs/1509.05051}{\texttt{arXiv:1509.05051}}.

\bibitem{TOTEM:2021zxa}
\hrefCMSnoop {}{{CMS and TOTEM Collaborations}, ``First search for exclusive
  diphoton production at high mass with tagged protons in proton-proton
  collisions at $\sqrt{s} = 13$~{TeV}'',} \textit{ Phys. Rev. Lett.} \textbf{
  129} (2022) 011801,
  \href{http://dx.doi.org/10.1103/PhysRevLett.129.011801}{\doi{10.1103/PhysRevLett.129.011801}},
  \href{http://www.arXiv.org/abs/2110.05916}{\texttt{arXiv:2110.05916}}.

\bibitem{CMS:2023jgd}
\hrefCMSnoop {}{{CMS and TOTEM Collaborations}, ``{Search for high-mass
  exclusive diphoton production with tagged protons in proton-proton collisions
  at $\sqrt{s}$ = 13 TeV}'',} \textit{ Phys. Rev. D} \textbf{ 110} (2024)
  012010,
  \href{http://dx.doi.org/10.1103/PhysRevD.110.012010}{\doi{10.1103/PhysRevD.110.012010}},
  \href{http://www.arXiv.org/abs/2311.02725}{\texttt{arXiv:2311.02725}}.

\bibitem{ATLAS:2023zfc}
\hrefCMSnoop {}{{ATLAS Collaboration}, ``{Search for an axion-like particle
  with forward proton scattering in association with photon pairs at ATLAS}'',}
  \textit{ JHEP} \textbf{ 07} (2023) 234,
  \href{http://dx.doi.org/10.1007/JHEP07(2023)234}{\doi{10.1007/JHEP07(2023)234}},
  \href{http://www.arXiv.org/abs/2304.10953}{\texttt{arXiv:2304.10953}}.

\bibitem{CHARM:1985anb}
\hrefCMSnoop {}{{CHARM} Collaboration, ``Search for axion-like particle
  production in 400~{GeV} proton-copper interactions'',} \textit{ Phys. Lett.
  B} \textbf{ 157} (1985) 458,
  \href{http://dx.doi.org/10.1016/0370-2693(85)90400-9}{\doi{10.1016/0370-2693(85)90400-9}}.

\bibitem{Riordan:1987aw}
\hrefCMSnoop {}{E.~M. Riordan { et~al.}, ``A search for short lived axions in
  an electron beam dump experiment'',} \textit{ Phys. Rev. Lett.} \textbf{ 59}
  (1987) 755,
  \href{http://dx.doi.org/10.1103/PhysRevLett.59.755}{\doi{10.1103/PhysRevLett.59.755}}.

\bibitem{Dolan:2017osp}
M.~J. Dolan\hrefCMSnoop {}{ { et~al.}, ``{Revised constraints and Belle II
  sensitivity for visible and invisible axion-like particles}'',} \textit{
  JHEP} \textbf{ 12} (2017) 094,
  \href{http://dx.doi.org/10.1007/JHEP12(2017)094}{\doi{10.1007/JHEP12(2017)094}},
  \href{http://www.arXiv.org/abs/1709.00009}{\texttt{arXiv:1709.00009}}.
  [Erratum: \DOI{10.1007/JHEP03(2021)190}.

\bibitem{Dobrich:2019dxc}
\hrefCMSnoop {}{B.~D{\"o}brich, J.~Jaeckel, and T.~Spadaro, ``{Light in the
  beam dump. Axion-Like Particle production from decay photons in proton
  beam-dumps}'',} \textit{ JHEP} \textbf{ 05} (2019) 213,
  \href{http://dx.doi.org/10.1007/JHEP05(2019)213}{\doi{10.1007/JHEP05(2019)213}},
  \href{http://www.arXiv.org/abs/1904.02091}{\texttt{arXiv:1904.02091}}.
  [Erratum: \DOI{10.1007/JHEP10(2020)046}.

\bibitem{NA64:2020qwq}
\hrefCMSnoop {}{{NA64} Collaboration, ``Search for axionlike and scalar
  particles with the {NA64} experiment'',} \textit{ Phys. Rev. Lett.} \textbf{
  125} (2020) 081801,
  \href{http://dx.doi.org/10.1103/PhysRevLett.125.081801}{\doi{10.1103/PhysRevLett.125.081801}},
  \href{http://www.arXiv.org/abs/2005.02710}{\texttt{arXiv:2005.02710}}.

\bibitem{Capozzi:2023ffu}
F.~Capozzi\hrefCMSnoop {}{ { et~al.}, ``{New constraints on ALP couplings to
  electrons and photons from ArgoNeuT and the MiniBooNE beam dump}'',} \textit{
  Phys. Rev. D} \textbf{ 108} (2023) 075019,
  \href{http://dx.doi.org/10.1103/PhysRevD.108.075019}{\doi{10.1103/PhysRevD.108.075019}},
  \href{http://www.arXiv.org/abs/2307.03878}{\texttt{arXiv:2307.03878}}.

\bibitem{Caputo:2022mah}
\hrefCMSnoop {}{A.~Caputo, H.-T. Janka, G.~Raffelt, and E.~Vitagliano,
  ``Low-energy supernovae severely constrain radiative particle decays'',}
  \textit{ Phys. Rev. Lett.} \textbf{ 128} (2022) 221103,
  \href{http://dx.doi.org/10.1103/PhysRevLett.128.221103}{\doi{10.1103/PhysRevLett.128.221103}},
  \href{http://www.arXiv.org/abs/2201.09890}{\texttt{arXiv:2201.09890}}.

\end{thebibliography}\endgroup

\cleardoublepage \appendix\section{The CMS Collaboration \label{app:collab}}\begin{sloppypar}\hyphenpenalty=5000\widowpenalty=500\clubpenalty=5000
\cmsinstitute{Yerevan Physics Institute, Yerevan, Armenia}
{\tolerance=6000
A.~Hayrapetyan, A.~Tumasyan\cmsAuthorMark{1}\cmsorcid{0009-0000-0684-6742}
\par}
\cmsinstitute{Institut f\"{u}r Hochenergiephysik, Vienna, Austria}
{\tolerance=6000
W.~Adam\cmsorcid{0000-0001-9099-4341}, J.W.~Andrejkovic, T.~Bergauer\cmsorcid{0000-0002-5786-0293}, S.~Chatterjee\cmsorcid{0000-0003-2660-0349}, K.~Damanakis\cmsorcid{0000-0001-5389-2872}, M.~Dragicevic\cmsorcid{0000-0003-1967-6783}, P.S.~Hussain\cmsorcid{0000-0002-4825-5278}, M.~Jeitler\cmsAuthorMark{2}\cmsorcid{0000-0002-5141-9560}, N.~Krammer\cmsorcid{0000-0002-0548-0985}, A.~Li\cmsorcid{0000-0002-4547-116X}, D.~Liko\cmsorcid{0000-0002-3380-473X}, I.~Mikulec\cmsorcid{0000-0003-0385-2746}, J.~Schieck\cmsAuthorMark{2}\cmsorcid{0000-0002-1058-8093}, R.~Sch\"{o}fbeck\cmsorcid{0000-0002-2332-8784}, D.~Schwarz\cmsorcid{0000-0002-3821-7331}, M.~Sonawane\cmsorcid{0000-0003-0510-7010}, W.~Waltenberger\cmsorcid{0000-0002-6215-7228}, C.-E.~Wulz\cmsAuthorMark{2}\cmsorcid{0000-0001-9226-5812}
% ...
\par}
\vskip\cmsinstskip
\dag:~Deceased\\
$^{1}$Also at Yerevan State University, Yerevan, Armenia\\
$^{2}$Also at TU Wien, Vienna, Austria\\
% ...
\end{sloppypar}
\end{document}